\documentclass[ALICE,manyauthors]{cernphprep}

\usepackage[comma,square,numbers,sort&compress]{natbib}
\usepackage{hyperref}
\usepackage{lineno}

\modulolinenumbers[5]

\newlength{\mylength}
\usepackage{bm}
\usepackage{graphicx}
\usepackage{dcolumn}
\usepackage{amssymb}
\usepackage{xspace}
\usepackage[normalem]{ulem}
\usepackage{color}

\begin{document}

\newcommand{\pbar}{\ensuremath{\bar{p}}\xspace}
\newcommand{\pp}{\ensuremath{\rm pp}\xspace}
\newcommand{\ks}{\ensuremath{\rm K_{S}^{0}}\xspace}
\newcommand{\lmd}{\ensuremath{\Lambda}\xspace}
\newcommand{\cme}{\ensuremath{\sqrt{s}}\xspace}
\newcommand{\bg}{\ensuremath{\beta\gamma}\xspace}
\newcommand{\pt}{\ensuremath{p_{\rm{T}}}\xspace}
\newcommand{\ptjet}{\ensuremath{p_{\rm{T,jet}}}\xspace}
\newcommand{\ptjetch}{\ensuremath{p_{\rm{T,ch.\ jet}}}\xspace}
\newcommand{\p}{\ensuremath{p}\xspace}
\newcommand{\dndeta}{\ensuremath{{\rm d}N/{\rm d}\eta}\xspace}
\newcommand{\snnt}[1]{\ensuremath{\sqrt{s_{\rm NN}} = #1 \text{\,TeV}}\xspace}
\newcommand{\snng}[1]{\ensuremath{\sqrt{s_{\rm NN}} = #1 \text{\,GeV}}\xspace}
\newcommand{\sppt}[1]{\ensuremath{\sqrt{s} = #1 \text{\,TeV}}\xspace}
\newcommand{\sppg}[1]{\ensuremath{\sqrt{s} = #1 \text{\,GeV}}\xspace}
\newcommand{\gevc}[1]{\ensuremath{#1 \text{\,GeV/$c$}}\xspace}
\newcommand{\mevcsq}[1]{\ensuremath{#1 \text{\,MeV/$c^2$}}\xspace}
\newcommand{\ppb}{p--Pb\xspace}
\newcommand{\pbpb}{Pb--Pb\xspace}
\newcommand{\ee}{e$^+$e$^-$\xspace}
\newcommand{\aaaa}{A--A\xspace}
\newcommand{\auau}{Au--Au\xspace}
 
\newcommand{\twotwo}{\ensuremath{2\rightarrow 2}\xspace}
\newcommand{\raa}{\ensuremath{R_{\rm AA}}\xspace}
\newcommand{\raas}{\ensuremath{R_{\rm AA}}s\xspace}
\newcommand{\dedx}{\ensuremath{{\rm d}E/{\rm d}x}\xspace}
\newcommand{\bolddedx}{\ensuremath{\mathbf{{\rm d}E/{\rm d}x}}\xspace}
\newcommand{\mdedx}{\ensuremath{\langle {\rm d}E/{\rm d}x \rangle}\xspace}

\newcommand{\pikandp}{\ensuremath{\pi^{\pm}, \rm K^{\pm},\text{and } \rm p(\bar{p})}\xspace}
\newcommand{\ptopi}{\ensuremath{({\rm p+ \bar{p}}) / (\pi^{+}+\pi^{-})}\xspace}
\newcommand{\ktopi}{\ensuremath{({\rm K}^{+}+{\rm K}^{-}) / (\pi^{+}+\pi^{-})}\xspace}
\newcommand{\chpi}{\ensuremath{\pi^{+}+\pi^{-}}\xspace}
\newcommand{\chk}{\ensuremath{{\rm K}^{+}+{\rm K}^{-}}\xspace}
\newcommand{\chp}{\ensuremath{{\rm p}+{\rm \bar{p}}}\xspace}

\newcommand{\mthetac}{\ensuremath{\langle \theta_{\rm{Ch}} \rangle}\xspace}
\newcommand{\thetac}{\ensuremath{\theta_{\rm{Ch}}}\xspace}
\newcommand{\ie}{{i.e.}}

\begin{titlepage}
\PHyear{2015}
\PHnumber{152}      
\PHdate{18 June}  

\title{Centrality dependence of the nuclear modification factor of charged
  pions, kaons, and protons in Pb-Pb collisions at $\mathbf{\sqrt{s_{\rm NN}}=2.76}$ TeV}
        
\ShortTitle{Nuclear modification factor of charged pions, kaons, and protons}   

\Collaboration{ALICE Collaboration\thanks{See Appendix~\ref{app:collab} for the list of collaboration members}}
\ShortAuthor{ALICE Collaboration} 

\begin{abstract}
Transverse momentum ($p_{\rm{T}}$) spectra of pions, kaons, and protons up to
$p_{\rm{T}} = 20 \text{\,GeV/$c$}$ have been measured in Pb--Pb collisions at
$\sqrt{s_{\rm NN}} = 2.76 \text{\,TeV}$ using the ALICE detector for six
different centrality classes covering 0-80\%.  The proton-to-pion and the
kaon-to-pion ratios both show a distinct peak at $p_{\rm{T}} \approx 3
\text{\,GeV/$c$}$ in central Pb--Pb collisions that decreases for more
peripheral collisions. For $p_{\rm{T}} > 10 \text{\,GeV/$c$}$, the nuclear
modification factor is found to be the same for all three particle species in
each centrality interval within systematic uncertainties of 10--20\%. This
suggests there is no direct interplay between the energy loss in the medium
and the particle species composition in the hard core of the quenched jet. For
$p_{\rm{T}} < 10 \text{\,GeV/$c$}$, the data provide important constraints for
models aimed at describing the transition from soft to hard physics.
\end{abstract}
\end{titlepage}
\setcounter{page}{2}

\section{\label{sec:1}Introduction}

In ultra-relativistic heavy-ion collisions, a strongly-interacting deconfined
medium of quarks and gluons is created. Experimental evidence for this state
of matter has been found both at the Relativistic Heavy-Ion Collider
(RHIC)~\cite{Arsene:2004fa,Adcox:2004mh,Back:2004je,Adams:2005dq} as well as
at the
LHC~\cite{Aamodt:2010pa,Aamodt:2010jd,Abelev:2012rv,Aad:2010bu,Chatrchyan:2011sx}. Transverse
momentum (\pt) spectra probe many different properties of this medium. At low
\pt ($\lesssim \gevc{2}$) the spectra provide information on bulk
production, while at high \pt ($\gtrsim \gevc{10}$) transport properties
of the medium can be studied via jet
quenching~\cite{Gyulassy:1990ye,Gyulassy:1993hr,Wang:2002ri}. The microscopic
QCD processes are different at low and high \pt, and it is an open question if
additional physics processes occur in the intermediate \pt region ($2 \lesssim
\pt \lesssim \gevc{10}$). In this paper, the centrality evolution of the
transverse momentum spectra of pions, kaons, and protons as a function of \pt
for \pbpb collisions at $\snnt{2.76}$ is presented. The focus is on
intermediate and high \pt, where these measurements allow comparison between
baryons and mesons, strange and non-strange particles, and the search for
particle mass-dependent effects.

For inclusive charged particle \pt spectra, jet quenching leads to a
suppression of high-\pt particle production at the
RHIC~\cite{Adcox:2001jp,Adler:2002xw,Adams:2003kv} and over an extended \pt
range, up to \gevc{100}, at the
LHC~\cite{Aamodt:2010jd,Abelev:2012hxa,CMS:2012aa,Aad:2015wga}. The
microscopic mechanism of jet quenching is not completely understood, and one of
the main goals of the experimental programs at the RHIC and the LHC is to
identify additional signatures associated with the jet quenching to constrain
theoretical modeling. Particle identification (PID) is of fundamental interest
since, due to the color Casimir factor, gluons interact two times stronger
with the medium than quarks~\cite{Wang:1998bha,Renk:2007rp} and it is known
from \ee studies of 3-jet events that gluons are more likely to fragment to
leading baryons than quarks are~\cite{Abreu:2000nw}. In addition, some models
for jet quenching predict large particle-species-dependent
effects~\cite{Sapeta:2007ad,Aurenche:2011rd,Bellwied:2010pr}. Measurements at
the RHIC, in particular for baryons, have so far been inconclusive due to the
limited \pt-range and the large systematic and statistical
uncertainties~\cite{Abelev:2006jr,Agakishiev:2011dc,Adare:2013esx}.

In the intermediate transverse momentum regime, the baryon-to-meson ratios,
e.g.\ the proton yield divided by the pion yield, measured by experiments at
the RHIC revealed a, so far, not well understood
enhancement~\cite{Adcox:2001mf,Adler:2003kg,Adams:2003am}. This so-called
``baryon anomaly'' could indicate the presence of new hadronization mechanisms
such as parton
recombination~\cite{PhysRevC.68.044902,Pop:2004dq,Brodsky:2008qp} that could
be significantly enhanced and/or extended out to higher \pt at the LHC due to
larger mini-jet production~\cite{Hwa:2006zq}. In recombination models, the
enhancement at intermediate \pt is an effect of the coalescence of lower \pt
quark-like particles that leads to a larger production of baryons than mesons.
In a model without new intermediate \pt physics, the rise of the
baryon-to-meson ratio is due to hydrodynamics and the decrease is solely a
consequence of the growing importance of fragmentation.

In a recent letter~\cite{Abelev:2014laa} ALICE reported the charged pions,
kaons, and proton \pt spectra for \pp and the most central and most peripheral
\pbpb collisions. The main observation was that, within statistical and
systematic uncertainties, the nuclear modification factor is the same for $\pt
> \gevc{10}$ for all three particle species. This suggests that there are no
significant particle-species-dependent effects related to the energy loss. In
this paper, the analysis used to obtain the measurements at high \pt is
presented in full detail, and the results for all centrality classes are
included. Recent measurements at low and intermediate \pt of identified
particle production and correlations in \ppb collisions have revealed
phenomena typically associated with fluid-like behavior in heavy-ion
collisions~\cite{Abelev:2013wsa,Abelev:2013haa,Khachatryan:2014jra}. This
raises questions if hydrodynamics and/or recombination can also be applied to
describe these small
systems~\cite{Shuryak:2013ke,Werner:2013ipa,Bozek:2013ska}. The centrality
evolution studies for \pbpb collisions can therefore also be seen as a
possible experimental interconnection between the smallest and the largest QCD
bulk systems.

The outline of this paper is as follows. In Sec.~\ref{sec:alice}, the data
analysis is described. The method using the energy loss in the TPC for
particle identification is laid out first and then the procedure using the
Cherenkov angle measured by the HMPID is presented. In Sec.~\ref{sec:results},
the final spectra are presented, and the particle ratios and nuclear
modification factors are discussed and compared with theoretical calculations
and results from previous experiments at lower center-of-mass energies.


\section{\label{sec:alice} Data analysis}

The results reported in this paper have been obtained with the central barrel
of the ALICE detector, which has full azimuthal coverage around mid­rapidity,
$|\eta|<0.8$~\cite{Aamodt:2008zz}. Different Particle IDentification (PID)
devices are used for the identification of \pikandp (see
Table~\ref{table:pbpbranges} for exact \pt ranges). Ordering by \pt, from
lowest to highest, the results are obtained using the specific energy loss,
\dedx, in the silicon Inner Tracking System (ITS), the \dedx in the Time
Projection Chamber (TPC), the time-of-flight measured by the Time-Of-Flight
(TOF) detector, the Cherenkov angle measured by the High Momentum Particle
Identification Detector (HMPID), and the TPC \dedx in the relativistic rise
region. The general performance of these devices is reported
in~\cite{Abelev:2014ffa}. Detailed description of the lower \pt analyses and
the resulting \pikandp \pt spectra in \pbpb collisions are already
published~\cite{Abelev:2013vea}. In this section, the method used to extract
these \pt spectra in the HMPID and the TPC \dedx relativistic rise analysis is
described in detail.

Due to the limited acceptance of the HMPID, the analysis has been performed
with the larger 2011 dataset where a centrality trigger was used, restricting
the HMPID results to 0-50\% central \pbpb collisions.

\subsection{\label{sec:highpT} TPC d\textit{E}/d\textit{x} relativistic rise analysis }

The relativistic rise of the \dedx in the TPC, where the average energy loss
increases as $\log{\bg}$ ($3 \ll \beta\gamma \ll 1000$), allows ALICE to
extend the PID of \pikandp up to $\pt = \gevc{20}$. This section will focus on
details of this analysis.

\subsubsection{Event and track selection}
\label{sec:eventsAndTracks}
The event and track selection follows closely that of the inclusive charged
particle analysis~\cite{Abelev:2012hxa}. The same spectrum normalization is
adopted so that the systematic uncertainties related to event and track
selection are common, allowing a precise comparison between the nuclear
modification factors for inclusive and identified charged particles. The
analysis with PID described here has additional systematic uncertainties
related to the particle identification that we will describe in
Sec.~\ref{sec:highptsyst}.

A total of $11 \times 10^{6}$ \pbpb collision events recorded in 2010 are used
in this analysis. The online (offline) trigger for minimum bias interactions
in \pbpb collisions requires signals in two (three) out of the three following
detector elements: the Silicon Pixel Detector (SPD) layers of the ITS and the
two forward scintillators (V0) located on opposite sides of the interaction
point. The centrality is determined from the measured amplitude in the V0
detector~\cite{Abbas:2013taa}.

Primary tracks are reconstructed in the ALICE TPC~\cite{Alme:2010ke} from
clusters in up to 159 pad rows, where each cluster consists of a group of
cells covering a few neighboring pads and time bins. The tracks used in the
analysis are restricted to $|\eta| < 0.8$ in order to be fully contained in
the TPC active volume. Furthermore, tracks are required to have at least one
hit in one of the two innermost SPD layers of the ITS, and the distance of
closest approach to the primary vertex is required to be less than 2 cm along
the beam axis and less than 7 standard deviations in the transverse plane
(${\approx}350\,\mu\text{m}$ for tracks with $\pt = \gevc{2}$, decreasing
slightly with \pt). The resulting relative \pt resolution for these tracks is
better than 5\% at $\pt = \gevc{20}$~\cite{Abelev:2012hxa}. The \pt spectra
have been corrected for this resolution using an unfolding procedure for $\pt
> \gevc{10}$~\cite{Abelev:2012hxa,Abelev:2013ala}. This correction is smaller
than 2\% at $\pt = \gevc{20}$.


\subsubsection{Particle identification at large transverse momentum}
\label{sec:dedx_pid_method}

\begin{figure}[htbp]
  \begin{center}
   \includegraphics[width=0.99\mylength]{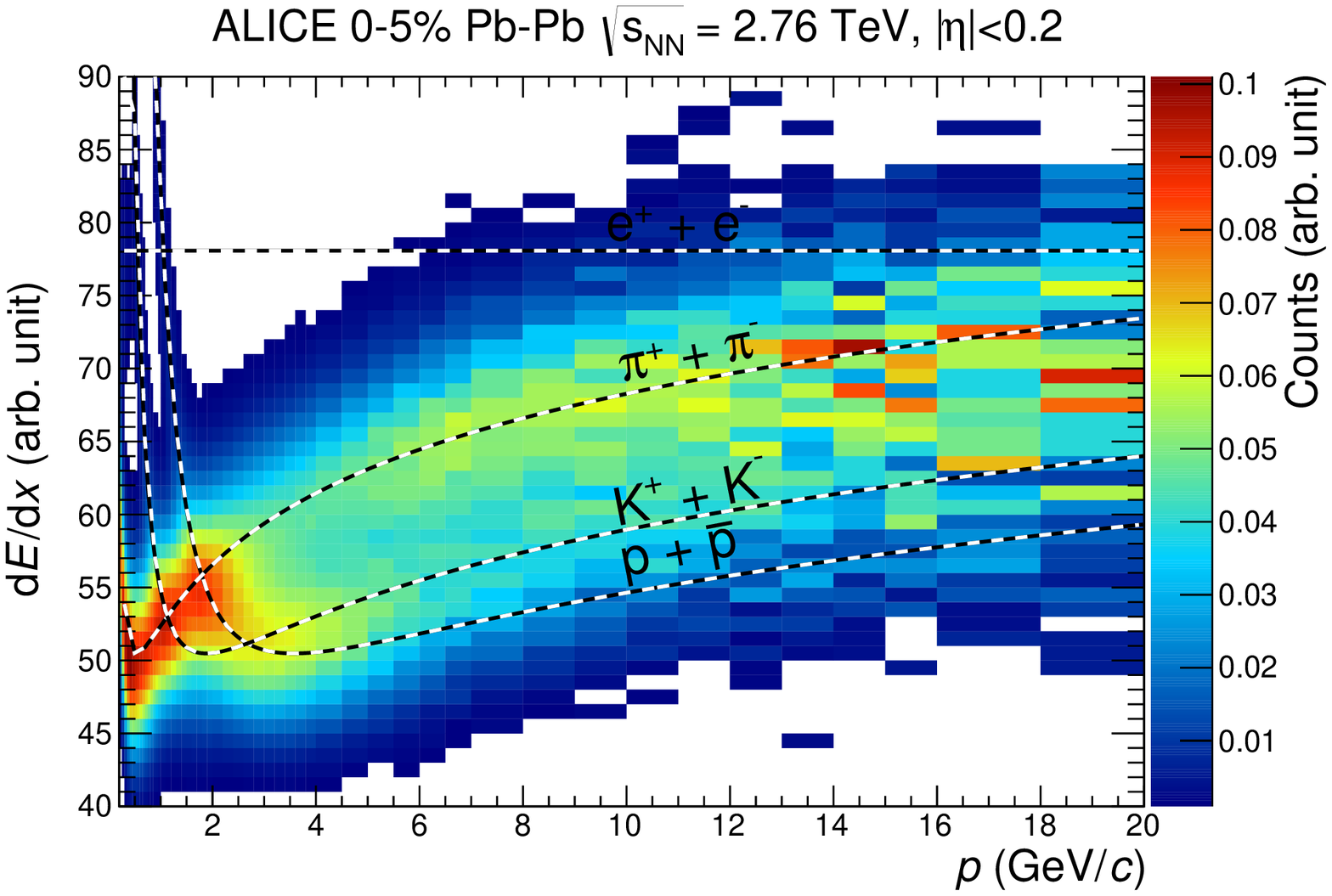}
   \includegraphics[width=0.99\mylength]{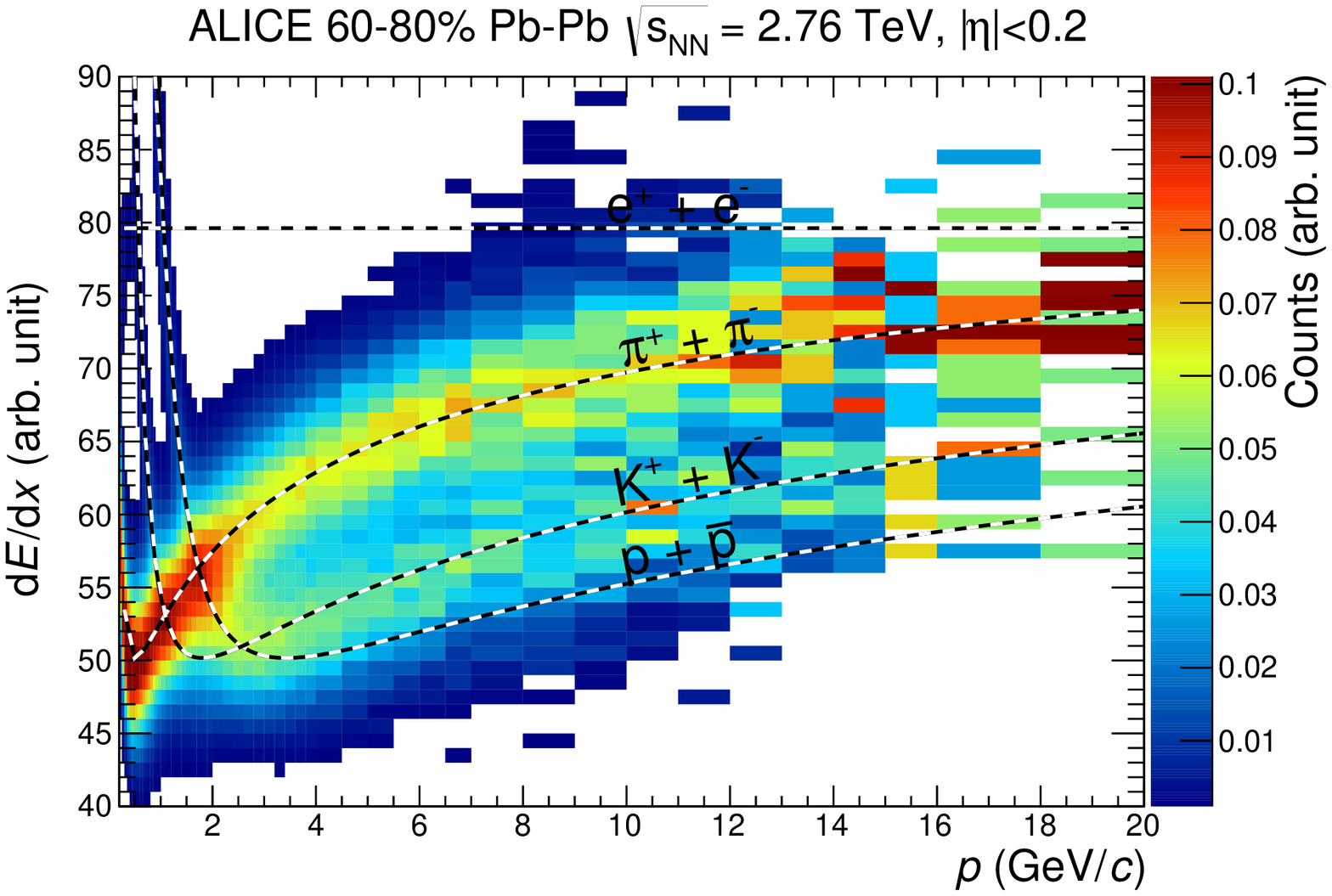}
    \caption{(Color online) The \dedx as a function of the momentum $p$ at
      mid-rapidity $|\eta|<0.2$ for 0-5\% (left panel) and 60-80\% (right
      panel) \pbpb collisions. In each momentum bin, the \dedx spectra have
      been normalized to have unit integrals and only bins with more than
      0.1\% of the counts are shown (making electrons not visible in this plot
      except at very low momentum). The curves show the final \mdedx responses
      for pions, kaons, and protons.}
  \label{fig:dedx_vs_p}
\end{center}
\end{figure}

\begin{figure}[htbp]
  \begin{center}
    \includegraphics[keepaspectratio,
      width=1.45\mylength]{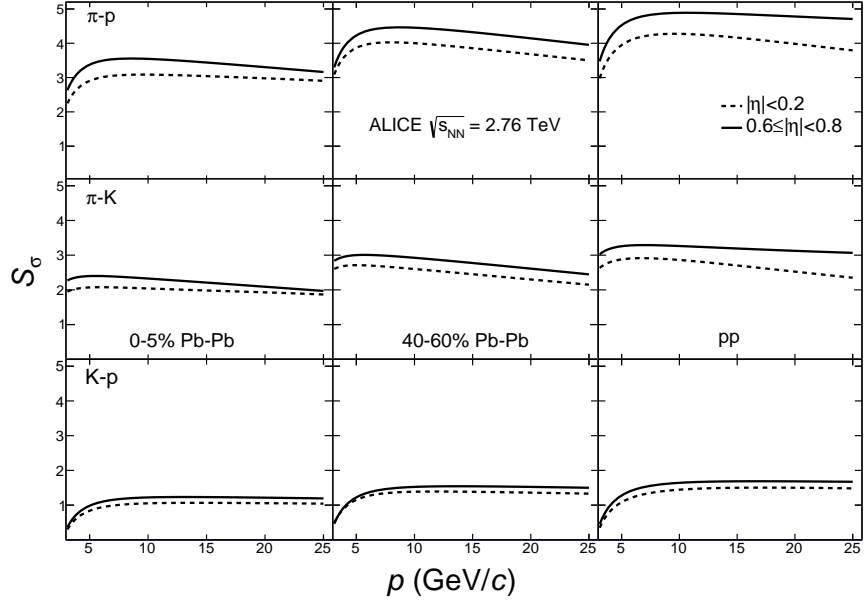}
    \caption{\label{fig:rtpc:2} Separation in number of standard deviations
      ($S_\sigma$) as a function of momentum between: pions and protons (upper
      panels), pions and kaons (middle panels), and kaons and protons (lower
      panels). Results are shown for 0-5\% (left panels) and 40-60\% (middle
      panels) \pbpb; and pp (right panels) collisions. Because the TPC
      response is track-length dependent, the separation is better for tracks
      at forward pseudorapidities (solid lines) than for those at smaller
      $\eta$ (dashed lines). The degradation in separation power in more
      central collisions is expected from occupancy effects -- in the most
      peripheral collisions an average of $149$ clusters are assigned to
      tracks with $\pt > \gevc{2}$, while in the most central collisions only
      $127$ clusters are assigned.}
    
  \end{center}
\end{figure}

Figure~\ref{fig:dedx_vs_p} shows the \dedx as a function of momentum $p$ in
0-5\% central \pbpb collisions. It is evident that particle identification in
the relativistic rise region requires precise knowledge of the \mdedx response
and resolution $\sigma$. To quantify this, and to motivate the detailed
studies in the following, the final response functions are used to estimate
the separation power, where, for example, the charged pion-to-kaon separation in
number of standard deviations, $S_\sigma$, is
\begin{equation}
\label{eq:rtpc:3}
S_\sigma = \frac{ \left\langle \frac{{\rm d}E}{{\rm d}x} \right\rangle_{\chpi} - \left\langle \frac{{\rm d}E}{{\rm d}x} \right\rangle_{\chk} }{0.5\left( \sigma_{\chpi} + \sigma_{\chk} \right) },
\end{equation}
that is, the absolute \mdedx difference normalized to the arithmetic average
of the resolutions. Fig.~\ref{fig:rtpc:2} shows that the separation power
between particle species is only a few standard deviations, making PID very
challenging, requiring optimization of the \dedx signal itself and the use of
external PID constraints to calibrate the response. In the following, these
analysis aspects will be covered in detail.

\subsubsection{The \dedx calibration}

The \dedx is obtained as a truncated mean, where the average is performed
considering only the 60\% lowest cluster charge values to remove the tail of
the Landau like cluster charge distribution. It is customary to use the
notation \dedx and talk about the Bethe--Bloch curve, even if the \dedx used in
the analysis is only the {\it truncated} mean and does not contain energy
losses deposited as sub-ionization-threshold excitations or the full
ionization from delta-electrons, discussed in detail
in~\cite{Bichsel:2006cs}. While the Bethe--Bloch specific energy loss depends
only on $\beta\gamma = p/m$, the one obtained from the detected truncated mean
also depends on other parameters such as the actual cluster sample length,
i.e., the pad length and/or track inclination over the pad. In the following,
we shall refer to the relationship between the two types of specific energy
losses as \emph{the transfer function} and it is this relationship that is
optimized in the \dedx calibration, and used also as input for the analysis
strategy discussed later.

Each of the up to 159 clusters used to reconstruct a track contains
information on the ionization energy loss in the TPC. To equalize the gain,
each individual readout channel has been calibrated using ionization clusters
produced by the decay of radioactive krypton, $^{86}_{36}{\rm Kr}$, released
into the TPC gas~\cite{Alme:2010ke}.

In \pp collisions the cluster integrated charge is used for calculating the
\dedx. The integrated charge is corrected for the tails of the charge
distribution that are below the readout threshold. Due to the large
probability for overlapping clusters in \pbpb collisions, the maximum charge
in the cluster is used to calculate the \dedx in this case. The maximum charge
is the largest charge in a cluster cell (pad and time bin). The maximum charge
has to be corrected for the drift-length dependent reduction due to diffusion
and the dependence on the relative pad position of the induced signal (the
measured maximum charge is largest if the cluster center is also the pad
center, and smallest if it is between two pads).

The performance and stability of the \dedx transfer function, with respect to
gain variations, is improved in the following two ways. Reconstructed space
points where the charge is deposited on a single pad, that are not used for
track fitting, are included in the \dedx calculation. An attempt is made, to
identify clusters below the readout threshold. If a row has no cluster
assigned to the track but clusters were assigned in both neighboring rows, it
is assumed that the cluster charge was below the readout threshold and a
virtual cluster is assigned with charge corresponding to the lowest
reconstructed charge cluster on the track. This virtual cluster is then
included in the calculation of the truncated mean. This is similar to
  the strategy adopted by ALEPH, but without changing the truncation
  range~\cite{Buskulic:1994wz}.

\begin{figure}[htb]
  \begin{center}
   \includegraphics[width=1.0\mylength]{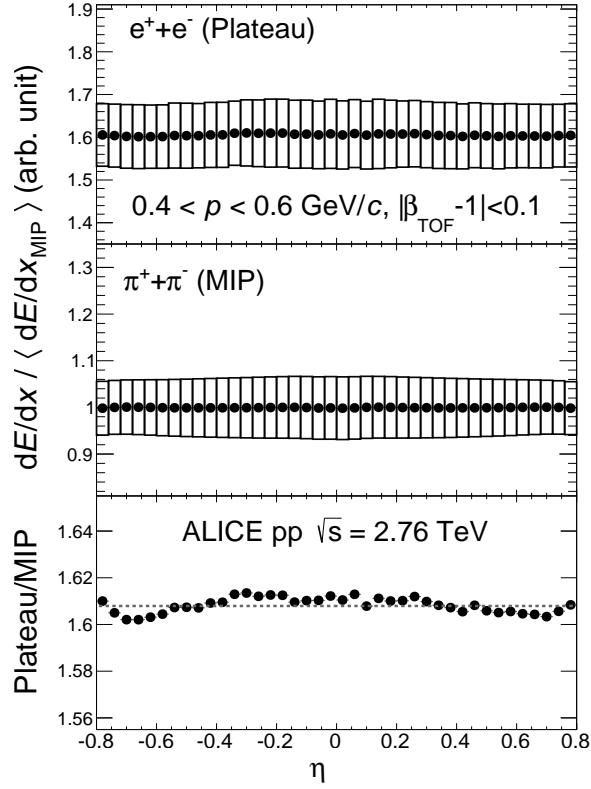}
    \caption{The \dedx as a function of $\eta$ for electrons on the Fermi
      Plateau (upper panel) and MIP pions (middle panel); the selection
      criteria are described in the text. The solid round markers indicate the
      average, \mdedx, and the height of the boxes is given by the standard
      deviation, $\sigma$. The lower panel shows the ratio between the
      Plateau and MIP \mdedx. The statistical uncertainty is smaller than the
      marker sizes. These results were obtained for \pp collisions at
      $\sqrt{s}=$2.76 TeV.}
  \label{fig:rtpc:0}
\end{center}
\end{figure}

The $\eta$ dependence of the \dedx is sensitive to corrections for the
track-length and the diffusion. There is also a small correction for
  the direct drift-length dependent signal attenuation, due to absorption, of
  ionization electrons by Oxygen~\cite{Alme:2010ke}. At $\eta=0$ the
ionization electrons drift the full 250~cm to the readout chambers and, as a
result, the signal is spread out, due to diffusion, making threshold effects
more prominent than for tracks with $\eta = 0.8$. At the same time, the sampled
track length is longer for track with $\eta = 0.8$ than with $\eta = 0$. The
\dedx calibration is validated using pions in the Minimum Ionizing Particle
(MIP) regime and electrons in the Fermi Plateau region. A clean sample of MIP
pions is selected via tracks with momenta $0.4 < p < \gevc{0.6}$ and energy
loss $0.8 < (\dedx) / \mdedx_{\rm MIP} < 1.2$. A clean electron sample is
obtained in the same momentum range via centrality dependent \dedx cuts (as
$S_\sigma$ depends on centrality) and by rejecting kaons using Time-Of-Flight
(TOF) information: $0.9 < \beta_{\rm TOF} < 1.1$. For both samples it is found
that the $\eta$-dependence of the \mdedx is negligible. We note that one
expects these two classes of tracks to have different sensitivity to threshold
corrections. The result of the validation test for pp collisions is shown in
Fig.~\ref{fig:rtpc:0}, which displays the \mdedx response as a function of
$\eta$ for electrons (upper panel) and pions (middle panel).

\subsubsection{Division into homogenous samples}
\label{sec:phicut}

\begin{figure} [htb]
  \begin{center}
    \includegraphics[width=1.0\mylength]{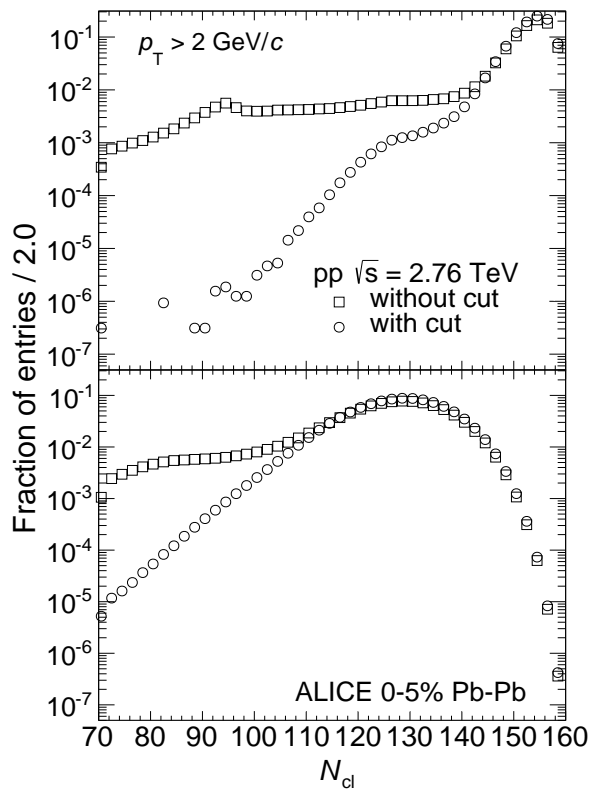}
    \caption{Number of clusters used in the \dedx calculation for
      $\pt>\gevc{2}$ without (squares) and with (circles) the geometric
      cut. Results are presented for \pp (upper panel) and central \pbpb
      (lower panel) collisions at $\sqrt{s_{\rm NN}}=$ 2.76 TeV. The minimum
      number of clusters on a track is 70.}
  \label{fig:rtpc:1}
\end{center}
\end{figure}

From studies of the transfer function, one expects a significant track-length
dependence. For the ``stiff'' high-\pt tracks used in this analysis, the
track-length in the transverse bending plane is rather similar, but there is a
significant $\eta$ dependence and the effect of this on the \dedx resolution
is visible in Fig.~\ref{fig:rtpc:0} for the pion MIPs. This motivates
performing the analysis in $|\eta|$ intervals: $|\eta|<0.2$,
$0.2\leq|\eta|<0.4$, $0.4\leq|\eta|<0.6$ and $0.6\leq|\eta|<0.8$ and then
combining the results.

Furthermore, tracks close to and/or crossing the TPC sector boundaries have
significantly fewer clusters assigned. Because the analyzed tracks are
``stiff'', those tracks close to the sector boundaries can be easily rejected
using a \textit{geometric} cut in the azimuthal track angle $\varphi$, which
excludes approximately 10\% of the tracks for $\pt>
\gevc{6}$. Figure~\ref{fig:rtpc:1} shows the effect of the geometric cut on
the distribution of the number of clusters per track. The cases before and
after the $\varphi$ cut are shown for \pp (upper panel) and central \pbpb
(lower panel) collisions. The large difference between the distributions for
\pp and central \pbpb is an occupancy effect and essentially independent of
\pt.  The cut significantly improves the \dedx performance by rejecting tracks
with less information (fewer clusters) in regions where the calibration is
more sensitive to complex edge behaviors that can have larger effects on
``stiff'' tracks. This also simplifies the analysis because in each $|\eta|$
interval, a single resolution parameter is sufficient to describe individual
particles species (e.g., all pions) in a given momentum bin.

\subsubsection{Obtaining the high-\pt yields}

\begin{figure}[htbp]
  \centering
  \includegraphics[keepaspectratio, width=2.0\mylength]{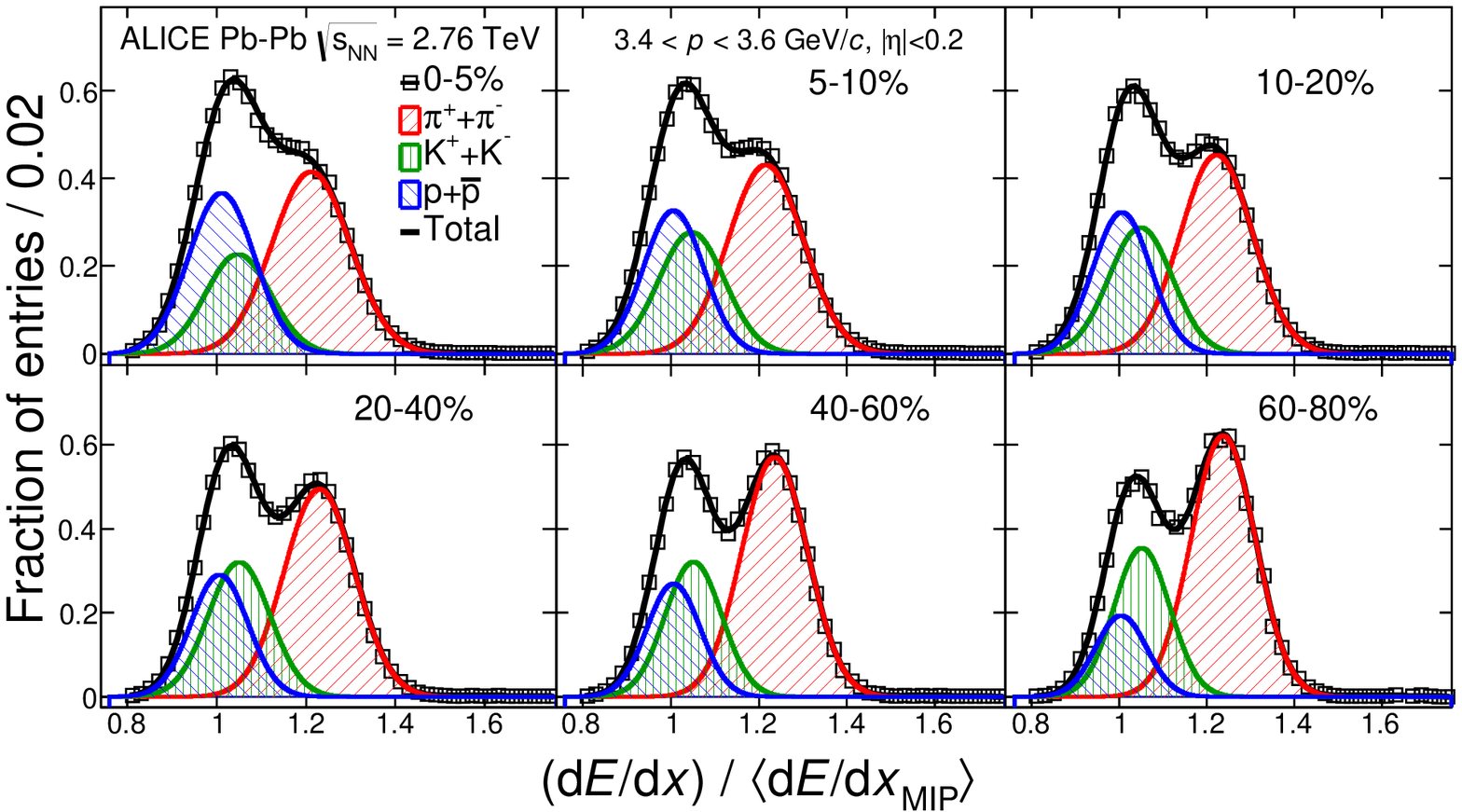}
  \includegraphics[keepaspectratio, width=2.0\mylength]{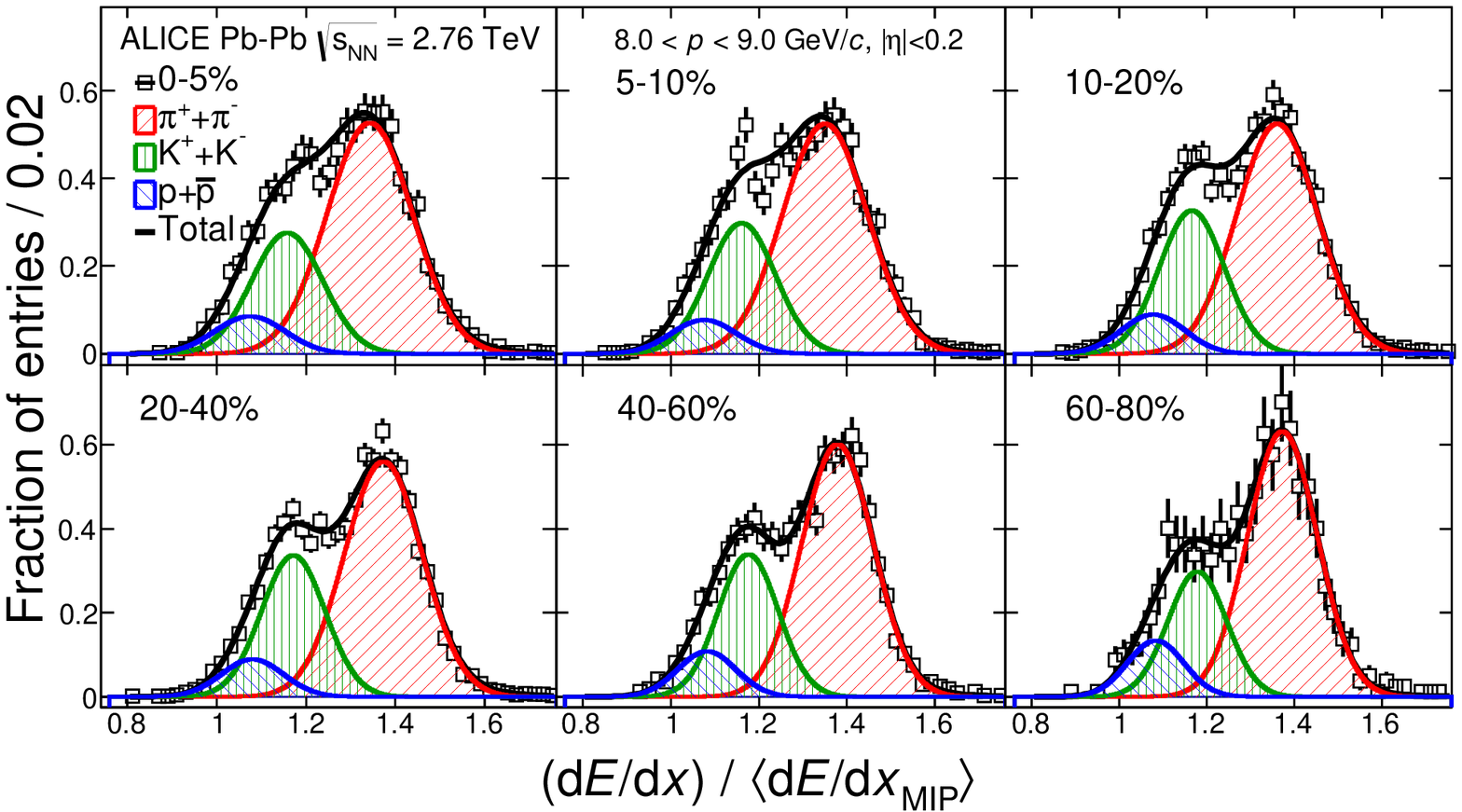}
  \caption{(Color online) Four-Gaussian fits (line) to the \dedx spectra
    (markers) for tracks having momentum in the range 3.4--\gevc{3.6} (upper
    figure) and 8.0--\gevc{9.0} (lower figure) with $|\eta|<0.2$. In each
    panel, the signals of pions (rightmost Gaussian), kaons, and protons
    (leftmost Gaussian) are shown as red, green, and blue dashed areas,
    respectively. The contribution of electrons is small ($<1\%$) and
    therefore not visible in the figure. Results for all six \pbpb centrality
    classes are presented. The \dedx spectra have all been normalized to have
    unit integrals.}
  \label{fig:rtpc:3}
\end{figure}

Since, as already mentioned, the event and track selection scheme is identical
to the one used for the inclusive charged particle
spectra~\cite{Abelev:2012hxa}, and each charged track has an associated TPC
\dedx measurement, the charged pion, kaon, and (anti)proton yields measured in
this analysis are normalized to the inclusive charged particle
spectra~\footnote{The $\varphi$ cut described in Sec.~\ref{sec:phicut} was not
  applied in the inclusive charged particle analysis, but as this cut is a
  geometric cut it is independent of particle species type and therefore does
  not affect this normalization.}. This highlights the unique direct
correspondence between the two analyses and guarantees that the results are
fully consistent even at the level of statistical uncertainties. The analysis
of the \dedx spectra is therefore aimed at extracting the relative yields of
\pikandp, referred to as the particle fractions in the following.

In a narrow momentum and $|\eta|$ interval, the \dedx distribution can be
described by a sum of four Gaussians ($\pi$, ${\rm K}$, $p$, and $e$), see
e.g.\ Fig.~\ref{fig:rtpc:3}, and the requirements for the analysis to be able
to extract the yields with high precision is that the means and widths of the
Gaussians are constrained. Additional external track samples such as protons
from $\Lambda$ decays are used to obtain the constraints. The method presented
in the following has been benchmarked using Monte Carlo (MC) simulations and
the closure tests, comparing reconstructed output with generated input, for
all yields show less than 2\% systematic deviations. From studies comparing
test beam data results with the ALICE specific MC implementation of the
energy-loss in the TPC, the MC is known to be precise and to take into account
all important detector effects~\cite{Christiansen:2009zza}, with the limit
that the test beam data was recorded under controlled conditions (fixed track
topology and large gas gain) and that ion tail effects are not included in the
MC simulations.

\subsubsection{Measurement of the TPC response: parameterization of the Bethe--Bloch and resolution curves}
The first step of the analysis is to extract the response parameterizations
used to constrain the fits. The Bethe--Bloch curve is parameterized as follows:
\begin{equation}
\label{eq:rtpc:2a}
\left\langle  \frac{{\rm d}E}{{\rm d}x} \right\rangle = a\left[ \frac{1+(\beta\gamma)^{2}}{(\beta\gamma)^{2}} \right] ^{e} + \frac{b}{c}\log\left[     \frac{(1+\beta\gamma)^{c}}{1+d'(1+\beta\gamma)^{c}}  \right],
\end{equation}
where $a, b, c, d$, and $e$ are free parameters (the variable $d'$ is used to
simplify the expression and is defined as $d'=\exp[c(a-d)/b]$ where $d$ is the
\mdedx in the Fermi Plateau regime, $\bg \gtrsim 1000$).

For $d' \ll 1$, as is the case here, the parameterization has a simple
behavior in different regions of \bg. For small $\bg$, $\bg \ll 3$--$4$,
$\left\langle \frac{{\rm d}E}{{\rm d}x} \right\rangle \approx \frac{a}{(\beta
  \gamma)^{2e}}$ , while on the logarithmic rise: $\left\langle \frac{{\rm
    d}E}{{\rm d}x} \right\rangle \approx a +b \log\left( 1+\beta\gamma
\right)$. The parameterization has been motivated by demanding this behavior
in the discussed $\bg$ limits, while at the same time requiring that each
parameter has a clear meaning. It uses $1+\bg$ to ensure that the logarithmic
term is always positive.

The relative resolution, $\sigma / \mdedx$, as a function of
\mdedx is parameterized with a second-degree polynomial, which was found to
describe the data well:
\begin{equation}
\label{eq:rtpc:2b}
\sigma / \mdedx = a_{0} + a_{1}\mdedx + a_{2}\mdedx^{2}.
\end{equation}

The TPC response (Bethe--Bloch and resolution curves) is determined for each
$\eta$ region. Due to the deterioration of the TPC \dedx performance with
increasing multiplicity, the curves differ significantly and have to be
extracted separately for \pp and each \pbpb centrality class.

\begin{figure} [htb]
  \begin{center}
    \includegraphics[width=1.0\mylength]{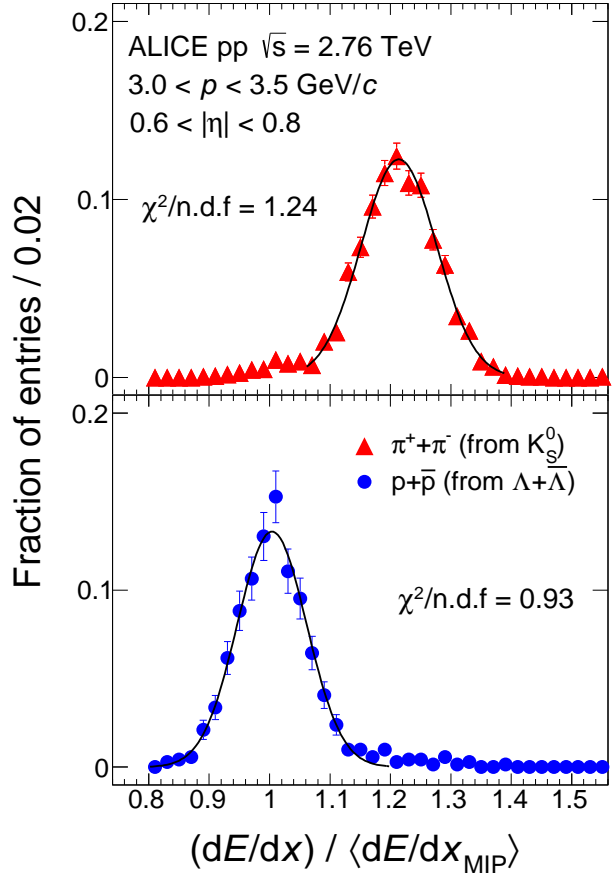}
    \caption{(Color online) \dedx spectra for secondary pions (top panel) and
      protons (bottom panel) identified via the reconstruction of the weak
      decay topology of ${\rm K_{S}^{0}}$ and $\Lambda$, respectively. A
      narrow invariant-mass cut reducing statistics was applied to select
      clean samples (but the pion sample still contains a small visible proton
      background). The curves are single Gaussian fits to the data and the
      reduced $\chi^2$ is calculated in the range indicated by the fit curves
      only.}
  \label{rtpc:pid:gauss}
\end{center}
\end{figure}

\begin{figure} [htb]
  \begin{center}
    \includegraphics[width=1.0\mylength]{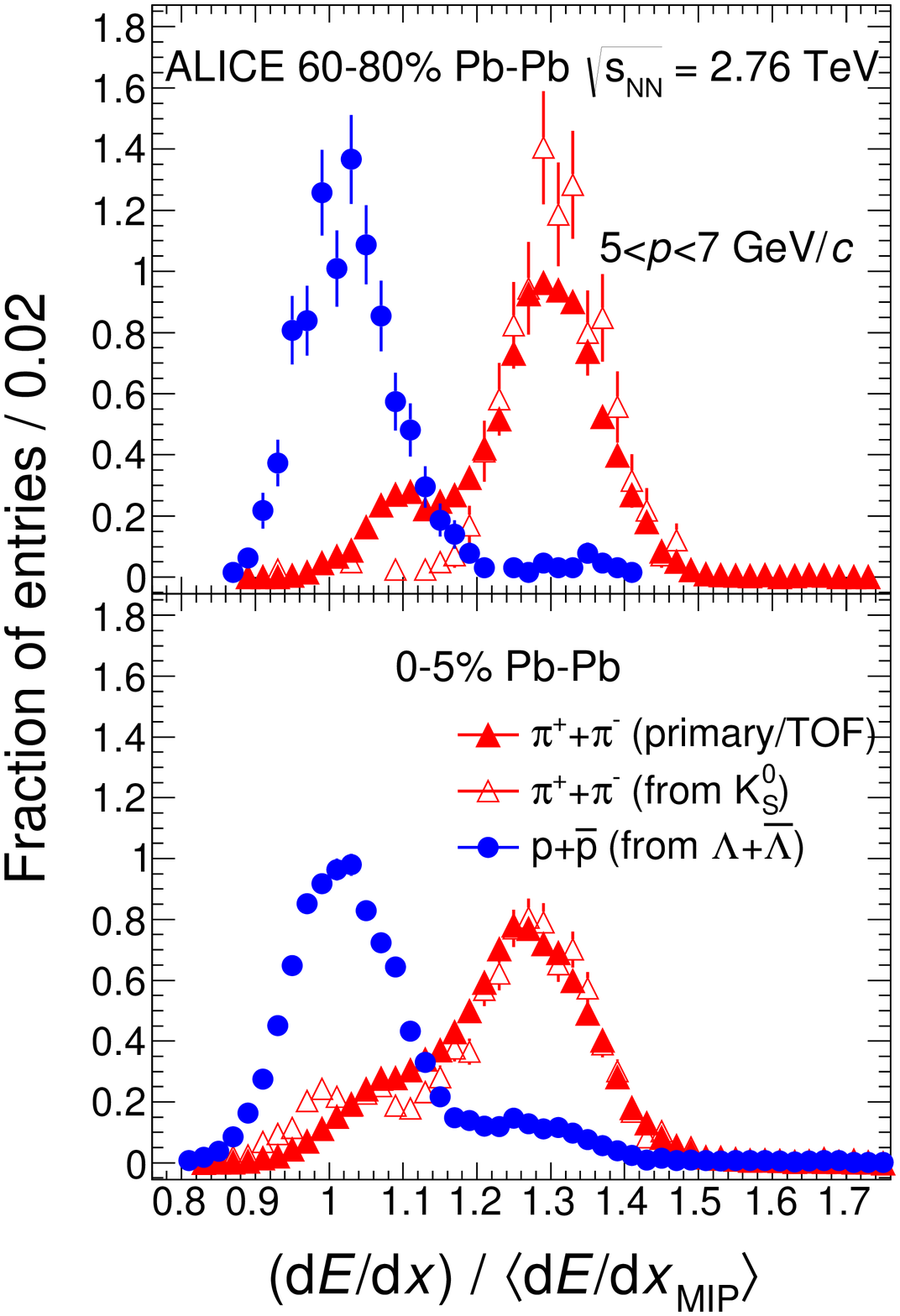}
    \caption{(Color online) \dedx spectra for secondary pions (open triangles)
      and protons (full circles) identified via the reconstruction of the weak
      decay topology of ${\rm K_{S}^{0}}$ and $\Lambda$, respectively. The
      spectra have been normalized to have the same integrals. The spectrum
      for primary pions (full triangles) is obtained by requiring $\beta_{\rm
        TOF} > 1$. Results for peripheral (upper panel) and central (lower
      panel) \pbpb collisions are shown. The tracks were chosen in the
      momentum (pseudorapidity) interval $5<p<\gevc{7}$
      ($0.6\leq|\eta|<0.8$). Note that most spectra also contain a
      well-understood background.}
  \label{rpc:pid:ext}
\end{center}
\end{figure}

The parameters $a,b,d,$ and $e$ are well determined using external PID
information. Secondary pion (proton) tracks identified via the reconstruction
of the weak decay topology of ${\rm K_{S}^{0}}$ ($\Lambda$) and data samples
with TOF enhanced ($\beta_{\rm TOF} > 1$) primary pions are used. The $V^0$
selection used in this analysis is similar to the one used in the dedicated
analysis~\cite{Abelev:2013xaa}. To verify that the \dedx response is Gaussian,
narrow invariant-mass cuts were applied to \pp data where the $V^0$
reconstruction is cleanest. Figure~\ref{rtpc:pid:gauss} shows single Gaussian
fits to the pion and proton peaks for such data and we note that the reduced
$\chi^2$ value is in the expected range for a valid fit model. In the
following, \mevcsq{10} wide invariant-mass cuts around the peaks was used to
select signal and reject background as a compromise between statistics and
purity. Using this information, the Bethe--Bloch function is constrained in
the $\beta\gamma$ interval of 3--60.  Figure~\ref{rpc:pid:ext} shows examples
of the TPC \dedx spectra for these samples in the momentum ($\eta$) range:
5--\gevc{7} ($0.6\leq|\eta|<0.8$) for the most central and most peripheral
\pbpb collisions analyzed. In this case, the proton candidate samples from the
$\Lambda$ decay are not pure samples, and have some contamination of pions
since the invariant mass peak region still contains considerable combinatorial
background. This contamination is seen in the asymmetry towards the higher
value of \dedx in the proton sample. In the case of the pion samples from the
$K_{S}^{0}$ decay, proton contamination creates the asymmetry towards the
lower value of \dedx in the spectra. Hence, in these cases, to obtain the mean
\dedx and resolution for each particle species, the asymmetric tail of the
Gaussians were not considered.\\ The Fermi Plateau is fixed using
electron-positron pairs from photon conversions (a photon conversion is
reconstructed similar to a V$^0$ decay and identified from the low invariant
mass). The same information is used to measure the \dedx resolution as a
function of \mdedx. The relative resolution around the MIP\footnote{The
  resolution depends on centrality and track length and is worse in central
  events and for smaller $|\eta|$.} is ${\approx}5.5$--$7.5\%$ and improves
with increasing \mdedx (primary ionization) in the relativistic rise region to
${\approx}4.5$--$5.5\%$. These data samples are henceforth referred to as the
external PID data.

In the relativistic rise region, the analysis is very stable because in this
region $\mdedx \approx a + b \log{\bg}$, so the \dedx separation between
particle species, e.g., protons and pions, is constant: $\mdedx_{\rm
  p}-\mdedx_{\pi} \approx a + b\log(p/m_{\rm p}) - (a + b\log(p/m_{\pi}))
\approx b\log(m_{\pi}/m_{\rm p})$. So as long as all particle species are in
this $\beta\gamma$ regime a simple extrapolation can be applied. For $\bg
\gtrsim 100$ the pions ($p \gtrsim \gevc{14}$) start to approach the Fermi
Plateau region and the \mdedx dependence on \bg is more complex. To address
this, a two dimensional fit to the \dedx vs $p$ distribution is performed. All
the parameters of the resolution function and the parameters $a,b,d,$ and $e$
of Eq.~\ref{eq:rtpc:2a} are fixed. The parameter $c$ and the yields of \chpi,
\chk and \chp in different momentum intervals are free parameters. This fit
method works fine if the corrections to the logarithmic rise, due to the
transition to the Plateau, are small, which restricts the current analysis to
$\pt < \gevc{20}$. With higher statistics and the use of cosmic muons as
additional constraints, we expect to be able to extend the method up to
\gevc{50}.

There is a final subtle point that should be mentioned here. The systematic
uncertainty on the yields from the \dedx method alone is rather large for
particles with ionization energy loss close to 1 MIP, but additional
information from other analyses can be used to constrain the results. One
would like to avoid using the actual lower \pt \pikandp measurements, as this
will introduce a direct bias in the final combined spectra
(Sec.~\ref{sec:results}). Instead, the neutral kaon yields are used to
constrain the charged kaons in \pbpb collisions\footnote{The assumption is
  that the invariant \pt spectra are the same. The charged kaon fraction
  ($f_{\chk}$) is obtained working backwards through Eq.~\ref{eq:rtpc:1} and
  Eq.\ref{eq:fraction}.}. The two dimensional fit is applied again, but the
parameter $e$, which mainly affects the proton \mdedx, is now allowed to vary
while the other parameters, $a$--$d$, are constrained and the charged kaon
yield in the fit is also restricted to be consistent with the neutral kaon
yield (the pion and proton yields are free). The effect of this refit is
largest in central collisions at low \pt ($<\gevc{4}$) and decreases with
centrality; at \gevc{3} the effect on the extracted kaon yield is 10\%
($<1$\%) for 0-5\% (60-80\%) collision centrality.

\begin{figure}[htb]
  \centering
  \includegraphics[keepaspectratio, width=1.38\mylength]{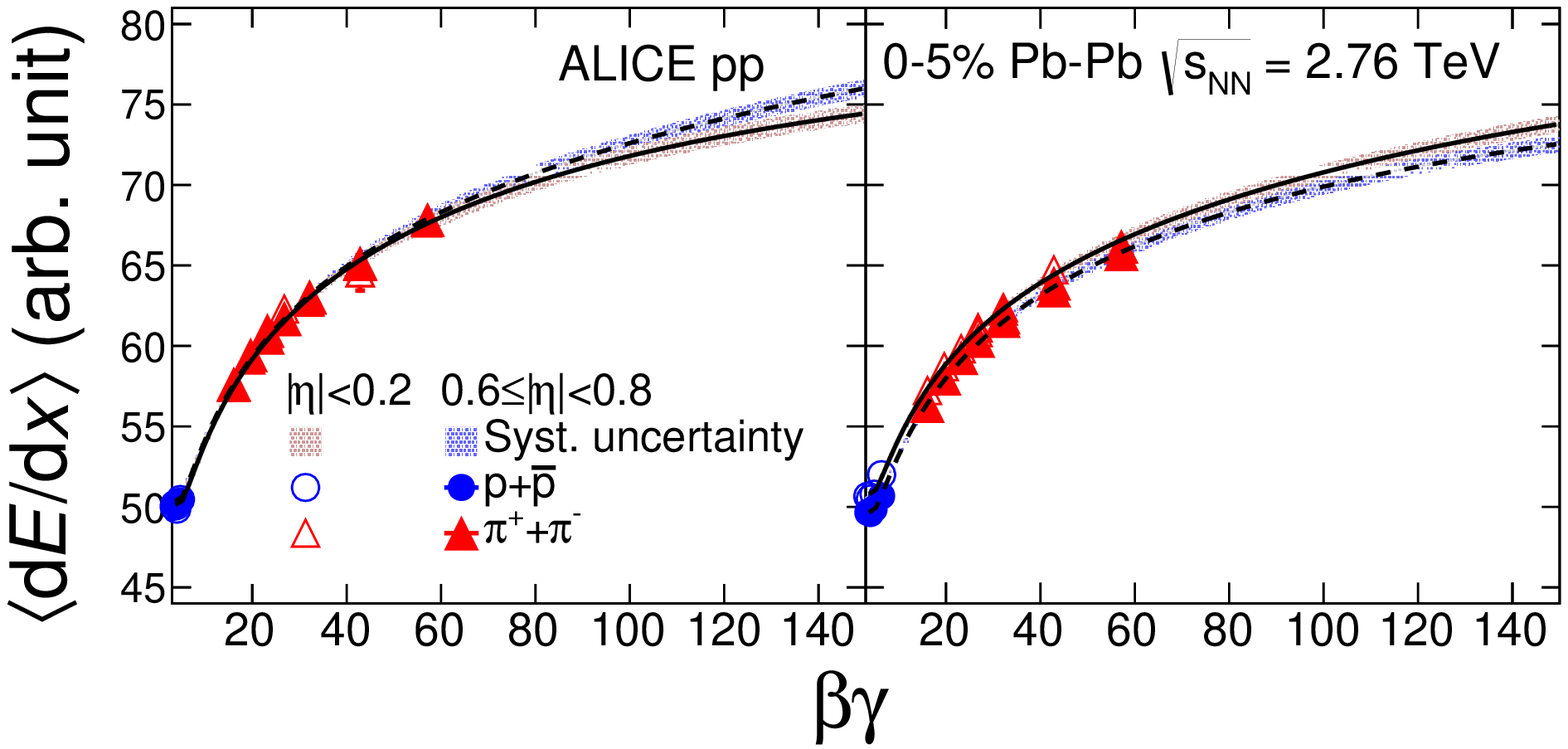}
  \includegraphics[keepaspectratio, width=1.38\mylength]{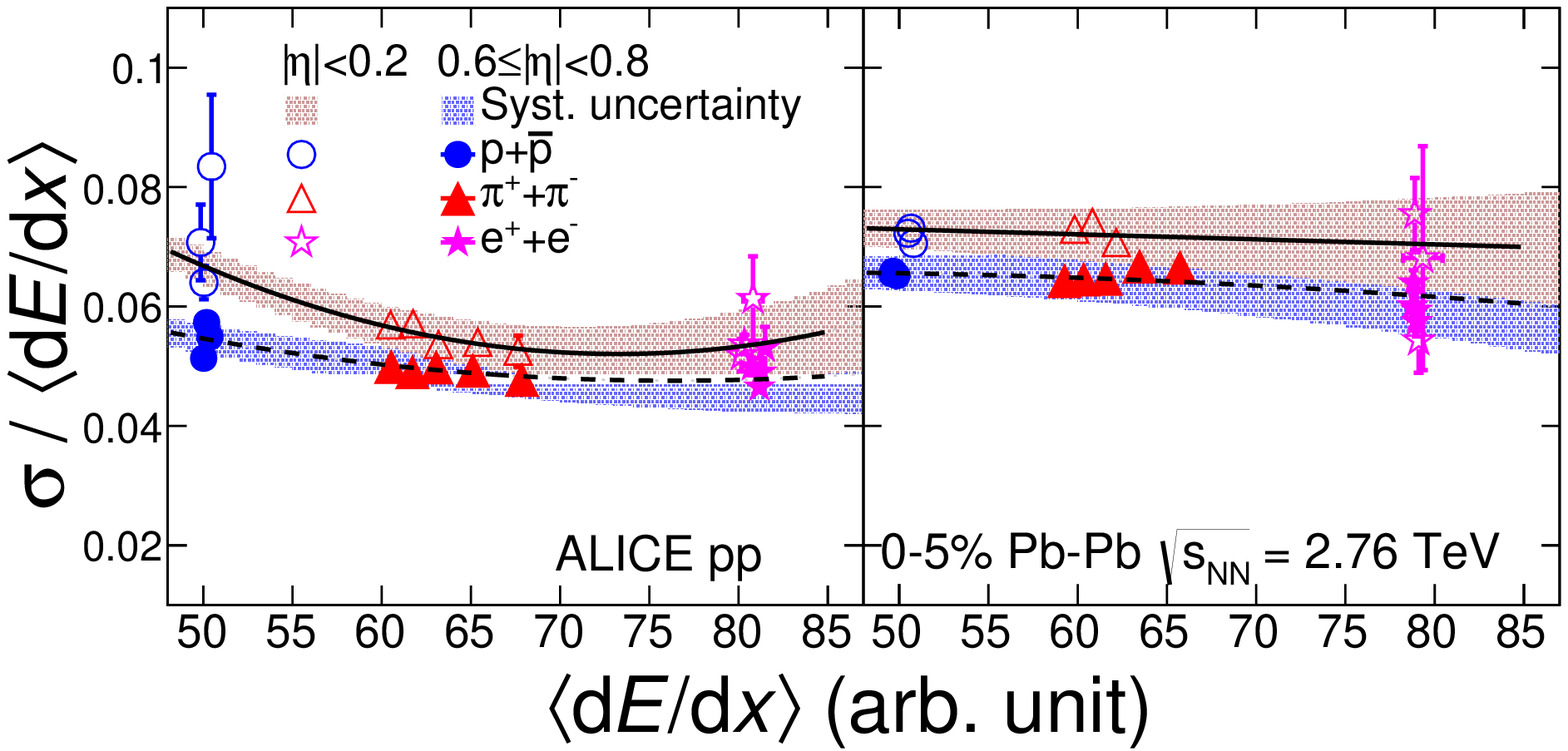}
  \caption{(Color online) Final Bethe--Bloch (upper figure) and resolution
    (lower figure) curves obtained as described in the text. Results are shown
    for \pp (left panels) and 0-5\% \pbpb (right panels) collisions. The
    Bethe--Bloch curve is shown in the region relevant for pions, kaons, and
    protons in this analysis. The external PID data samples of pions, protons and electrons are
    used to obtain the parameterizations, these data are plotted as
    markers. The shaded areas represent the systematic uncertainty of the
    parameterizations.}
  \label{fig:rtpc:2a}
\end{figure}

\begin{table}[htbp]
  \begin{center}
    \begin{tabular}{cccc}
      \hline \hline
      \textrm{Parameter}&
      \textrm{\pbpb 0-5\%}&
      \textrm{\pbpb 60-80\%}&
      \textrm{\pp}\\
      \hline
      $a$ & 33.9--35.4   & 32.9--33.1   & 32.5--33.3  \\
      $b$ & 7.66--7.89   & 8.58--9.01   & 8.52--8.77  \\
      $c$ & 2.18--7.18   & 1.25--2.38   & 1.65--43.0  \\
      $d$ & 78.0--78.5   & 80.0--80.6   & 80.6--80.7  \\
      $e$ & 1.22--1.30   & 1.37--1.39   & 1.43--1.55  \\
      \hline \hline
    \end{tabular}
    \caption{\label{tab:rtpc:table1} Parameters obtained for the Bethe--Bloch
      function (Eq.~\ref{eq:rtpc:2a}) for central and peripheral \pbpb
      collisions and \pp collisions. Results are given as the range found for
      the four $|\eta|$ intervals.}
  \end{center}
\end{table}
  
Figure~\ref{fig:rtpc:2a} shows the final parameterizations of the Bethe--Bloch
and resolution curves for \pp and the most central \pbpb collisions. The
values obtained for the external PID data are also
shown. Table~\ref{tab:rtpc:table1} shows the values of the parameters of
Eq.~\ref{eq:rtpc:2a} for different centrality classes and \pp collisions. All
parameters except $c$ are close for the four $|\eta|$ intervals and similar
across systems. As previously mentioned, the parameter $c$ is related to the
transition in the logarithmic rise to the Plateau and the large difference
mainly reflects that the parameter is statistically not well constrained for
some of the datasets. For the \pp dataset, where the largest variation is
observed, we obtain similar results within statistical uncertainties if $c=2$
is used for all $|\eta|$-slices .

The separation power, $S_\sigma$, obtained with the final parameterizations for
\pp, 0-5\% \pbpb, and 40-60\% \pbpb collisions are shown in
Fig.~\ref{fig:rtpc:2}. As expected, the performance is the best for low
multiplicity events and decreases as the multiplicity increases and the
separation is better for the longest tracks ($0.6\leq|\eta|<0.8$). For
$p>\gevc{6.0}$ the $S_\sigma$ separation is nearly constant as expected because
of the logarithmic relativistic rise (as $\sigma \propto \mdedx$ a small
decrease of the separation is observed). The separation power plays an
important role in the determination of the systematic uncertainties described
in Sec.~\ref{sec:highptsyst}.

\subsubsection{Extraction of the particle fractions}

All the following results are for the sum of positive and negative pions,
kaons, and protons. Positive and negative yields were found to be comparable
at the 5\% level or better for all six centrality classes and \pp collisions.

Having determined the Bethe--Bloch and resolution curves as described in the
previous section, it is now straightforward to fit the \dedx spectra using the
sum of four Gaussian distributions for pions, kaons, protons, and
electrons. For each momentum interval, the \mdedx position and width of each
Gaussian are fixed. Figure~\ref{fig:rtpc:3} shows examples of these fits for
the momentum intervals 3.4--\gevc{3.6} and 8--\gevc{9}. The electrons are
hardly visible in any of the fits as the yield is below 1\% of the total. For
$\pt > \gevc{10}$, it is no longer possible to separate electrons from pions,
and the relative fraction of electrons is assumed to remain constant above
this \pt.  There is a small contamination of primary muons in the pions due to
the similar mass (and therefor similar \mdedx). High-\pt muons are
predominantly the result of semi-leptonic decays of hadrons containing heavy
quarks and for those decays one expects muon and electron branching ratios to
be similar, so the electron yield (fraction) is subtracted from the pions to
correct for the muon contamination.  This correction changes the pion yield by
less than 1\% in the full \pt range in agreement with MC simulations based on
the PYTHIA generator~\cite{Sjostrand:2006za}. Since this \dedx analysis is not
optimized for electrons and the contamination is extrapolated to high \pt,
half of the correction is assigned as a systematic uncertainty. The
contamination of (anti)deuterons in the (anti)proton sample is negligible
($<1\%$).

\begin{figure}[htbp]
  \centering
  \includegraphics[keepaspectratio, width=2.0\mylength]{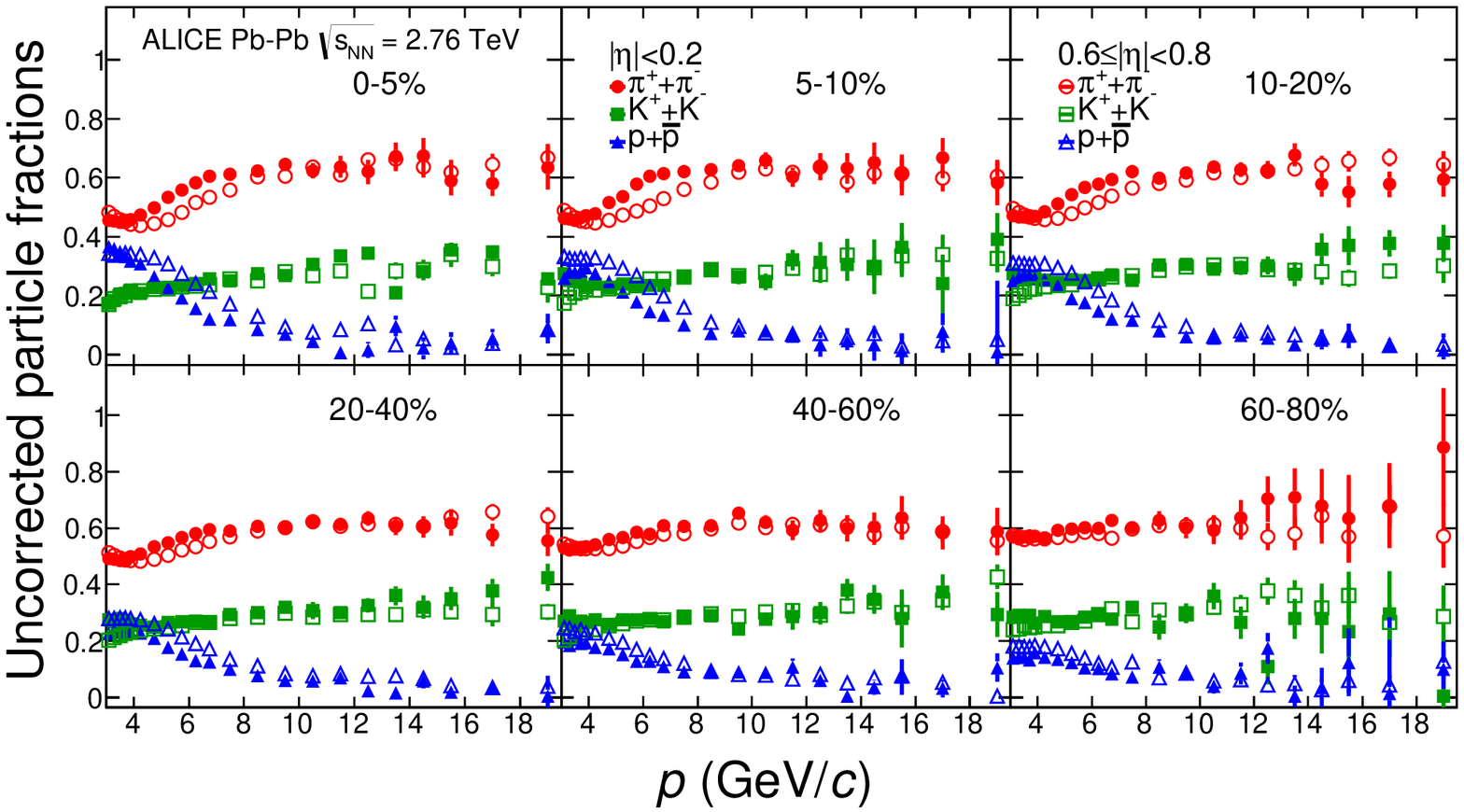}
  \includegraphics[keepaspectratio, width=2.0\mylength]{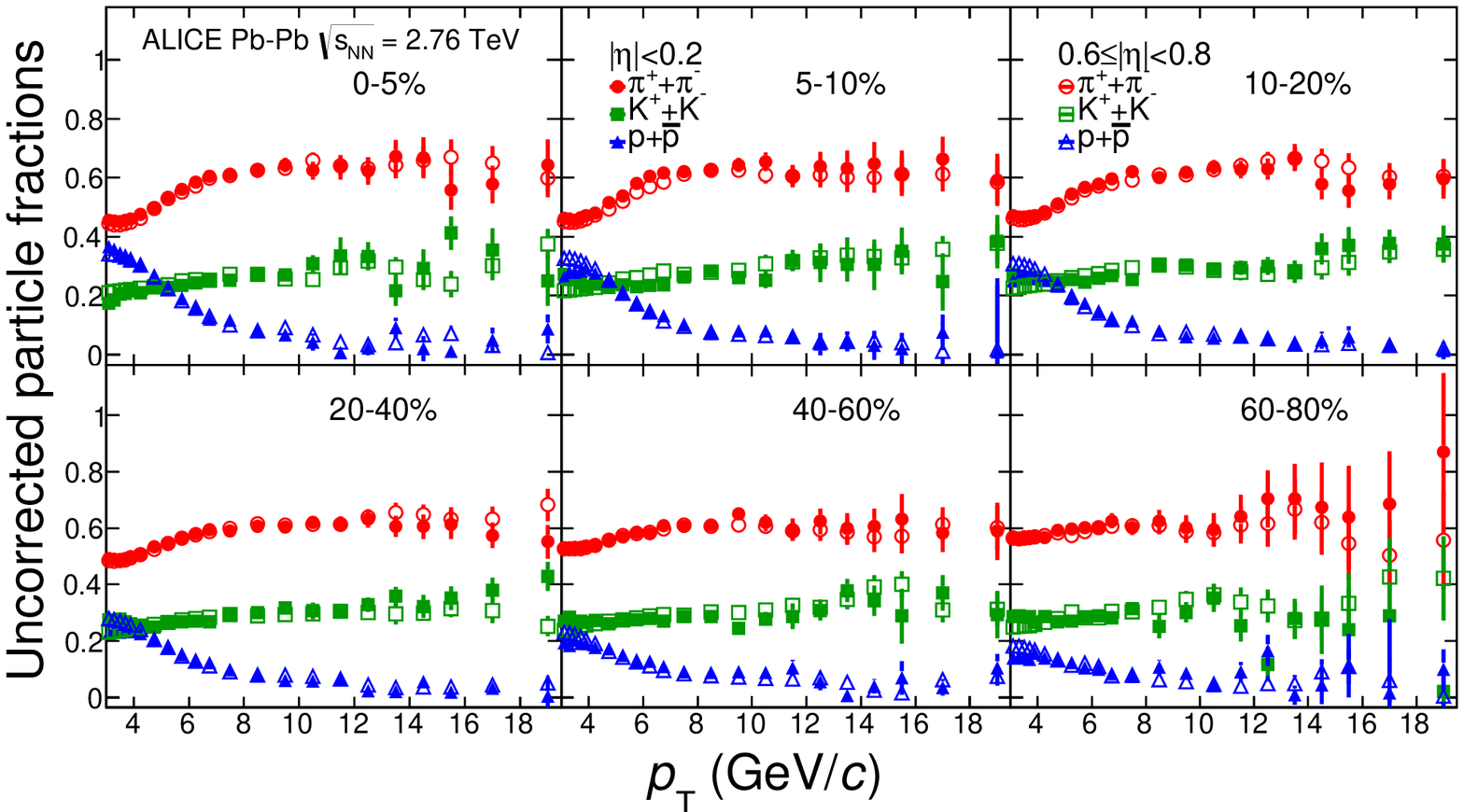}
  \caption{(Color online) Uncorrected particle fractions as a function of
    momentum (upper figure) and as a function of \pt (lower figure) for $|\eta|<0.2$ (full
    markers) and $0.6\leq|\eta|<0.8$ (empty markers). Charged pions, kaons,
    and (anti)protons are plotted with circles, squares, and triangles,
    respectively. The error bars indicate the statistical uncertainty.
    Results for six centrality classes are presented.}
  \label{fig:rtpc:5}
\end{figure}

The particle fractions, \ie, the contribution of charged pions ($f'_{\chpi}$),
kaons ($f'_{\chk}$), and (anti)protons ($f'_{\chp}$) to the yield of inclusive
charged particles, obtained as a function of momentum, are plotted in
Fig.~\ref{fig:rtpc:5} (upper figure) as a function of centrality for the two
extreme $|\eta|$ intervals. One observes a significant $\eta$ dependence of
pion and proton fractions for $p < \gevc{10}$. 

The extracted fractions as a function of transverse momentum are obtained
bin-by-bin using a weighting procedure
\begin{equation}
\label{eq:fraction}
f_{\rm id} (\langle \pt \rangle_{i}) = \sum_{j} f'_{\rm id} (\langle p \rangle_{j}) {\rm R}(\langle p \rangle_{i}, \langle \pt \rangle_{j}),
\end{equation}
where $f_{\rm id}$ ($f'_{\rm id}$) is given in bins of $\pt$ ($p$) and $\rm R$
is a response matrix reflecting the relation between $p$ and \pt bins. This
averaging introduces some smoothing of the fractions as neighboring \pt
fractions have contributions from the same $p$ fractions, but the analysis is
done in narrow $|\eta|$ intervals so only a few momentum bins contribute and the
fractions depend only weakly on $p$; therefore, we consider the systematic
effect of this procedure negligible. The fractions $f_{\rm id}$ are shown in
the lower panel of Fig.~\ref{fig:rtpc:5}. The transformation has little effect
for $|\eta|<0.2$, as expected, but we now observe that for $0.6 \leq |\eta| <
0.8$ the results are consistent with particle ratios being constant at
mid­rapidity. We find that all four pseudo-rapidity intervals are consistent
and the final fractions used to obtain the spectra in the next section are
computed as the weighted average of the four pseudorapidity intervals.

\subsubsection{Spectra}
\label{sec:spectra}
The invariant yields are obtained from the particle fractions using the
relation
\begin{equation}
\label{eq:rtpc:1}
\frac{{\rm d}^{2}N_{\rm id}}{{\rm d}\pt{\rm d}y}={\rm J}_{\rm id} \frac{\epsilon_{\rm ch}}{\epsilon_{\rm id}} f_{\rm id} \times \frac{{\rm d}^{2}N_{\rm ch}}{{\rm d}\pt{\rm d}\eta}.
\end{equation}
The first expression on the right hand side is the input from the PID
analysis, where ($\epsilon_{\rm ch}$) $\epsilon_{\rm id}$ is the efficiency
for (inclusive) identified charged particles and ${\rm J}_{\rm id}$ is the
Jacobian correction (from pseudorapidity $\eta$ to rapidity $y$) and $f_{\rm
  id}$ is the fractional yield. The second expression is the fully corrected
transverse momentum spectrum of inclusive charged particles that has already
been published by ALICE~\cite{Abelev:2012hxa}.\\

The relative efficiency correction, $\epsilon_{\rm ch}/\epsilon_{\rm id}$, was
found to be consistent within $\pm3\%$ for all centrality classes and \pp
collisions, and for event generators: PYTHIA~\cite{Sjostrand:2006za},
PHOJET~\cite{Bopp:1998rc}, and HIJING~\cite{PhysRevD.44.3501}. Thus, an
average correction was used and a systematic uncertainty of 3\% was
assigned. At high \pt the correction is nearly constant and on the order of
0.95. It is below 1 because the inclusive charged particle spectra contain
weakly decaying baryons such as $\Sigma^+$ that are not reconstructed with the
charged particle selection for primary particles. The proton and pion spectra
have been corrected for feed-down from weak decays using MC simulations for
the relative fraction of secondaries scaled to those extracted from
Distance-of-Closest-Approach MC template fits to
data~\cite{Abelev:2013vea}. For \pt $\approx$ 2 (3)\,GeV/$c$, the correction
is approximately 0.3\% ($4$\%) for the pion (proton) yield and decreasing with
increasing \pt. Scaling between data and MC has a limited precision and could
be different at higher \pt. To be conservative, half of the correction is
therefore assigned as a systematic uncertainty. This contribution to the
systematic uncertainty is still small, as shown in
Table~\ref{tab:rtpc:table2}.

\begin{figure}[htbp]
  \centering
\includegraphics[keepaspectratio, width=2.0\mylength]{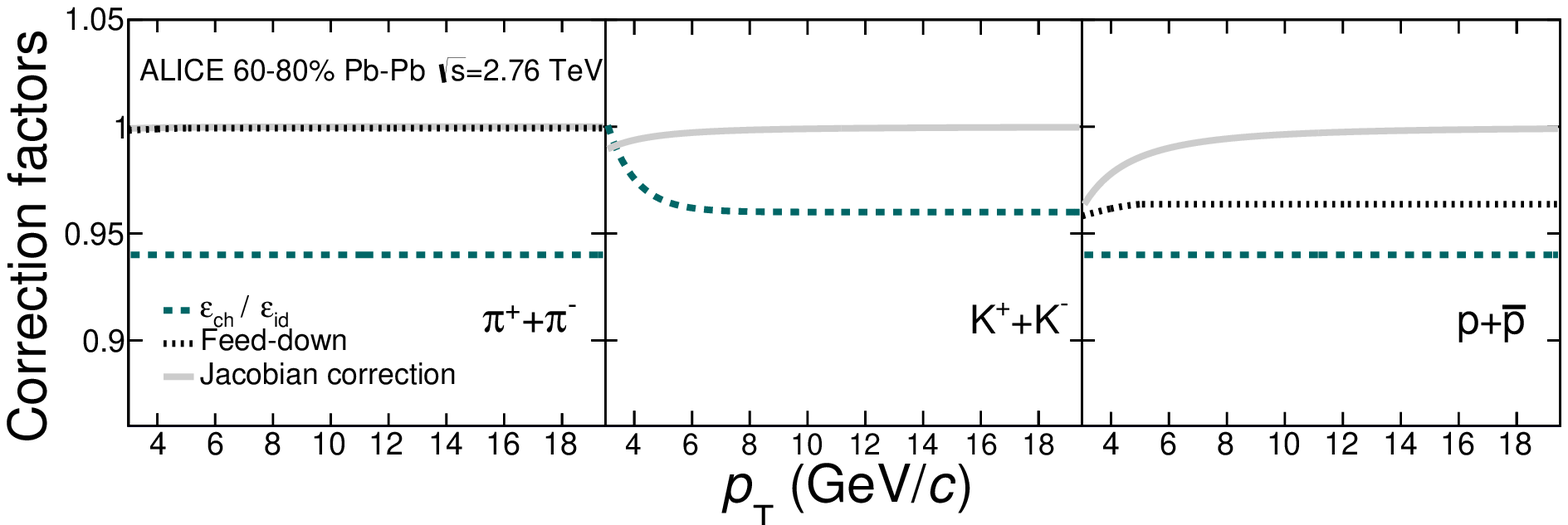}
\includegraphics[keepaspectratio, width=2.0\mylength]{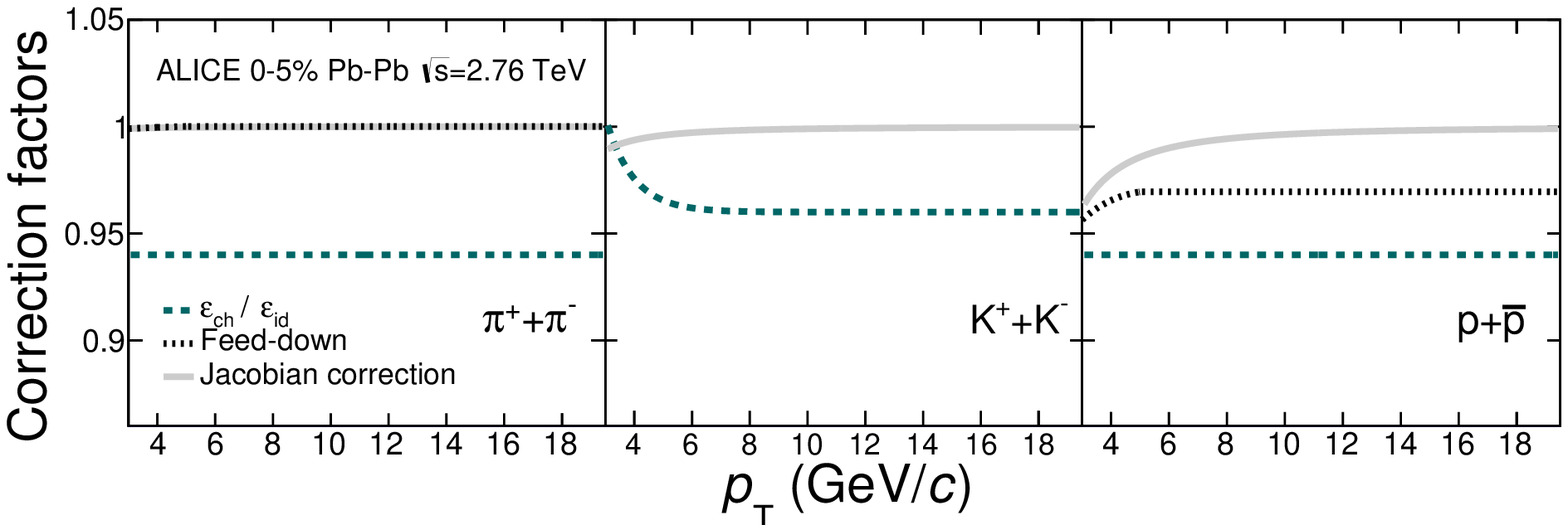}
  \caption{(Color online) Correction factors as a function of \pt. These are
    applied to the fractions of pions (left panels), kaons (middle panels),
    and protons (right panels). Results are presented for peripheral (upper
    figure) and central (lower figure) \pbpb collisions. The correction to the
    pion fraction due to the muon contamination is not drawn, but is $\leq
    1\%$.  Only pions and protons are corrected for feed-down.}
  \label{fig:corrections:1}
\end{figure}

The efficiency and feed-down corrections are plotted in
Fig.~\ref{fig:corrections:1} as a function of \pt for central and peripheral
\pbpb collisions. The Jacobian correction from $\eta$ to $y$, which has to be
included for the lower \pt bins, is also shown and the largest effect is
observed for protons, as expected. At \pt$\approx \gevc{3}$, the correction is
${\approx}5\%$, ${\approx}1\%$, and ${\ll}1$\% for protons, kaons, and pions,
respectively.

\subsubsection{Systematic uncertainties}
\label{sec:highptsyst}
The systematic uncertainty on the invariant yields has three main components:
event and track selection, efficiency correction of the fractions, and the
fraction extraction. Contributions from the event and track selection are
taken directly from the inclusive charged particle
result~\cite{Abelev:2012hxa}. The systematic uncertainties for the corrections
have been covered in the previous sections and are summarized in
Table~\ref{tab:rtpc:table2}.

\begin{figure}[htbp]
  \center
\includegraphics[keepaspectratio, width=2.0\mylength]{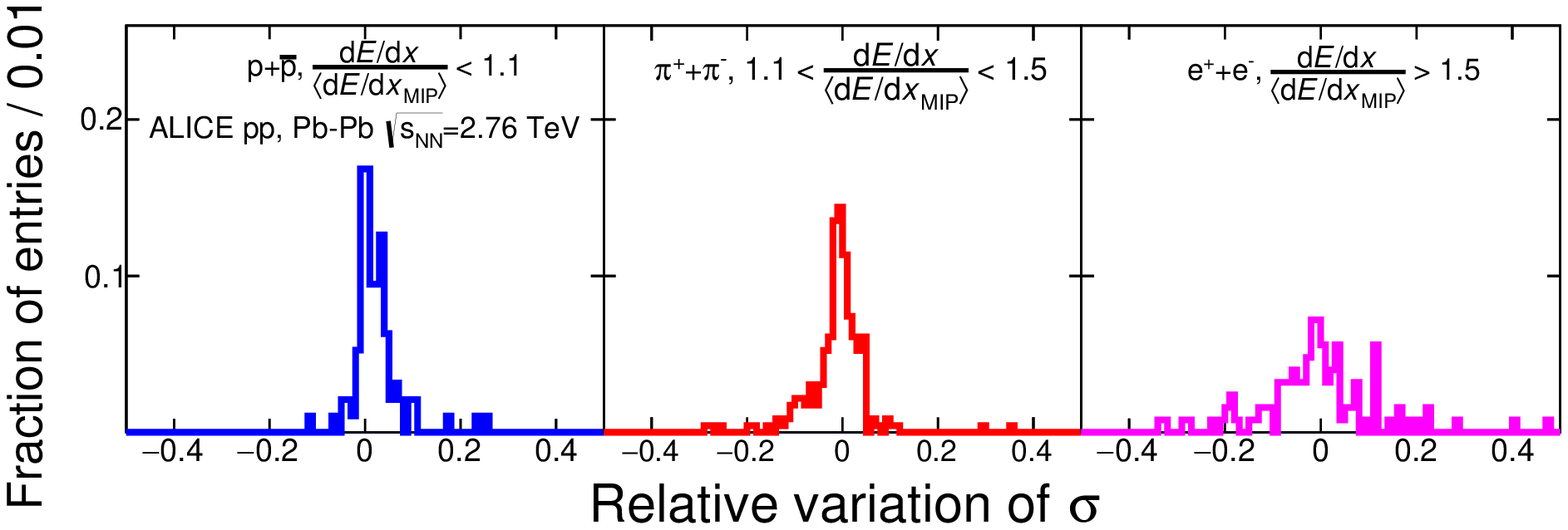}   
\includegraphics[width=1.38\mylength]{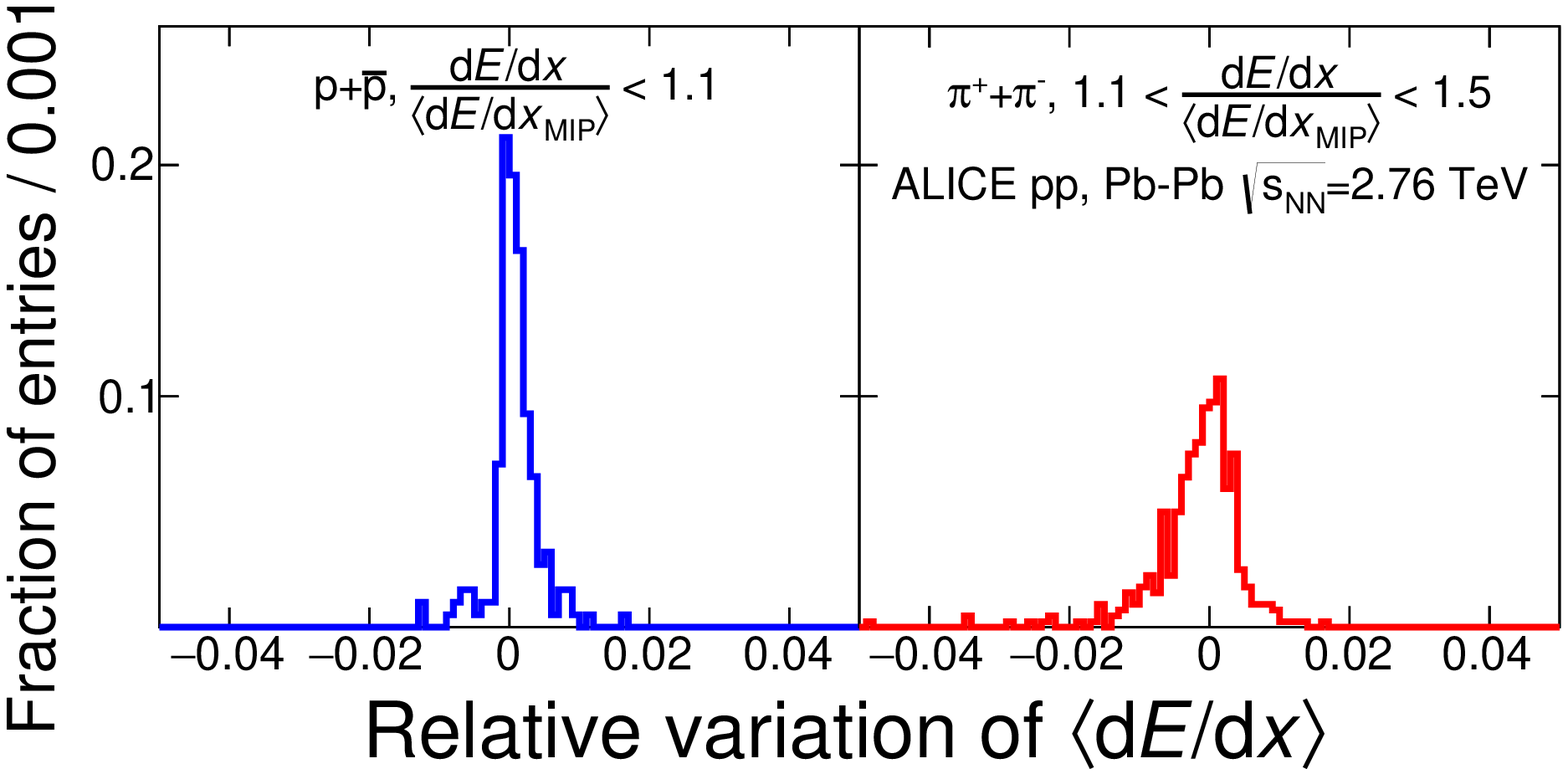}
  \caption{(Color online) Upper figure: Relative variation of the width
    parameterizations with respect to the measured values in different $\dedx
    / \langle \dedx_{\rm MIP} \rangle$ intervals. Lower figure: Relative
    variation of the Bethe--Bloch \mdedx parameterization with respect to the
    measured values in different $\dedx / \langle \dedx_{\rm MIP} \rangle$
    intervals. The distributions were constructed using all the available
    data, six centrality classes and \pp collisions with four sub-samples
    ($|\eta|$ intervals) each. }
  \label{fig:syst:1}
\end{figure}

The systematic uncertainty on the fractions is mainly due to the uncertainties
in the parameterization of the Bethe--Bloch and resolution curves used to
constrain the fits. This systematic uncertainty can be due to calibration
effects such that, for example, the \mdedx does not depend on \bg alone, it
can be related to the parameterizations not being able to describe the data
properly, or it can be due to the statistical precision of the external PID
data sets. To evaluate the uncertainty due to these effects, the deviation of
the fitted curves from the actual measured means and widths of the \dedx
spectra obtained from the analysis of the external pion, proton, and electron
samples are used. Figure~\ref{fig:syst:1} shows the relative variations; all
the available data were used for constructing the distributions, \ie, each of
the six centrality classes and \pp collisions have four sub-samples of tracks
at different $|\eta|$. It was found that the precision of all these data sets
is similar, so the final variation in systematic uncertainties for the same
observable for different centrality classes and \pp collisions is caused by
the different separation power shown in Fig.~\ref{fig:rtpc:2}. The results for
the width (Fig.~\ref{fig:syst:1} upper panel) are shown for $\rm p+\bar{p}$,
$\pi^{+}+\pi^{-}$, and $e^{+}+e^{-}$, corresponding to the different samples and
covering different $\mdedx / \langle \dedx_{\rm MIP} \rangle$ ranges.  In a
given $\mdedx / \langle \dedx_{\rm MIP} \rangle$ interval, the standard
deviation of the distribution was taken as the systematic uncertainty
associated with the extraction of the widths. An analogous analysis was done
for the Bethe--Bloch curve, an example of which is shown in the lower panel of
Fig.~\ref{fig:syst:1}.

In peripheral collisions, an additional contribution originating from the
statistical uncertainty in the fits to the external PID data has to be taken
into account for the Bethe--Bloch curve. The total systematic uncertainty is
assigned as the quadratic sum of both contributions and is the band shown
around the parameterizations in Fig.~\ref{fig:rtpc:2a}.

\begin{figure}[htbp]
  \center
  \includegraphics[keepaspectratio, width=2.0\mylength]{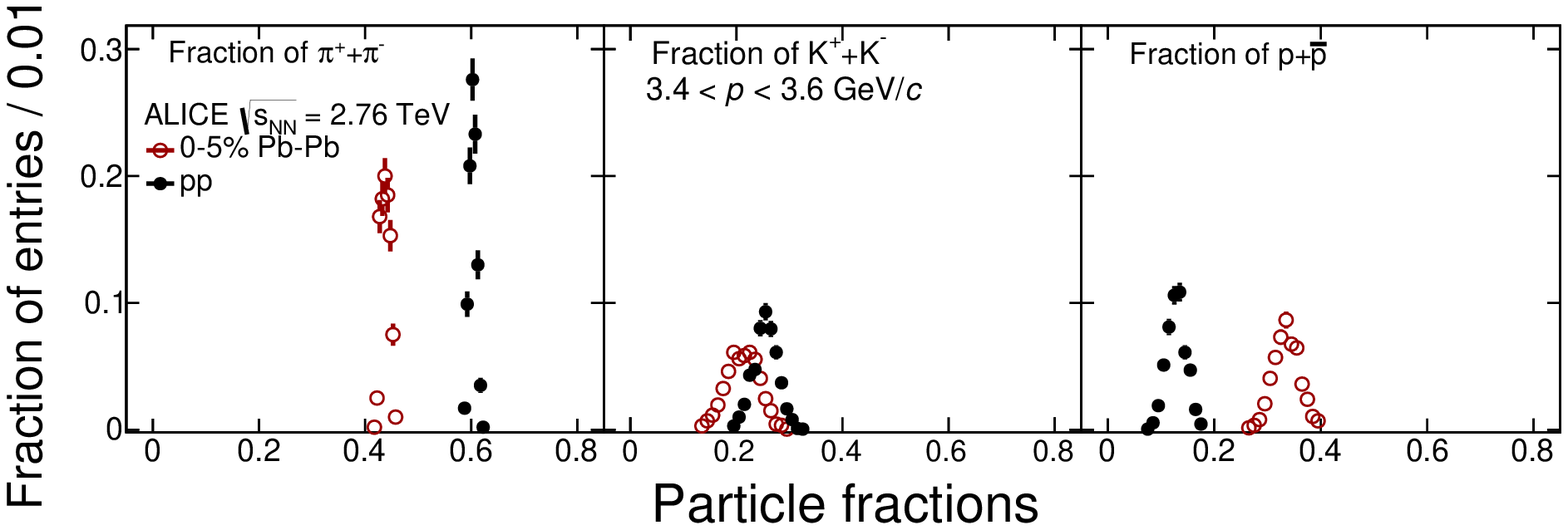}   
  \includegraphics[width=1.38\mylength]{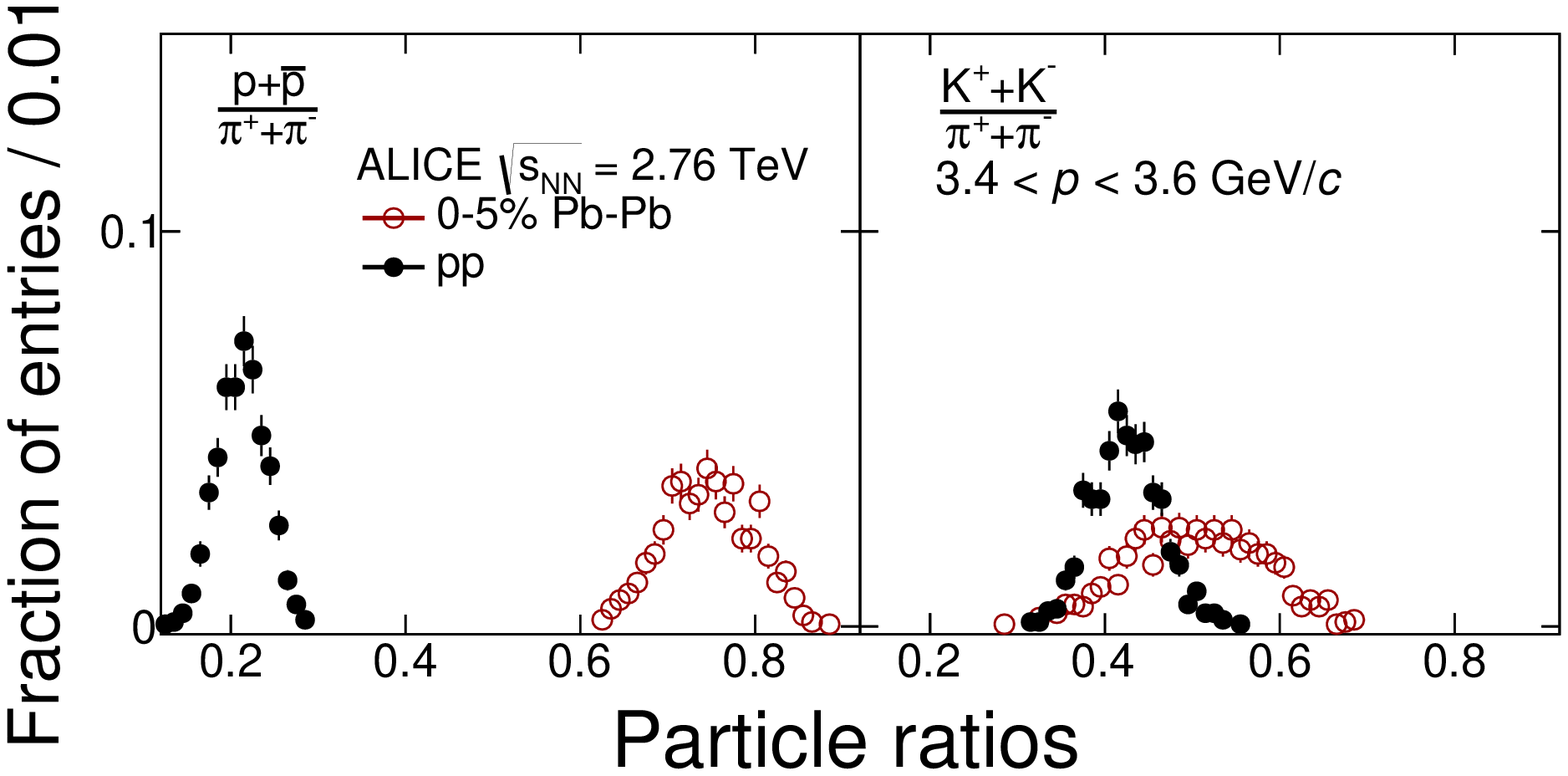}
  \caption{\label{fig:syst:4} (Color online). An example of the systematic
    uncertainty estimation in 0-5\% \pbpb and \pp collisions for $3.4 < p \leq
    \gevc{3.6}$. Upper figure: From left to right: the variation of extracted
    fractional yields for pions (left panel), kaons (middle panel), and
    protons (right panel) when the fixed values for the \mdedx and the
    resolution are randomly varied. Lower figure: the corresponding variation
    of the particle ratios.}
\end{figure}

The propagation of the uncertainties to the particle fractions is done by
refitting the \dedx spectra, while randomly varying the constrained
parameters, \mdedx and $\sigma$, within the uncertainty for the
parameterizations assuming a Gaussian variation centered at the nominal
value. For each \pt bin, all the \mdedx and $\sigma$ values are randomly varied
and refitted 1000 times resulting in fraction distributions like those shown
in Fig.~\ref{fig:syst:4}. The systematic uncertainties assigned to the
particle fractions are the standard deviation of the associated
distributions. By using the same method for the particle ratios
(Fig.~\ref{fig:syst:4} lower panel), the correlation in the fit between the
extracted yields for the two different particle species are directly taken
into account. At high \pt, the variation becomes dominated by statistical
fluctuations due to the limited amount of data. But, as the fractions are
nearly constant there (see Fig.~\ref{fig:rtpc:5}) and the separation is also
nearly constant (see Fig.~\ref{fig:rtpc:2}), a constant absolute systematic
uncertainty is assigned for $\pt > \gevc{8}$.

A summary of the different contributions to the systematic uncertainty is
shown in Table~\ref{tab:rtpc:table2} for all centrality classes and for two
representative \pt regions. For pions, the dominant contribution comes from the
event and track selection, which amounts to 7--8\% over the whole \pt range
while the PID systematic uncertainty stays between 1--2\%. For kaons and
protons, the PID systematic uncertainty is the largest. The systematic
uncertainty decreases with increasing separation and is smaller where the
fractions are larger, see Fig.~\ref{fig:rtpc:5}. For protons at $\pt =
\gevc{3}$, the two effects largely compensate (the fractional yields increase
for more central collisions) to keep the systematic uncertainty nearly
constant. For kaons, at the same \pt, there is a strong centrality dependence
because the fractional yields also are lower for more central collisions. For
the lower multiplicity intervals (\pp and 60-80\% centrality) this trend is
broken because of the significant statistical uncertainty in the parameterized
curves.

At high \pt (${\approx}\gevc{10}$) the PID systematic uncertainty for kaons
stays between 7--8\% for \pbpb collisions and is around 5\% for \pp
collisions. For protons, the contribution is 16--20\% (except for 60-80\% \pbpb
collisions where it is 29\% due to a much larger statistical uncertainty in
the fits to the external PID data).

\begin{table}[htb]
\centering
\small
\footnotesize
\begin{tabular}{lcccccccccc}
\hline \hline & \multicolumn{2}{c}{ \uline{$\pi^{+}+\pi^{-}$} } &
\multicolumn{2}{c}{ \uline{${\rm K^{+}+K^{-}}$} } & \multicolumn{2}{c}{
  \uline{${\rm p+\bar{p}}$} } & \multicolumn{2}{c}{ \uline{$\rm
    K/\pi$}} & \multicolumn{2}{c}{ \uline{$\rm
    p/\pi$} } \\ \pt (GeV/$c$) & 2.0 & 10 & 3.0 & 10 & 3.0 & 10
& 3.0 & 10 & 3.0 & 10 \\ \hline \hline

\multicolumn{11}{c}{\bf{\pbpb collisions (0-5\%)}} \\ \hline  
(a) & 8.4\% & 8.1\% & 8.2\% & 8.1\%  & 8.2\% & 8.1\% &  \multicolumn{4}{c}{ - } \\
(b) & \multicolumn{2}{c}{$<0.1$\%} & \multicolumn{2}{c}{ - } & 2.1\% &  1.5\%  & \multicolumn{2}{c}{ $<0.1$\% } &  2.1\% &  1.5\% \\
(c) & 0.1\% & 1.7\% &  \multicolumn{4}{c}{ - } & 0.6\% & 1.7\% & 0.6\% & 1.7\% \\
(d) &  1.5\% & 2.2\% & 18\% & 8.4\%  & 9.8\% & 17\% & 22\% & 10\% & 11\%  & 16\%\\ \hline \hline

    \multicolumn{11}{c}{\bf{\pbpb collisions (5-10\%)}} \\ \hline
(a) & 8.4\% & 8.2\% & 8.2\% & 8.2\%  & 8.2\% & 8.2\% &  \multicolumn{4}{c}{ - } \\
(b) & \multicolumn{2}{c}{$<0.1$\%} & \multicolumn{2}{c}{ - } & 2.1\% &  1.5\%  & \multicolumn{2}{c}{ $<0.1$\% } &  2.1\% &  1.5\% \\
(c) & 0.2\% & 1.5\% &  \multicolumn{4}{c}{ - } & 0.6\% & 1.5\% & 0.6\% & 1.5\% \\
(d) &  1.4\% & 2.2\% & 16\% & 8.0\%  & 9.5\% & 16\% & 18\% & 10\% & 9.8\%  & 15\%\\ \hline \hline

    \multicolumn{11}{c}{\bf{\pbpb collisions (10-20\%)}} \\ \hline
(a) & 8.3\% & 8.1\% & 8.2\% & 8.1\%  & 8.2\% & 8.1\% &  \multicolumn{4}{c}{ - } \\
(b) & \multicolumn{2}{c}{$<0.1$\%} & \multicolumn{2}{c}{ - } & 2.2\% &  1.8\%  & \multicolumn{2}{c}{ $<0.1$\% } &  2.2\% &  1.8\% \\
(c) & 0.3\% & 1.3\% &  \multicolumn{4}{c}{ - } & 0.6\% & 1.3\% & 0.6\% & 1.3\% \\
(d) &  1.5\% & 2.3\% & 16\% & 8.9\%  & 10\% & 20\% & 16\% & 11\% & 9.2\%  & 18\%\\ \hline \hline

    \multicolumn{11}{c}{\bf{\pbpb collisions (20-40\%)}} \\ \hline
(a) & 8.4\% & 8.2\% & 8.2\% & 8.2\%  & 8.2\% & 8.2\% &  \multicolumn{4}{c}{ - } \\
(b) & \multicolumn{2}{c}{$<0.1$\%} & \multicolumn{2}{c}{ - } & 2.1\% &  1.6\%  & \multicolumn{2}{c}{ $<0.1$\% } &  2.1\% &  1.6\% \\
(c) & 0.2\% & 1.3\% &  \multicolumn{4}{c}{ - } & 0.5\% & 1.3\% & 0.5\% & 1.3\% \\
(d) &  1.5\% & 2.2\% & 15\% & 8.4\%  & 10\% & 17\% & 16\% & 11\% & 10\%  & 17\%\\ \hline \hline

    \multicolumn{11}{c}{\bf{\pbpb collisions (40-60\%)}} \\ \hline
(a) & 8.7\% & 8.5\% & 8.6\% & 8.5\%  & 8.6\% & 8.5\% &  \multicolumn{4}{c}{ - } \\
(b) & \multicolumn{2}{c}{$<0.1$\%} & \multicolumn{2}{c}{ - } & 1.9\% &  1.6\%  & \multicolumn{2}{c}{ $<0.1$\% } &  1.9\% &  1.6\% \\
(c) & 0.3\% & 1.1\% &  \multicolumn{4}{c}{ - } & 0.5\% & 1.1\% & 0.5\% & 1.1\% \\
(d) &  1.4\% & 2.1\% & 14\% & 8.0\%  & 11\% & 17\% & 15\% & 10\% & 11\%  & 17\%\\ \hline \hline

    \multicolumn{11}{c}{\bf{\pbpb collisions (60-80\%)}} \\ \hline
(a) & 10\% & 9.7\% & 9.8\% & 9.7\%  & 9.8\% & 9.7\% &  \multicolumn{4}{c}{ - } \\
(b) & \multicolumn{2}{c}{$\leq 0.1$\%} & \multicolumn{2}{c}{ - } & 2.0\% &  1.8\%  & \multicolumn{2}{c}{ $\leq 0.1$\% } &  2.0\% &  1.8\% \\
(c) & 0.3\% & 0.8\% &  \multicolumn{4}{c}{ - } & 0.4\% & 0.8\% & 0.4\% & 0.8\% \\
(d) &  1.4\% & 2.4\% & 16\% & 7.1\%  & 20\% & 29\% & 16\% & 8.9\% & 18\%  & 22\%\\ \hline \hline

    \multicolumn{11}{c}{\bf{pp collisions}} \\ \hline
(a) & 7.4\% & 7.6\% & 7.4\% & 7.6\%  & 7.4\% & 7.6\% &  \multicolumn{4}{c}{ - } \\
(b) & \multicolumn{2}{c}{$\leq 0.1$\%} & \multicolumn{2}{c}{ - } & 2.0\% &  1.8\%  & \multicolumn{2}{c}{ $\leq 0.1$\% } &  2.0\% &  1.8\% \\
(c) & 0.4\% & 0.6\% &  \multicolumn{4}{c}{ - } & 0.5\% & 0.6\% & 0.5\% & 0.6\% \\
(d) &  1.1\% & 1.7\% & 16\% & 5.7\%  & 24\% & 17\% & 16\% & 6.8\% & 25\%  & 13\%\\ \hline 
(e)  &  \multicolumn{6}{c}{ 3.0\% } & \multicolumn{4}{c}{ 4.2\% }  \\
\hline \hline

\end{tabular}
\caption{\label{tab:rtpc:table2} Summary of the systematic uncertainties for
  the charged pion, kaon, and (anti)proton spectra and for the particle
  ratios. The different contributions are (a) event and track selection,
  (b) feed-down correction, (c) correction for muons, (d) parameterization of
  Bethe--Bloch and resolution curves, and (e) efficiency correction (same for
  all systems). Note that ${\rm K}/\pi = ({\rm
    K^{+}+K^{-}})/(\pi^{+}+\pi^{-})$ and ${\rm p}/\pi = ({\rm
    p+\bar{p}})/(\pi^{+}+\pi^{-})$.  }
\end{table}

\subsection{HMPID analysis of Pb-Pb data}
\label{sec:hmpid}

\begin{figure}[htb]
  \centering
  \includegraphics[width=1.0\mylength]{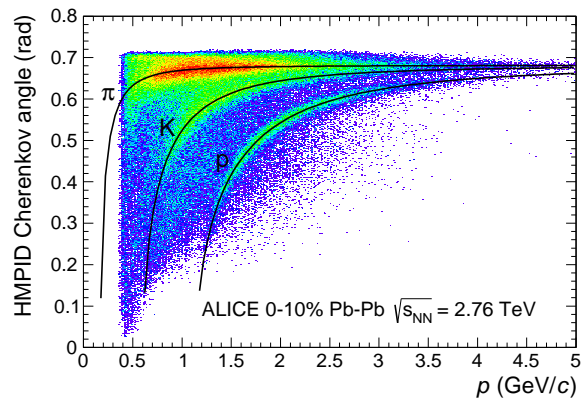}
  \caption{Cherenkov angle measured in the HMPID as a function of the momentum
    $p$ in 0-10\% central \pbpb collisions.  The solid lines represent the
    theoretical curves for each particle species. The $z$-axis indicated by
    the color scale is logarithmic.}
  \label{AnglevsMom}
\end{figure}

The HMPID is used in order to constrain the uncertainty of the charged pion,
kaon, and (anti)proton measurements in the transition region between the TOF
and TPC relativistic rise methods (in the region around $\pt =
\gevc{3}$). Thus, it both improves the precision of the measurement and
validates the other methods in the region where they have the worst PID
separation.

The HMPID \cite{hmpid:tdr} detector consists of seven identical proximity
focusing RICH (Ring Imaging Cherenkov) counters. Photon and charged particle
detection is provided by a Multi-Wire Proportional Chamber (MWPC) coupled to a
CsI photocathode segmented into pads of size $0.8{\times}0.84$\,cm$^2$ (the
probability to obtain an amplified signals for an incident photon, the quantum
efficiency, is~$\approx$~25\% for $\lambda_{\rm{ph}}$ = 175\,nm). The
amplification gas is CH$_4$ at atmospheric pressure with an anode-cathode gap
of 2\,mm; the operational voltage is 2050\,V corresponding to a gain of
${\approx}4\times$10$^4$. It is located at about 5 m from the beam axis,
covering a limited acceptance of $|\eta| < 0.5$ and $1.2^{\circ} < \varphi <
58.5^{\circ}$.

The HMPID analysis uses the 2011 \pbpb data with around $7.8 \times 10^{6}$
central triggered events (0-10\% centrality) and $5 \times 10^{6}$
semi-central triggered events (10-50\% centrality\footnote{To match centrality
  classes with the high-\pt analysis only spectra for 0-40\% will be shown in
  this paper. Results for 20-30\%, 30-40\%, and 40-50\% are available on
  HepData.}).  The event and track selection is similar to the one described
in Sec.~\ref{sec:eventsAndTracks}, but in addition it is required that the
tracks are propagated and matched to a primary ionization cluster in the MWPC
gap of the HMPID detector (denoted matched cluster in the following). The
matching efficiency, including spurious matches, is ${\approx}95\%$ (see
$\varepsilon_{\text{match}}$ below). The matching criteria are tightened to
reject the fake cluster-track matches, which account for ${\approx}30$--$40\%$
(see $C_{\text{distance}}$ later), so that only tracks matched with their
corresponding primary ionization cluster are identified. The PID in the HMPID
is done by measuring the Cherenkov angle, $\thetac$ \cite{hmpid:tdr}, given by
\begin{equation}
\cos\thetac = \frac{1}{n\beta} \Rightarrow \thetac = \arccos \left(\frac{\sqrt{p^{2} + m^{2}}}{np}\right),
\label{costheta}
\end{equation}
where $n$ is the refractive index of the radiator used (liquid C$_{6}$F$_{14}$
with $n$ = 1.29 at temperature $T = 20\,^\circ\text{C}$ for photons with
energy 6.68\,eV). Figure~\ref{AnglevsMom} shows the Cherenkov angle as a
function of the momentum for central \pbpb collisions.

The measurement of the single photon $\thetac$ angle in the HMPID requires
knowledge of the track impact position and angle. These are estimated from the
track extrapolation from the central tracking devices up to the radiator
volume, where the Cherenkov photons are emitted. Only one matched cluster is
associated to each extrapolated track, selected as the closest cluster to the
extrapolated track point on the cathode plane, with a charge above
${\approx}120$~ADC. The cut on the charge excludes clusters from electronic
noise ($\sigma_{pedestal} \approx 1$\,ADC) and photons. The matching
efficiency is defined for tracks extrapolated to the HMPID acceptance as
\begin{equation}
\varepsilon_{\text{match}} = \frac{N(\text{Extrapolated with matched cluster})}{N(\text{Extrapolated})}.                                
\end{equation}
This efficiency is ${\approx}95\%$ and independent of momentum, particle
species, and event multiplicity.

\begin{figure}[htb]
  \centering
  \includegraphics[width=1.38\mylength]{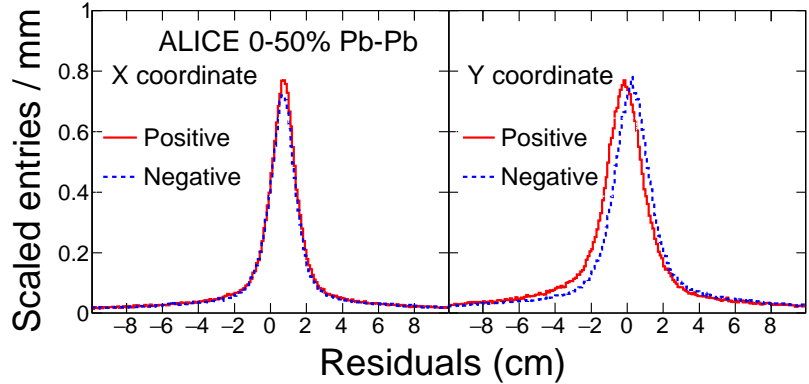}
  \caption{(Color online) Distribution of the $X$ (left panel) and $Y$ (right
    panel) residuals between the matched cluster position and the closest
    extrapolated track point at the HMPID chamber plane (HMPID module 2), for
    positive and negative tracks with $\pt > \gevc{1.5}$ in \pbpb collisions
    (0-50\% centrality). The histograms have been scaled to have a similar
    maximum value. The small shift between positive and negative tracks
    in the $Y$ residuals is due to a radial residual misalignment and an
    imperfect estimate of the energy loss in the material traversed by the
    track and is not corrected for in the calculation of the residual
    distance. }
  \label{Residual}
\end{figure}

\begin{figure}[htb]
  \centering
  \includegraphics[width=1.0\mylength]{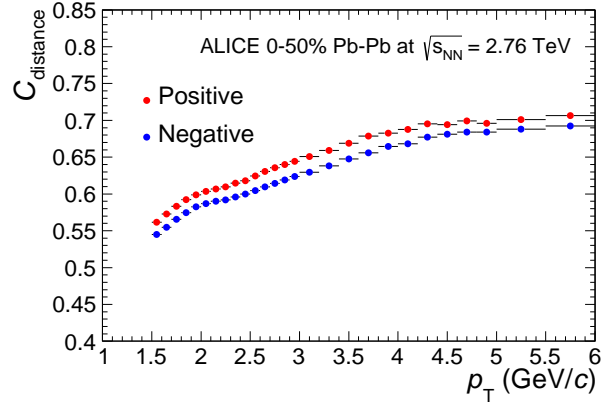}
  \caption{(Color online) The distance cut correction,
    $C_{\text{distance}}$, as a function of \pt for positive (red)
    and negative (blue) tracks, respectively, in \pbpb collisions (0-50\%
    centrality).}
  \label{CutCorr}
\end{figure}

In Fig.~\ref{Residual}, the residuals distribution between the track
extrapolation and the matched cluster position in local chamber coordinates,
$X$ and $Y$, for tracks with $\pt > \gevc{1.5}$ is shown. The distributions
have a resolution of $\sigma_{res} \approx~2~\text{cm}$. To reject fake
cluster-match associations in the detector, mainly the situation when there is
no correct signal to match as for example the particle was absorbed or
deflected in the material between the TPC and the HMPID detector, a selection
on the distance, $\sqrt{X^2 + Y^2}$, computed on the cathode plane between the
track extrapolation and the matched cluster is applied. This distance has to be
less than 5~cm. This represents the best compromise between the loss of
statistics and the probability of an incorrect association, where the latter
becomes negligible (${<}0.1\%$) even in the most central collisions, as
estimated from MC simulations. The distance cut leads to a correction factor
\begin{equation}
C_{\text{distance}} = \frac{N(\text{Extrapolated with matched cluster distance
    $< 5$ cm})}{N(\text{Extrapolated with matched cluster})},
\end{equation}
for each momentum bin and does not depend on event
multiplicity. Fig.~\ref{CutCorr} shows this correction factor as a function of
\pt for positive and negative tracks integrated over the centrality classes
(0-50\%).

\begin{figure}[htbp]
  \centering
  \includegraphics[width=1.38\mylength]{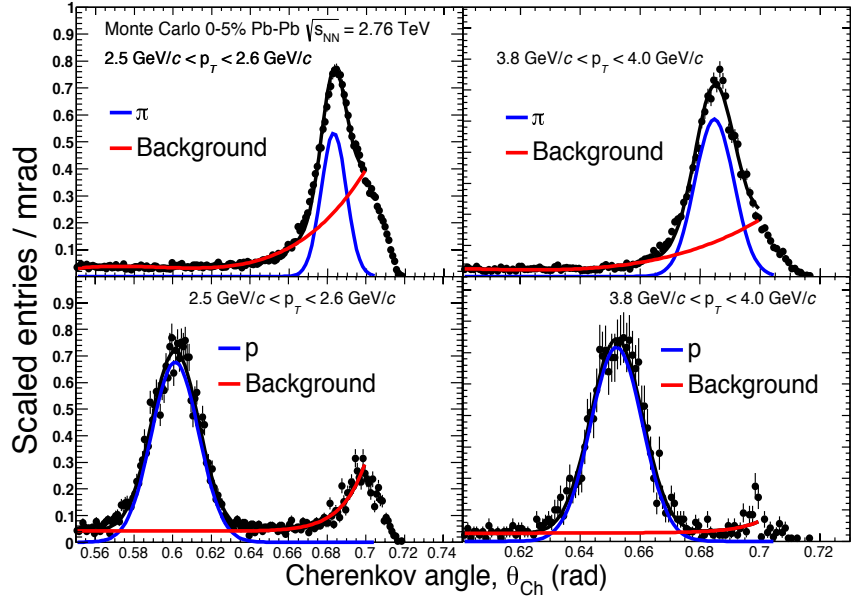}
  \caption{(Color online) Fit to the \thetac-distributions of pions (upper
    panel) and protons (lower panel) obtained in MC simulations for two
    different momentum bins. The histograms have been scaled to have a similar maximum value.}
  \label{MCbackground}
\end{figure}

Starting from the photon cluster coordinates on the photocathode, a
back-tracking algorithm calculates the corresponding emission angle. The
Cherenkov photons are selected by the Hough Transform Method
(HTM)~\cite{Cozza:2001zz}, which for each track transforms the coordinates of
photon hits into emission angles. The angle interval with the most hit
candidates is selected and $\thetac$ is computed as the weighted mean of the
single photon angles. In central \pbpb collisions, where the total number of
signals in the HMPID chambers is large, it is possible that the angle is
constructed based on hits not corresponding to the Cherenkov photons
associated with the track. This results in a significant reduction of the PID
efficiency in the most central collisions. Figure~\ref{MCbackground} gives an
example of the same effect in MC simulations. The response function consists
of a Gaussian distribution for correctly assigned rings (signal) plus a
distribution strongly increasing with the Cherenkov angle for incorrectly
assigned rings (background). The signals from other tracks and photons in the
same event are uniformly distributed on the chamber plane, and so the
background rises with \thetac since the probability of finding background
clusters increases. The background contribution decreases with increasing
track momentum because higher momentum tracks give rise to a larger number of
Cherenkov photons and have a smaller inclination angle, producing rings that
are more likely to be fully contained inside the acceptance. As a result of
this, the probability of incorrectly associating an angle computed from
background clusters to the track decreases. The shoulder in the distribution
starting at 0.7 rad is a boundary effect due to the finite geometrical
acceptance of the chamber.

\begin{figure}[htbp]
  \centering
  \includegraphics[width=2.\mylength]{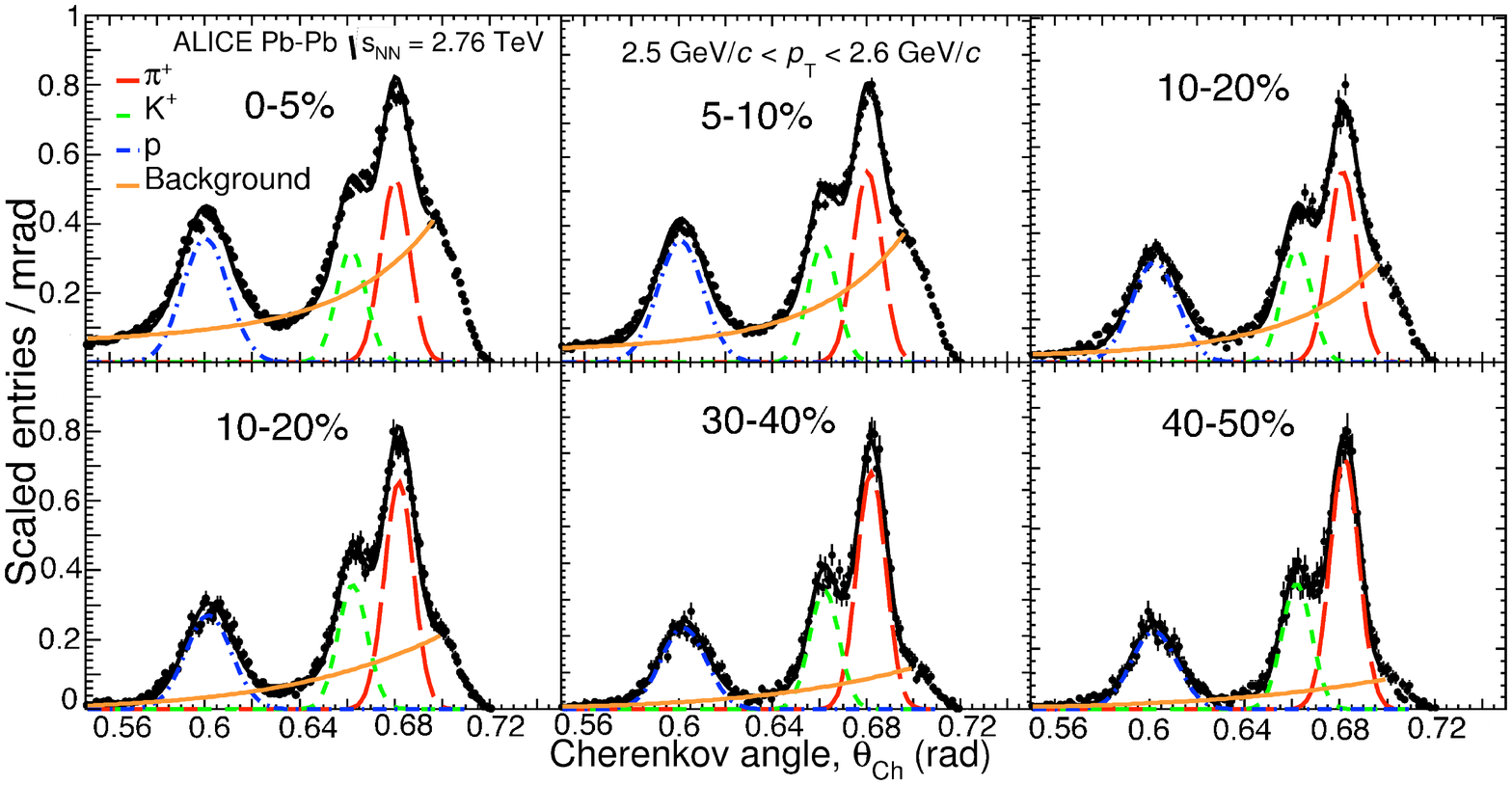}
  \includegraphics[width=2.\mylength]{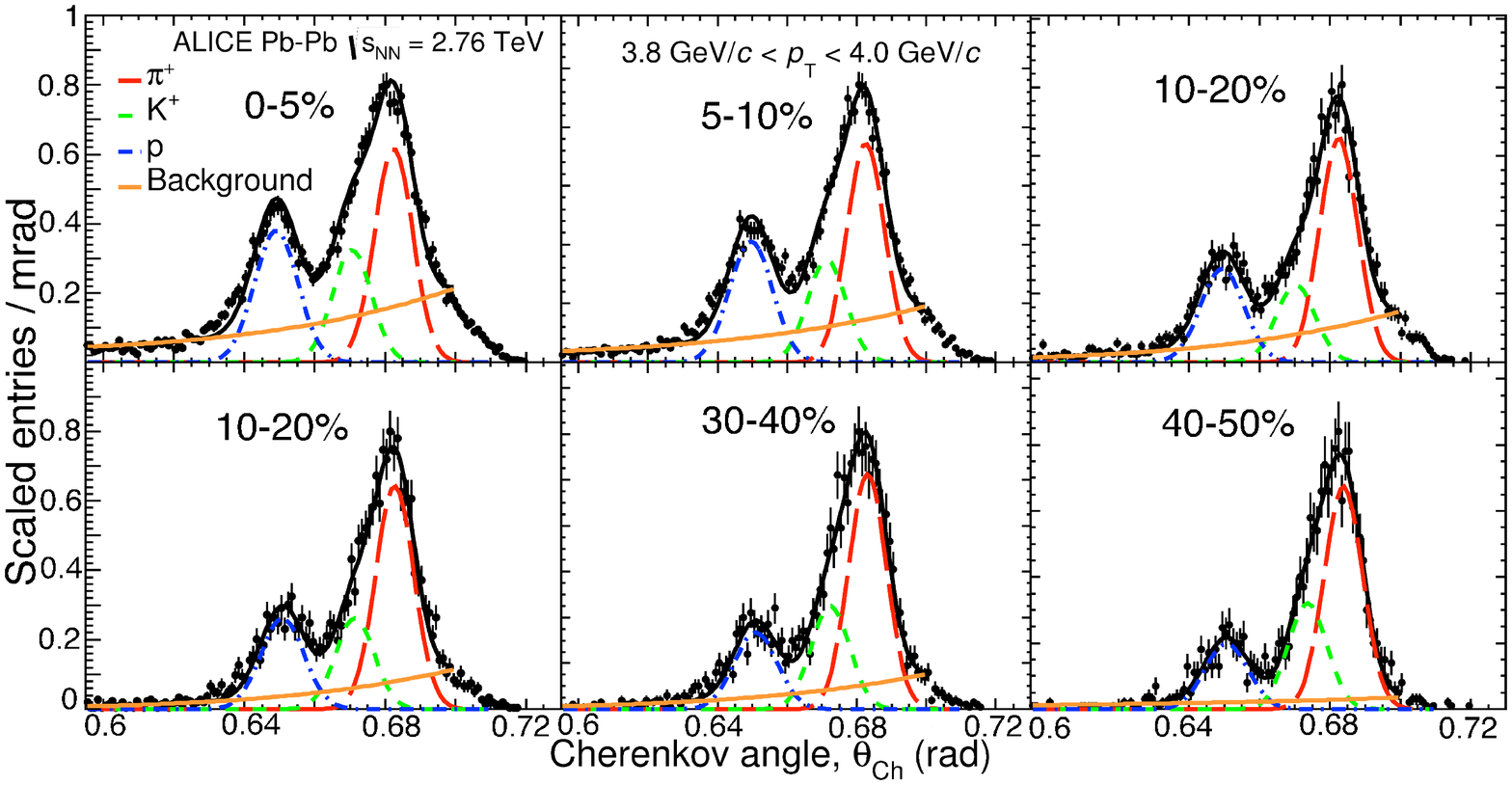}
  \caption{(Color online) Distributions of the Cherenkov angle measured in the
    HMPID for positive tracks having \pt in the range 2.6--\gevc{2.7} (upper
    figure) and in the range 3.8--\gevc{4.0} (lower figure), for six different
    centrality classes, 0-5\%, 5-10\%, 10-20\%, 20-30\%, 30-40\%, and
    40-50\%. The histograms have been scaled to have a similar maximum
    value. The shoulder in the distributions starting at 0.7 rad is a boundary
    effect due to the finite chamber geometrical acceptance.}
  \label{allFits}
\end{figure}

Figure~\ref{allFits} gives examples of the reconstructed Cherenkov angle
distributions in two narrow \pt intervals for different centrality classes;
the reconstructed angle distribution is fitted with a sum of three Gaussian
distributions, corresponding to the signals from pions, kaons, and protons,
plus a distribution associated with the misidentified tracks that is modeled
with a 6th-degree polynomial function that minimizes the reduced $\chi^{2}$ of
the fit.

\begin{figure}[htb]
  \centering
  \includegraphics[width=1.0\mylength]{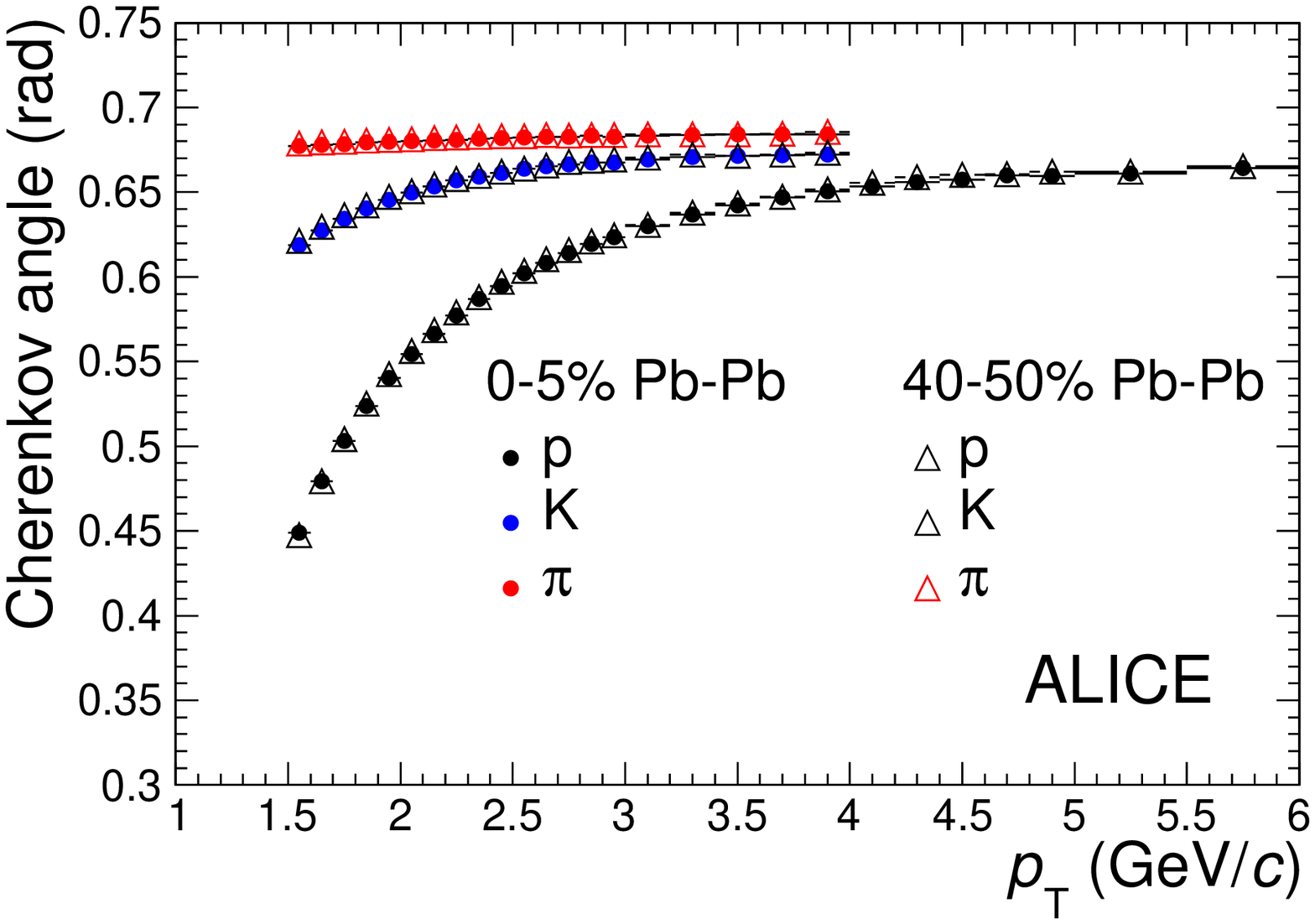}
  \includegraphics[width=1.0\mylength]{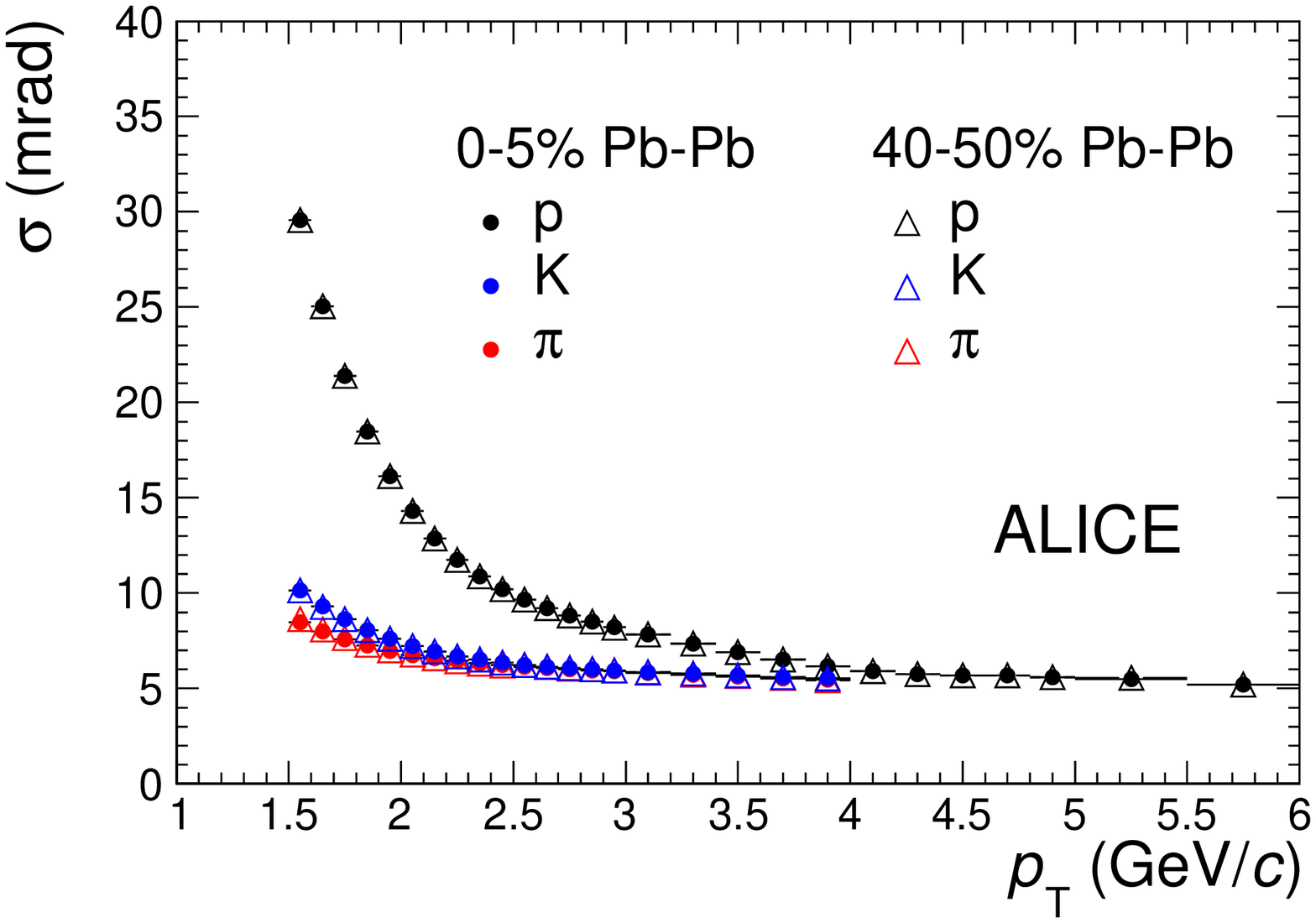}
  \caption{(Color online) Mean Cherenkov angle (upper panel) and standard
    deviation (lower panel) values for pions, kaons, and protons obtained by
    the three-Gaussian fitting procedure as a function of \pt for 0-5\% and
    40-50\% centrality \pbpb collisions. The data points from the two
    different centrality classes overlap such that the difference is smaller
    than the size of the symbols used.}
  \label{MeanSigma}
\end{figure}

\begin{figure}[htb]
  \centering
  \includegraphics[width=1.38\mylength]{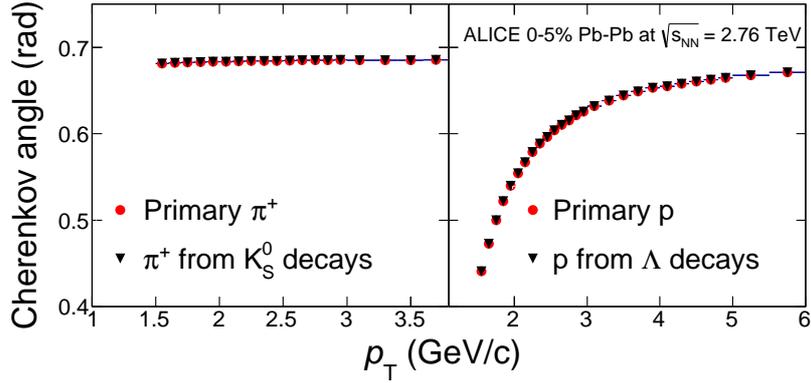}
  \caption{(Color online) Comparison of the mean Cherenkov angle values
    obtained by the three-Gaussian fitting procedure and those evaluated from
    the V$^0$s study for pions (left panel) and protons (right panel) in the
    most central \pbpb collisions.}
  \label{MeanAngleDatavsV0}
\end{figure}

The fitting is performed in 2 steps. In the first step, the initial parameters
are based on the expected values. For the signal, the means $\mthetac_i$ are
obtained from Eq.~\ref{costheta}, tuning the refractive index to match the
observed Cherenkov angles, and the sigma values $\sigma_i$ are taken from the
MC distribution in the given transverse momentum bin. The initial shape of the
6th-degree polynomial background is taken from MC simulations. Furthermore,
the signal parameters are constrained to the ranges: [$\mthetac_i$ -
  $\sigma_i$,$\mthetac_i$ + $\sigma_i$] for the means, and [$\sigma_i$ -
  0.1$\cdot\sigma_i$, $\sigma_i$ + 0.1$\cdot\sigma_i$] for the widths. After
this first step, the \pt dependence of each parameter is fitted with a
continuous function. In the second step, the fitting is repeated with only the
yields as free parameters and constraining the mean and sigma values to the
continuous functions.
The means and widths constrained in this way are all found to be independent
of centrality as shown in Fig.~\ref{MeanSigma} for 0-5\% and 40-50\%
centrality classes. In Fig.~\ref{MeanAngleDatavsV0}, a comparison is shown
between the mean values of the Cherenkov angle obtained from the fitting
procedure with those obtained using a clean sample of protons and pions
identified from $\Lambda$ and $\rm{K}^{0}_{S}$ decays.

\begin{figure}[htb]
  \centering
  \includegraphics[width=1.0\mylength]{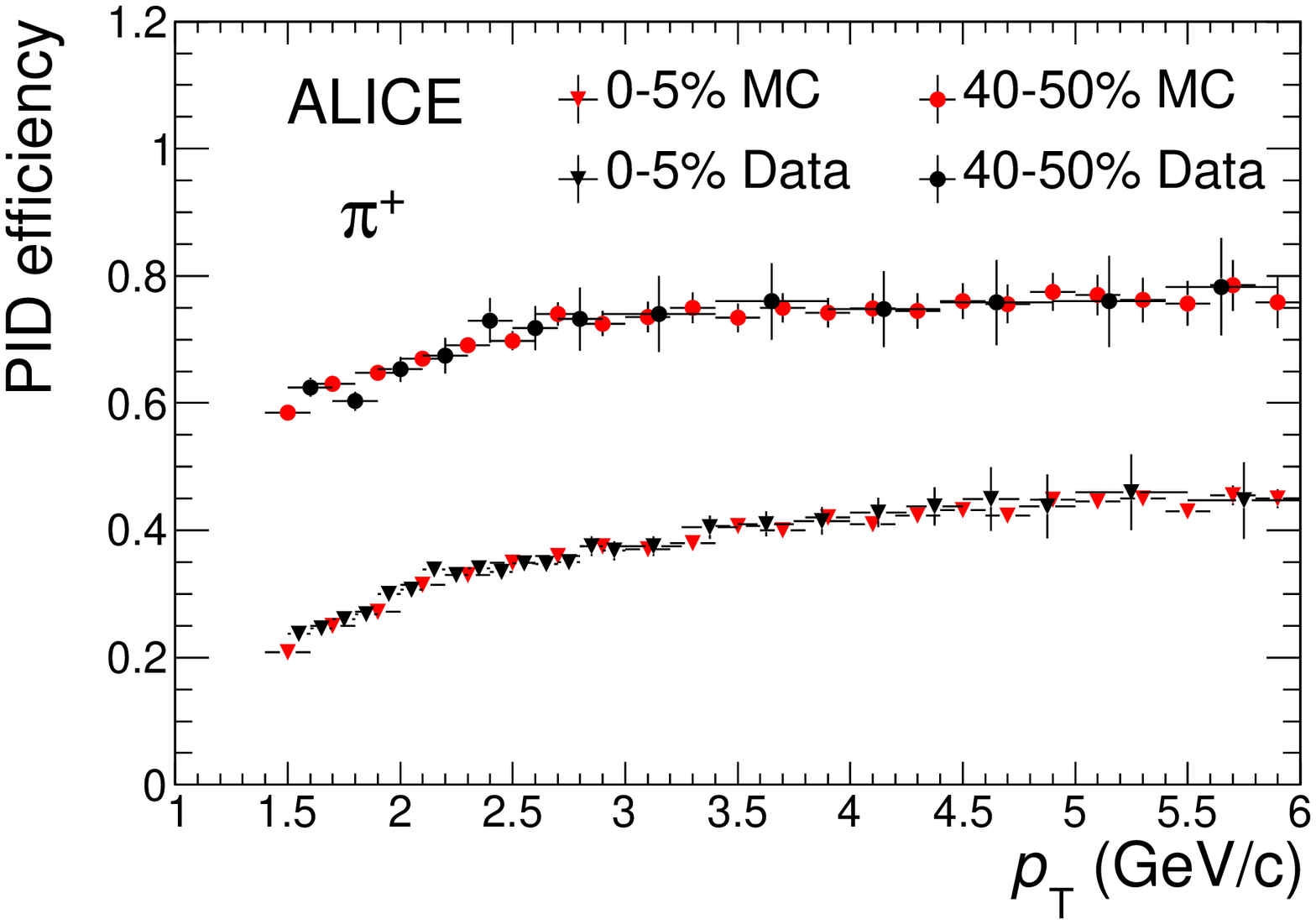}
  \includegraphics[width=1.0\mylength]{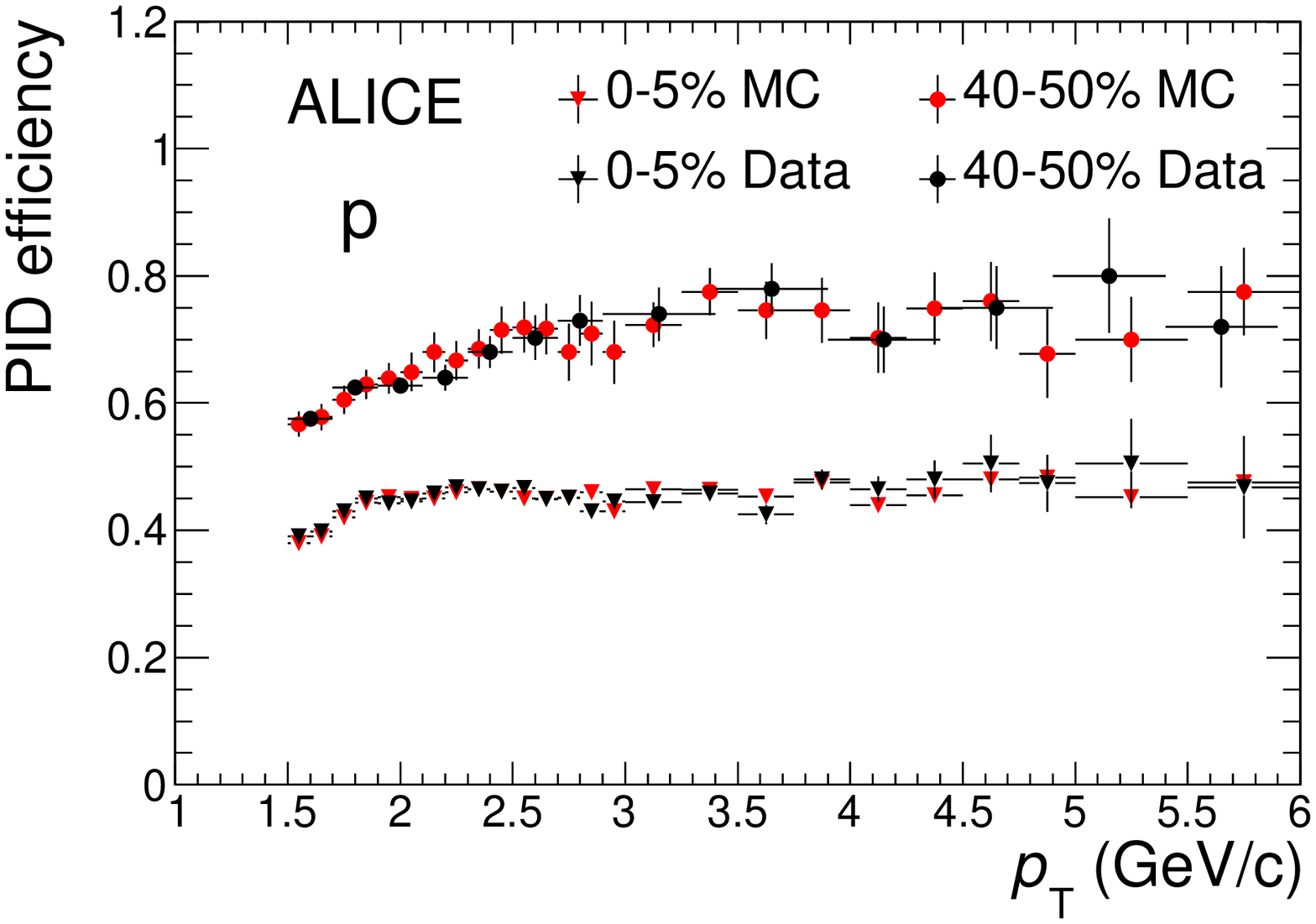}
  \caption{(Color online) Identification efficiency for pions (upper panel)
    and protons (lower panel) selected exploiting V$^0$ decay properties,
    compared with the MC results for primary tracks for 0-5\% and 40-50\%
    centrality classes.}
  \label{pidEffMC_data}
\end{figure}

To correct for the incorrectly assigned Cherenkov rings, a PID efficiency is
used. This efficiency has to be derived from a dataset containing identified
particles of a single species, so one can use MC or $V^0$ daughters. For such a
clean set of particles that passes the distance cut, e.g.\ MC pions as in
Fig.~\ref{MCbackground}, the PID efficiency is
\begin{equation}
\varepsilon_{\text{PID}} = \frac{N(\text{signal})}{N(\text{signal and background})},                                
\end{equation}
where the signal is the integral of the Gaussian fit function. The PID
efficiency has been evaluated from MC simulations that reproduce the
background observed in the data well.  A data-driven cross check of the
efficiency has been performed using a clean sample of $V^0$ daughter
tracks. The comparison between data and MC is shown in
Fig.~\ref{pidEffMC_data} for 0-5\% and 40-50\% centrality classes, and shows
good agreement. We also observe that, as expected, the efficiency decreases
for more central collisions due to the occupancy effects mentioned above. The
maximum value of the PID efficiency is ${\approx}80\%$ at $\pt \sim
\gevc{6}$ in the 40-50\% centrality class. As an additional check of the PID
efficiency, the ratio between the raw yields extracted from the fit (signal)
corrected by the PID efficiency and the total entries in the original
histogram (signal and background) has been evaluated for each \pt bin for all
centralities. The ratio is consistent with unity when the systematic
uncertainties listed in Table~\ref{tab:hmpid:systematics} are taken into account.\\

The systematic uncertainty for the HMPID analysis has contributions from
tracking and PID. These uncertainties have been estimated by individually
changing the track selection cuts and the parameters of the fit function used
to extract the raw yields. The means of the Gaussian functions have been
changed by $\pm \sigma$. Similarly, the widths of the Gaussian functions have
been varied by $\pm 10\%$, accounting for the maximum expected variation of
the resolution as a result of the different running conditions of the detector
during data acquisition that can have an impact on the performance. When the
means are changed, the widths are fixed to the default value, and vice
versa. The parameter variation is performed for all three particles
species. In addition, the uncertainty on the association of the track to the
matched cluster is obtained by varying the value of the distance cut required
for the match by $\pm$1 cm. These contributions do not vary with the collision
centrality. To estimate the uncertainty due to the incomplete knowledge of the
shape of the background distribution, an alternative background function,
depending on tan($\theta$) and derived from geometrical considerations in case
of orthogonal tracks \cite{hmpid:tdr}, has been used:
\begin{equation}
f(\theta) = a + b\times \tan\theta + c\times[\tan\theta(1 + \tan^{2}\theta)]^{d},
\label{tanTheta}
\end{equation}
where $a, b, c$, and $d$ are free parameters.  The corresponding systematic
uncertainty reaches a maximum value at low momenta for the most central
collisions ($\approx$ 15\% for pions and $\approx$ 8\% for kaons and
protons). The systematic uncertainty decreases with \pt because, as previously
explained, the background contribution decreases with increasing track
momentum. A summary of the different contributions to the systematic
uncertainty for the HMPID \pbpb analysis is given in
Table~\ref{tab:hmpid:systematics}.
  
\begin{table*}[htb]
\centering
\begin{tabular}{lcccccc}
\hline
\hline
 Effect         &  \multicolumn{2}{c}{$\pi^\pm$}  &   \multicolumn{2}{c}{K$^\pm$}  &  \multicolumn{2}{c}{p and $\rm\overline{p}$}   \\
\hline
\pt range (GeV/$c$) & 2.5  & 4  & 2.5 & 4 & 2.5 & 4 \\
PID  & 6\% & 12\% & 6\% & 12\% & 4\% & 5\% \\
Tracking efficiency  & \multicolumn{2}{c}{6\%} & \multicolumn{2}{c}{6\%} & \multicolumn{2}{c}{7\%} \\
Distance cut correction  & 6\% & 2\% & 6\% & 2\% & 4\% & 2\% \\ 
Background (Pb-Pb 0-5\%)  & 10\% & 4\% & 5\% & 3\% & 5\% & 3\% \\ 
Background (Pb-Pb 5-10\%)  & 7\% & 4\% & 3\% & 2\% & 3\% & 2\% \\ 
Background (Pb-Pb 10-20\%)  & 6\% & 4\% & 3\% & 2\% & 3\% & 2\% \\ 
Background (Pb-Pb 20-30\%)  & 5\% & 3\% & 3\% & 2\% & 2\% & 2\% \\ 
Background (Pb-Pb 30-40\%)  & 3\% & 1\% & 2\% & 1\% & 2\% & 1\% \\ 
Background (Pb-Pb 40-50\%)  & 2\% & 1\% & 2\% & 1\% & 2\% & 1\% \\ 
\hline
\hline
\end{tabular}
\caption{Main sources of systematic uncertainties for the HMPID \pbpb
  analysis.}
\label{tab:hmpid:systematics}
\end{table*}


\section{\label{sec:results}Results and discussion}%

\begin{figure}[htbp]
\centering
\includegraphics[width=2.0\mylength]{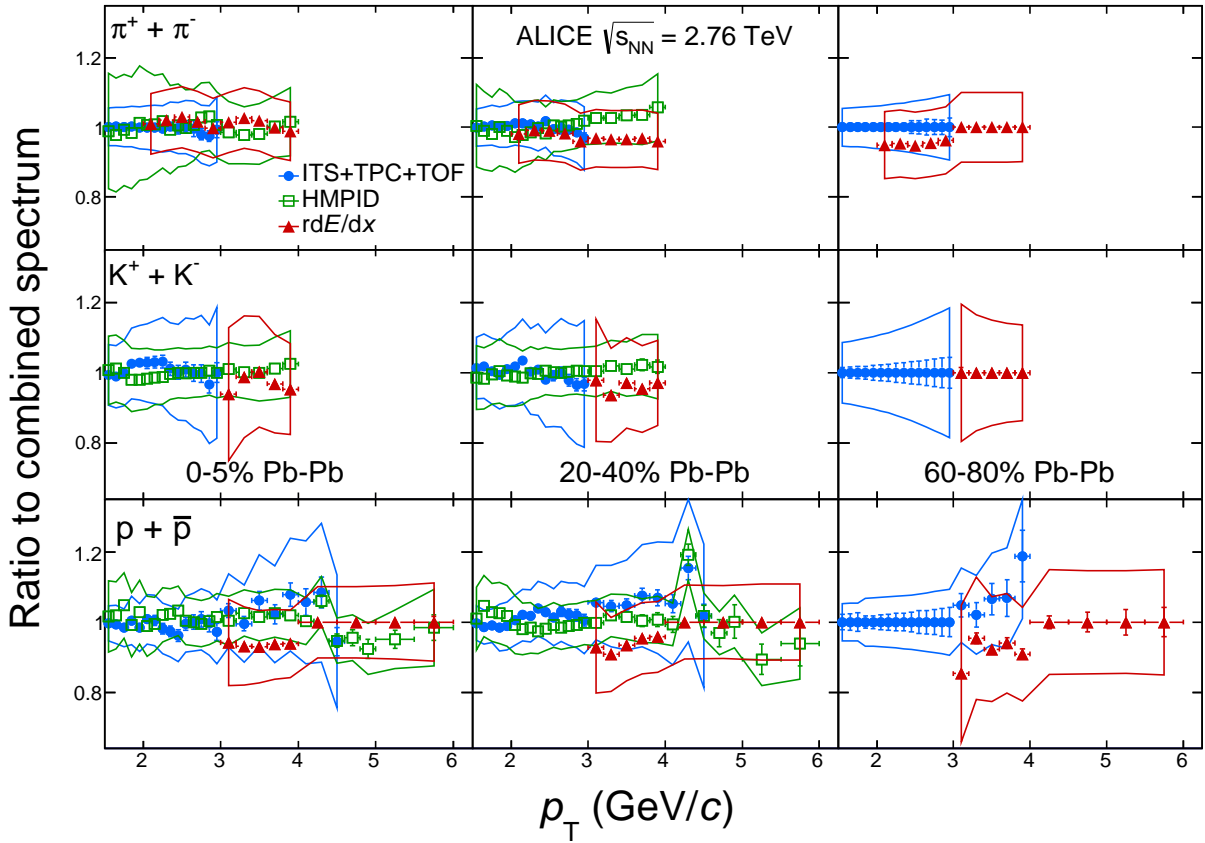}
\caption{(Color online). The ratio of individual spectra to the combined
  spectrum as a function of \pt for pions (upper panels), kaons (middle
  panels), and protons (lower panels). From left-to-right the columns show
  0-5\%, 20-40\%, and 60-80\% (where there are no HMPID results). Only the \pt
  range where the analyses overlap is shown. For $\pt > \gevc{4}$, no
  combination is done and the TPC \dedx relativistic rise results are used
  directly, which gives rise to a small discontinuity for protons at this
  \pt. The ITS+TPC+TOF spectra are the results published
  in~\cite{Abelev:2013vea}. The statistical and independent systematic
  uncertainties are shown as vertical error bars and as a band, respectively,
  and only include those on the individual spectra.}
\label{fig:combination}
\end{figure}

The measurement of charged pion, kaon, and (anti)proton transverse momentum
spectra has been performed via several independent analyses, each one focusing
on a sub-range of the total \pt distribution, using individual detectors and
specific techniques to optimize the signal extraction (see
Table~\ref{table:pbpbranges}). The results were combined in the overlapping
ranges using a weighted average with the independent systematic uncertainties
as weights (a $3\%$ common systematic uncertainty due to the TPC tracking is
added directly to the combined spectrum). The statistical uncertainties are
much smaller and therefore neglected in the combination weights. For $\pt >
\gevc{4}$ only the TPC \dedx relativistic rise analysis is used for all
species. Figure~\ref{fig:combination} shows the ratio of individual spectra to
the combined spectrum for the 0-5\%, 20-40\%, and 60-80\% central \pbpb data,
illustrating the compatibility between the different analyses. In the
centrality intervals where the HMPID measurements are available, they improve
the systematic uncertainty of the kaon and proton yields by approximately a
factor of two in the \pt region where it is later observed that the peaks of
the kaon-to-pion and the proton-to-pion ratios are located (see
Fig.~\ref{fig:results:2} and Fig.~\ref{fig:results:3}). We note that the final
charged pion spectra are consistent with the neutral pion spectra scaled by a
factor of two within statistical and systematic
uncertainties~\cite{Abelev:2014ypa}.

\begin{table}[htbp]
  \begin{center}
    \begin{tabular}{ c  c  c  c }
      \hline \hline
      & ITS+TPC+TOF & HMPID & TPC \dedx rel. rise \\
      \hline
      $\pi^{\pm} $ & 0.1 -- 3.0 & 1.5 -- 4.0 & 2.0 -- 20.0 \\
      K$^{\pm}$ & 0.2 -- 3.0 & 1.5 -- 4.0 & 3.0 -- 20.0 \\
      p($\bar{\rm p})$ & 0.3 -- 4.6 & 1.5 -- 6.0 & 3.0 -- 20.0 \\
      K/$\pi$ & 0.2 -- 3.0 & 1.5 -- 4.0 & 3.0 -- 20.0 \\
      p/$\pi$ & 0.3 -- 3.0 & 1.5 -- 4.0 & 3.0 -- 20.0 \\
      \hline \hline
    \end{tabular}
    \caption{The \pt ranges (GeV/$c$) used in the combination of the most
      central results. In \pp and peripheral \pbpb collisions the separation
      power is different and in some cases the \pt ranges therefore change a
      little.}  \centering
    \label{table:pbpbranges}
  \end{center}
\end{table}

\begin{figure}[htb]
  \begin{center}
    \includegraphics[keepaspectratio,width=2.0\mylength]{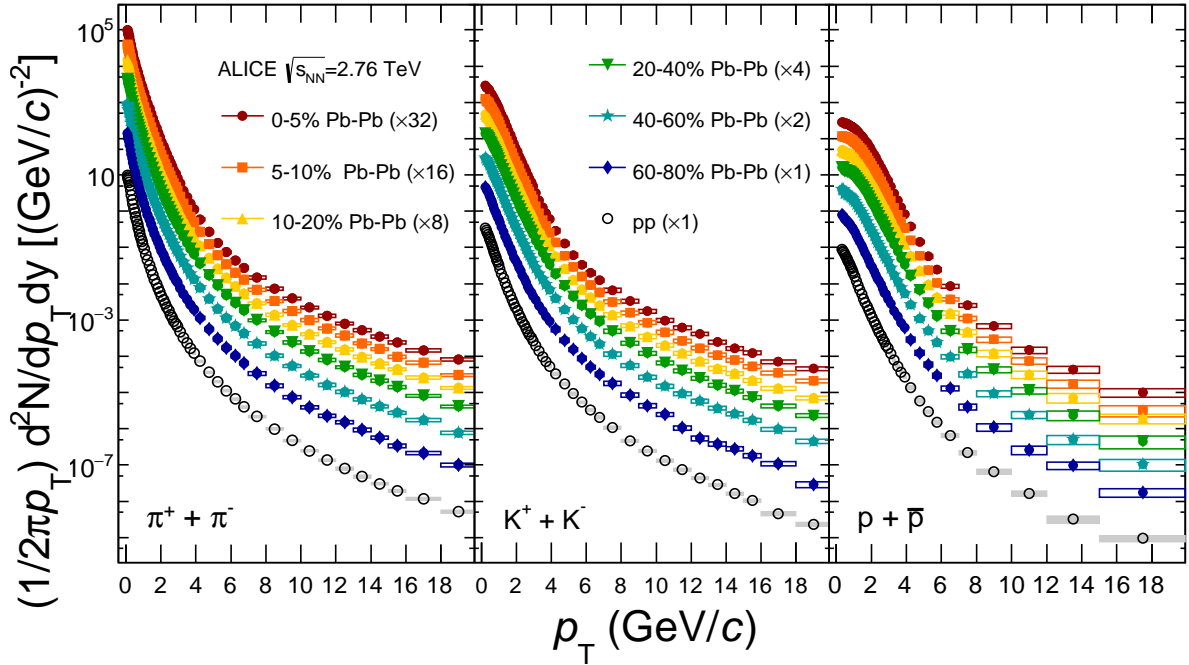}
    \caption{\label{fig:results:1} (Color online). Transverse momentum spectra
      of charged pions (left panel), kaons (middle panel), and (anti)protons
      (right panel) measured in \pbpb and \pp collisions at $\sqrt{s_{\rm
          NN}}=2.76$ TeV. The systematic and statistical error are plotted as
      color boxes and vertical error bars (hard to see), respectively. The
      spectra have been scaled by the factors listed in the legend for
      clarity.}
  \end{center}
\end{figure}

The final combined transverse momentum distributions for the three particle
species are shown in Fig.~\ref{fig:results:1}. For $\pt < \gevc{3}$, a
hardening of the spectra is observed going from peripheral to central
events. This effect is mass dependent and is characteristic of hydrodynamic
flow as discussed in~\cite{Abelev:2013vea}. For high \pt ($> \gevc{10}$) the
spectra follow a power-law shape as expected from perturbative QCD (pQCD)
calculations.
In the following, the high-\pt results are first discussed before going on to
the intermediate \pt region.

\subsection{The high-\pt results}

\begin{figure}[htb]
  \begin{center}
    \includegraphics[keepaspectratio, width=2.0\mylength]{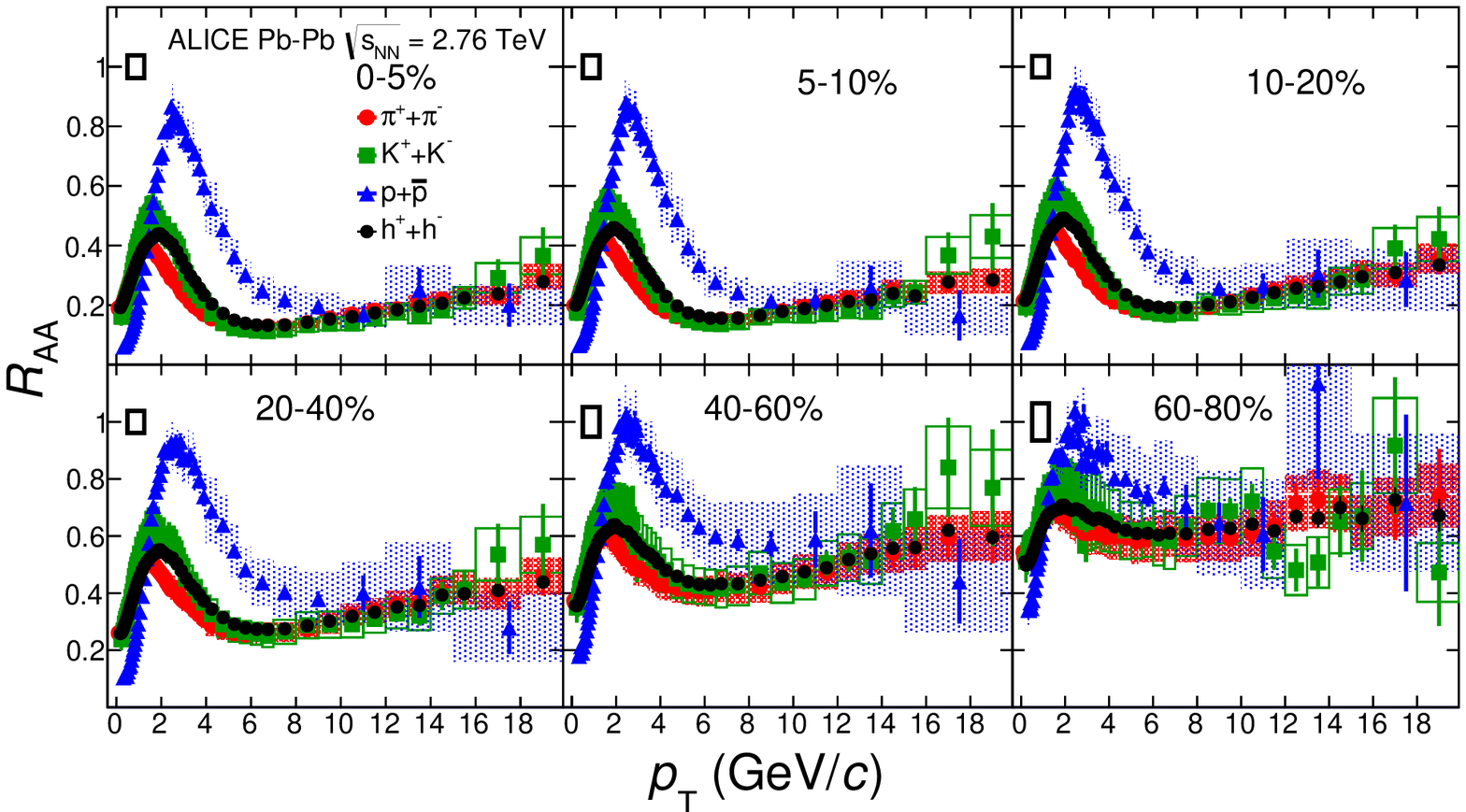}
    \caption{\label{fig:results:raa:all} (Color online). The nuclear
      modification factor \raa as a function of \pt for different particle
      species. Results for different collision centralities are
      shown. Statistical and PID systematic uncertainties are plotted as
      vertical error bars and boxes around the points, respectively. The total
      normalization uncertainty (\pp and \pbpb) is indicated in each panel by
      the vertical scale of the box centered at \pt = \gevc{1} and \raa =
      1~\cite{Abelev:2012hxa}.}
  \end{center}
\end{figure}

To study jet quenching at high \pt, the nuclear modification factor, \raa, is
constructed. The \raa is 
\begin{equation}\label{eq:res:1}
R_{\rm AA} =  \frac{{\rm d}^{2}N_{\rm id}^{\rm AA}/{\rm d}y{\rm d}\pt}{\langle T_{\rm AA} \rangle {\rm d}^{2}\sigma_{\rm id}^{\rm pp}/{\rm d}y{\rm d}\pt},
\end{equation}
where $N_{\rm id}^{\rm AA}$ and $\sigma^{\pp}_{\rm id}$ are the charged
particle yield in nucleus-nucleus (A--A) collisions and the cross section in
\pp collisions, respectively, and $\langle T_{\rm AA} \rangle$ is the nuclear
overlap function. The latter is obtained from a Glauber
model~\cite{Abelev:2013qoq} and is related to the average number of binary
nucleon-nucleon collisions ($N_{\rm coll}$) and the inelastic nucleon-nucleon
cross section as $\langle T_{\rm AA} \rangle=\langle N_{\rm coll} \rangle /
\sigma^{\rm NN}_{\rm inel}$.

Figure~\ref{fig:results:raa:all} shows the \raa for all centrality
classes. The results show that for all centrality classes any particle species
dependence of the nuclear modification for $\pt > \gevc{10}$ is small,
compared with the large suppression ($\raa \ll 1$). This suggests that jet
quenching does not produce signatures that affect the particle species
composition for the leading particles. The results presented in this paper are
all done at the particle level, while for some models that motivated these
studies the predictions are done for jets, e.g.\ the Sapeta--Wiedemann
model~\cite{Sapeta:2007ad}. It is not obvious how to compare the results
presented here with such calculations. In the following we therefore discuss
how inclusive \pt spectra compare with inclusive jet \pt spectra. In
particular, it is examined if the results are likely to be affected by a
quenched jet fragmentation bias (if quenched jets emit less high-\pt particles
than unquenched ones) or a surface bias (if unquenched jets from the surface
dominate).

At the LHC, by studying dijets in \pbpb collisions and selecting on the dijet
asymmetry, one can study samples with large asymmetries where one knows, based
on comparisons with \pp results, that at least the subleading jet has suffered
a large energy loss~\cite{Aad:2010bu,Chatrchyan:2011sx}. The study of the
Fragmentation Functions (FFs) for these quenched jets has shown that for
charged tracks with $\pt > \gevc{4}$ they are similar to those observed in \pp
collisions for subleading jets with $\ptjet >
\gevc{50}$~\cite{Chatrchyan:2012gw}, in agreement with what one also finds for
inclusive jets~\cite{Chatrchyan:2014ava}. This rules out a large fragmentation
bias (for lower jet \pt see below) and suggests that any surface bias is the
same as for inclusive jets. To understand the jet \pt covered by the results
presented here, one can now, thanks to the similarity of the FFs in \pp and
\pbpb collisions, rely on NLO pQCD calculations for \pp collisions. The FFs
found to best describe the inclusive charged particle
spectra~\cite{d'Enterria:2013vba} are the Kretzer
distributions~\cite{Kretzer:2000yf}. NLO pQCD calculations using the Kretzer
FFs suggest that more than half of the particles with \pt between 10 and
\gevc{20} are from gluon jets and that the typical jet \pt is roughly a factor
of 2-3 larger than the hadron \pt ($\langle z \rangle = p_{\rm T, hadron} /
p_{\rm T, jet} \approx 0.4$)~\cite{d'Enterria:2013vba}\footnote{The
  publication contains only calculations for \sppg{900} and \sppt{7} that have
  been averaged as an approximate estimate for the energy of \sppt{2.76} shown
  here since the energy dependence is not that strong.}. The conclusions for
jets with $\ptjet > \gevc{50}$ is therefore expected to be directly applicable
also for the highest-\pt particles studied here. ALICE has studied charged
jets in \pbpb collisions where it was found that requiring minimum one track
with $\pt > \gevc{10}$ in a jet gives the same fragmentation bias of the jet
reconstruction efficiency in \pbpb collisions as in PYTHIA for $20 < \ptjetch
< \gevc{110}$~\cite{Abelev:2013kqa}, so there is no evidence even for lower
\pt jets that there is a different fragmentation bias in \pbpb collisions than
in \pp collisions. The results in Fig.~\ref{fig:results:raa:all} therefore
indicate that for jets with final $\pt$ of order $25$ to \gevc{50}, jet
quenching does not produce large particle-species-dependent effects in the
hard core of the jet where leading particle production mainly occurs.

To be able to set stronger constraints, one needs theoretical modeling. As the
\raa for charged pions, kaons, and protons reported here for $\pt > \gevc{10}$
are all compatible to the \raa for inclusive charged
particles~\cite{Abelev:2012hxa} and neutral pions~\cite{Abelev:2014ypa} we
refer to these papers for comparisons with models without large particle-species-dependent effects. When compared with models which include large particle-species-dependent effects, the results indicate that the jet quenching
mechanism does not involve direct exchange of quantum numbers with the medium,
and there are also no indications of a modified color structure of the
fragmentation~\cite{Sapeta:2007ad} or that the probe is excited to other color
states~\cite{Aurenche:2011rd}. Models in which the hadronization of jet
fragments occurs in the medium also appear to be ruled
out~\cite{Bellwied:2010pr}. It seems that the medium quenches the jet as a
whole rather than directly interacting with its fragments. Such a picture has
recently been proposed~\cite{CasalderreySolana:2012ef}, arguing that the
medium typically cannot resolve the structure inside the hard core of the jet
such that all fragments lose energy coherently.

\begin{figure}[htb]
  \begin{center}
    \includegraphics[keepaspectratio, width=2.0\mylength]{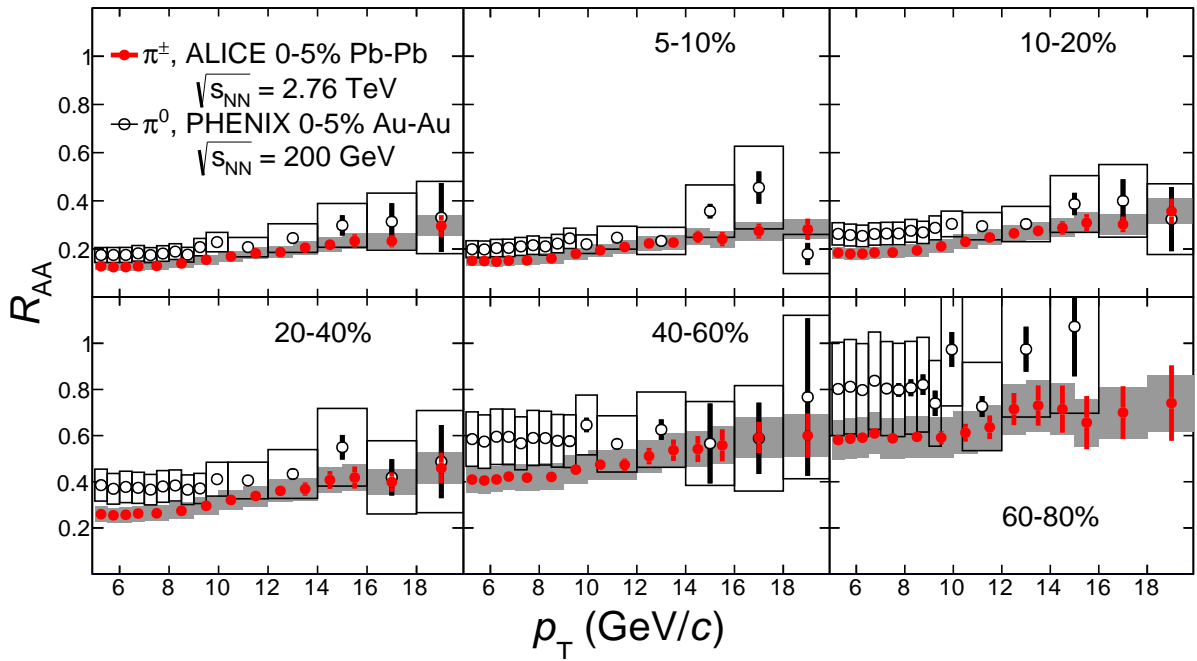}
    \caption{\label{fig:results:raa_pions} (Color online). The nuclear
      modification factor \raa as a function of \pt for charged pions, compared
      with PHENIX results for neutral pions~\cite{Adare:2012wg}. Results for
      different collision centralities are shown. Statistical and PID systematic
      uncertainties are plotted as vertical error bars and boxes around the
      points, respectively.}
  \end{center}
\end{figure}

In Fig.~\ref{fig:results:raa_pions}, the \raa for charged pions, the most
precise measurement in this work and the one least sensitive to radial flow,
is compared with the \raa for neutral pions measured by
PHENIX~\cite{Adare:2012wg} at the RHIC\footnote{The results have been obtained
  from the tables at the PHENIX website and the 5-10\% data set has been
  constructed from the 0-5\% and 0-10\%.}. The ALICE results are
systematically below the PHENIX values for $\pt < \gevc{10}$ but consistent
within systematic uncertainties for larger \pt. We note that the relative
centrality evolution is similar at the two center-of-mass
energies. In~\cite{Christiansen:2013hya}, a simple study of the \raa at $\pt =
\gevc{10}$ found that the energy loss is ${\approx}40\%$ larger at the LHC
than at the RHIC in all centrality classes (it scales as $\sqrt{\dndeta}$ for
a fixed initial geometry).

\begin{figure}[htbp]
  \begin{center}
    \includegraphics[keepaspectratio, width=2.0\mylength]{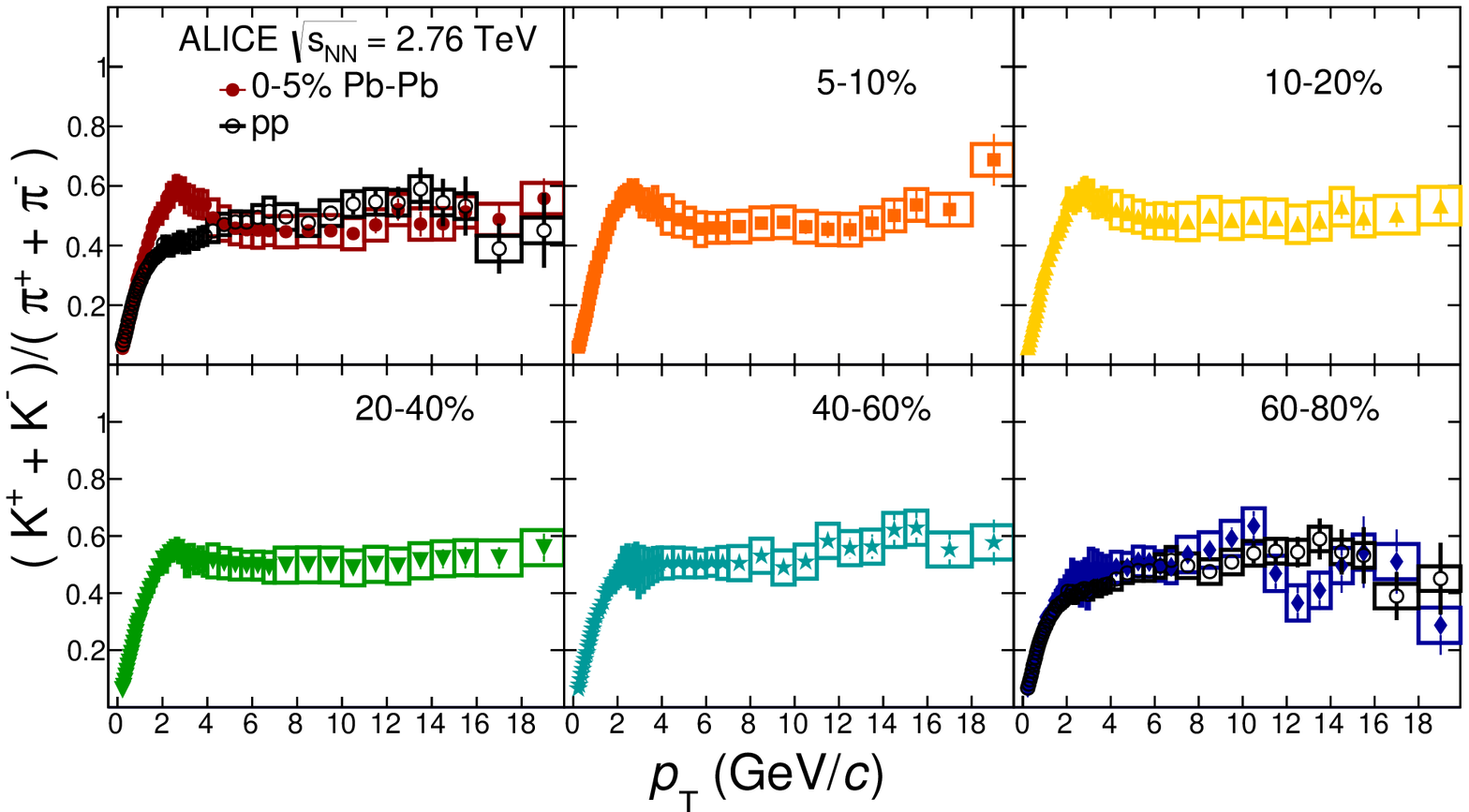}\\
    \includegraphics[keepaspectratio, width=2.0\mylength]{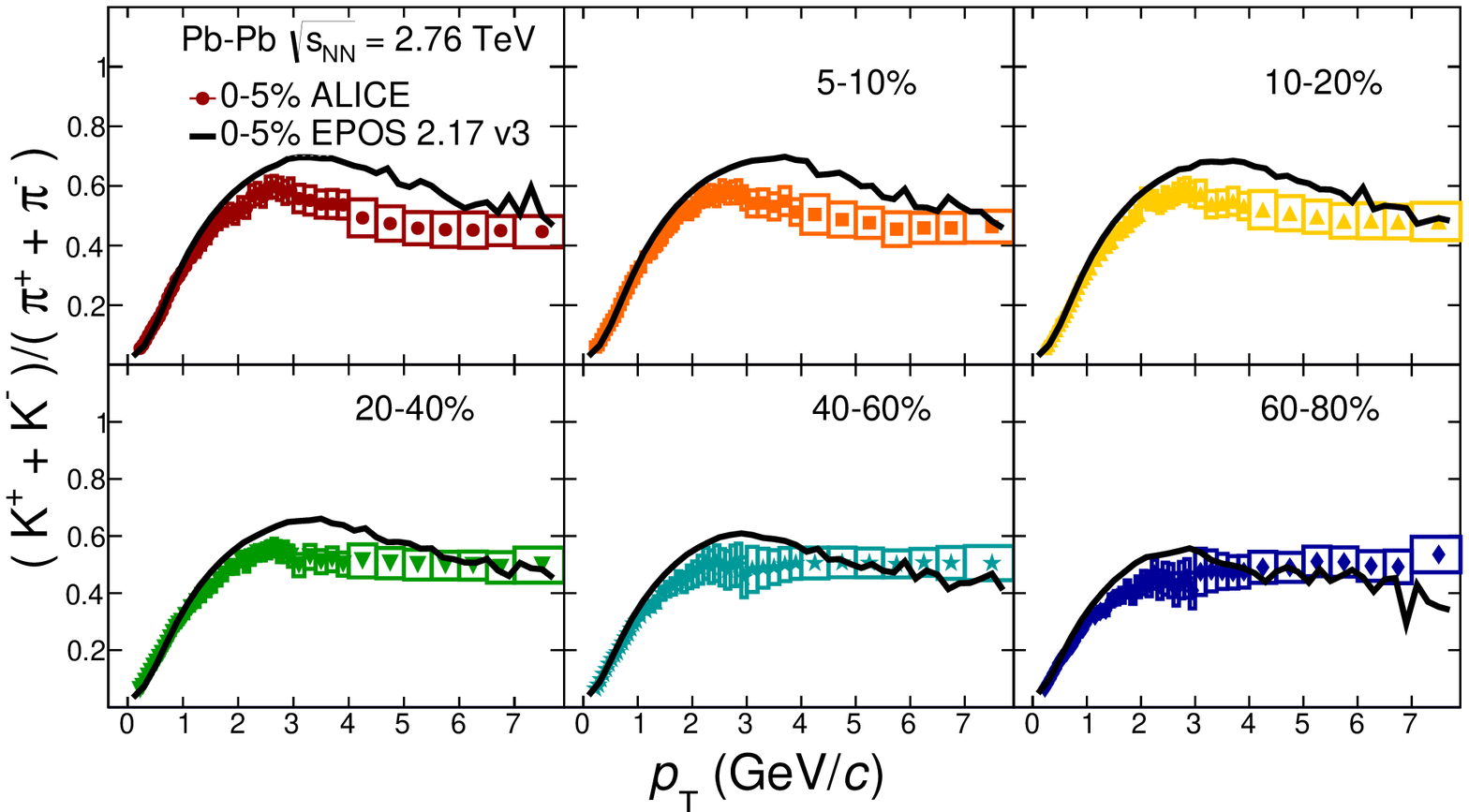}
    \caption{\label{fig:results:2} (Color online). Charged kaon to charged
      pion ratio as a function of transverse momentum (solid markers). The
      upper figure shows the full \pt-range with the \pp results (open
      markers) overlaid in the most central and the most peripheral centrality
      class. In the lower figure the \pbpb results for $\pt < \gevc{8}$ are
      compared with EPOS model 2.17-3 (line). The systematic and statistical
      error are plotted as color boxes and vertical error bars,
      respectively. }
  \end{center}
\end{figure}

\begin{figure}[htbp]
  \begin{center}
    \includegraphics[keepaspectratio, width=2.0\mylength]{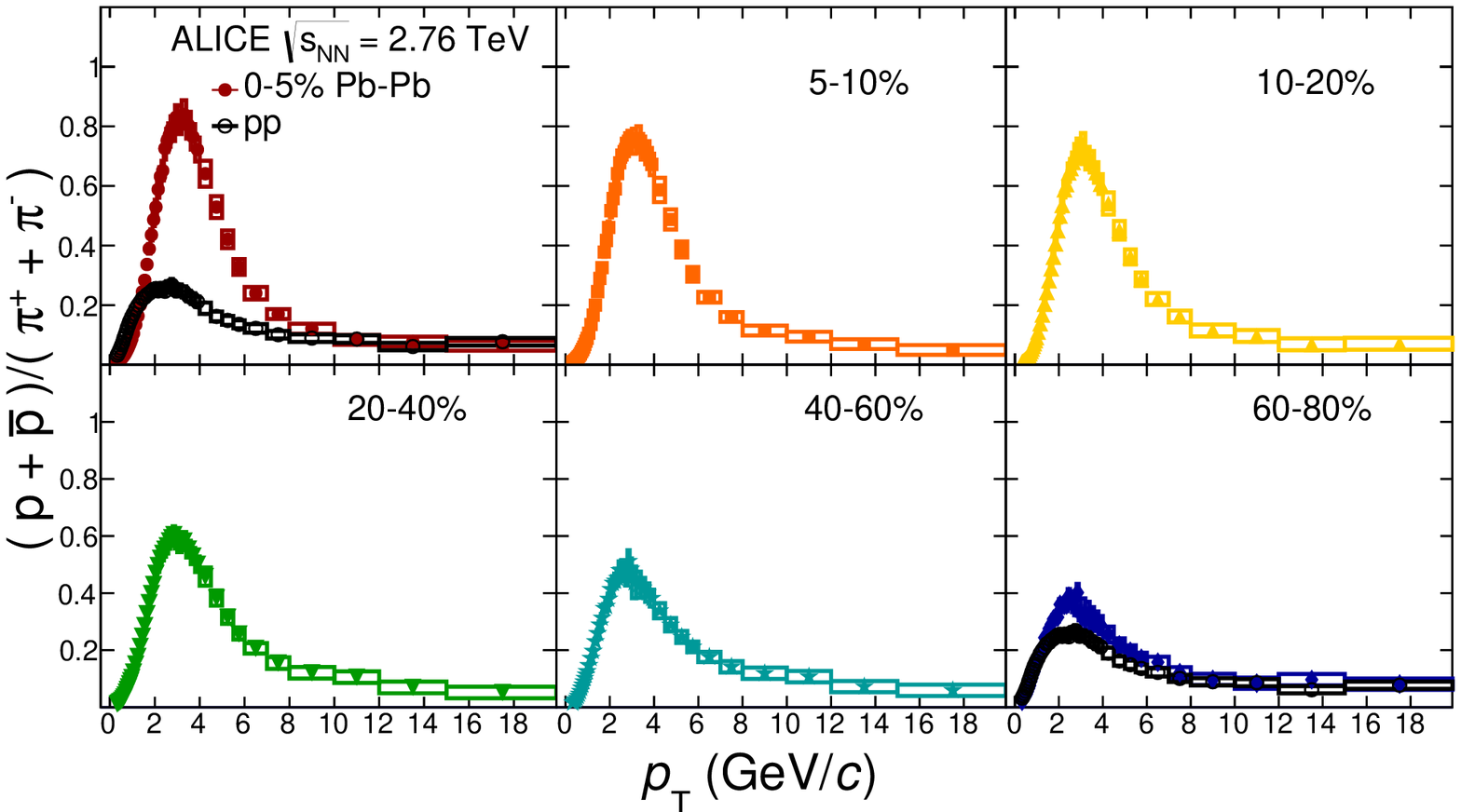}\\
    \includegraphics[keepaspectratio, width=2.0\mylength]{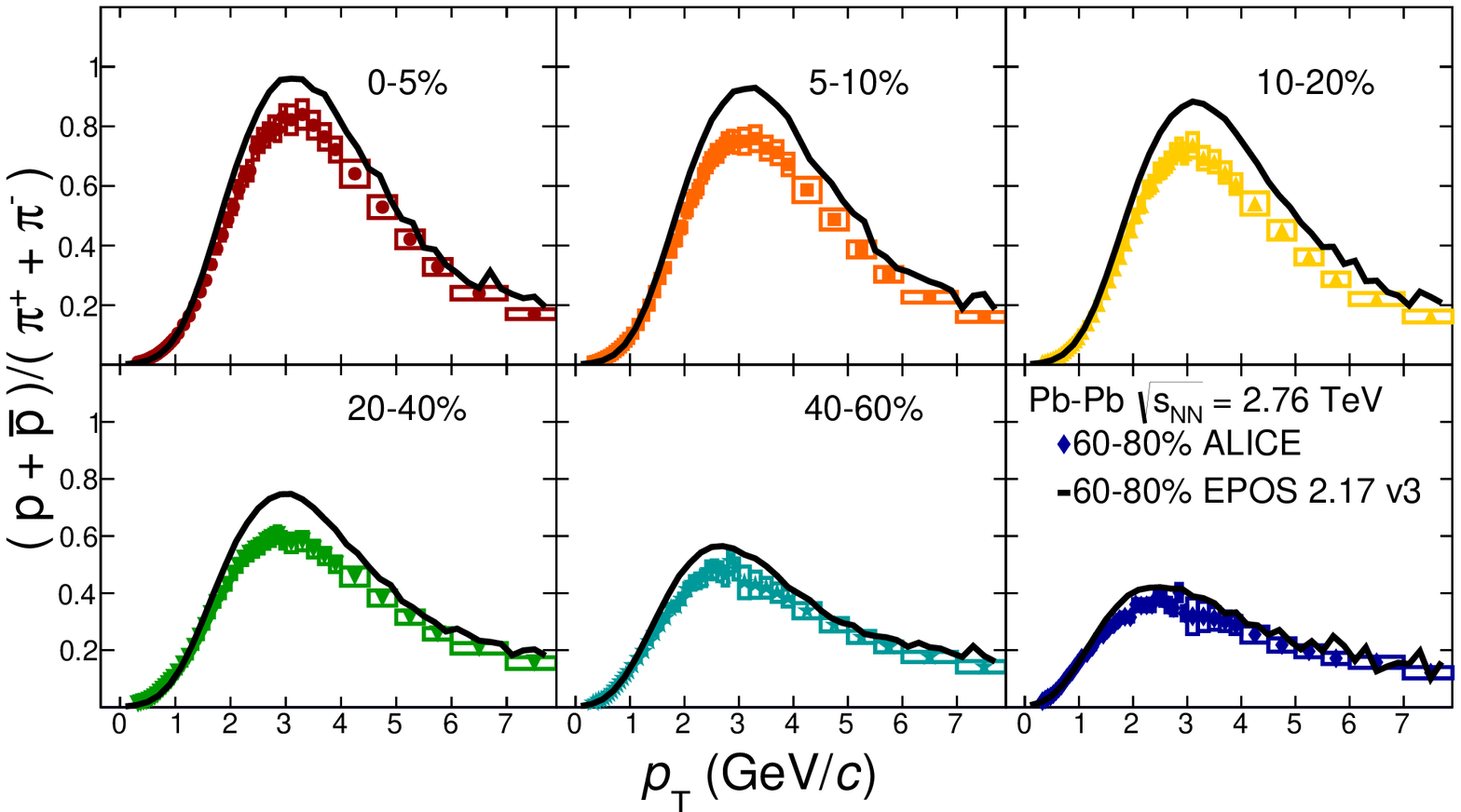}
    \caption{\label{fig:results:3} (Color online). (Anti)proton to charged
      pion ratio as a function of transverse momentum (solid markers). The
      upper figure shows the full \pt-range with the \pp results (open
      markers) overlaid in the most central and the most peripheral centrality
      class. In the lower figure the \pbpb results for $\pt < \gevc{8}$ are
      compared with EPOS model 2.17-3 (line). The systematic and statistical
      error are plotted as color boxes and vertical error bars, respectively.}
  \end{center}
\end{figure}
  
\begin{figure}[htb]
  \begin{center}
    \includegraphics[keepaspectratio, width=1.38\mylength]{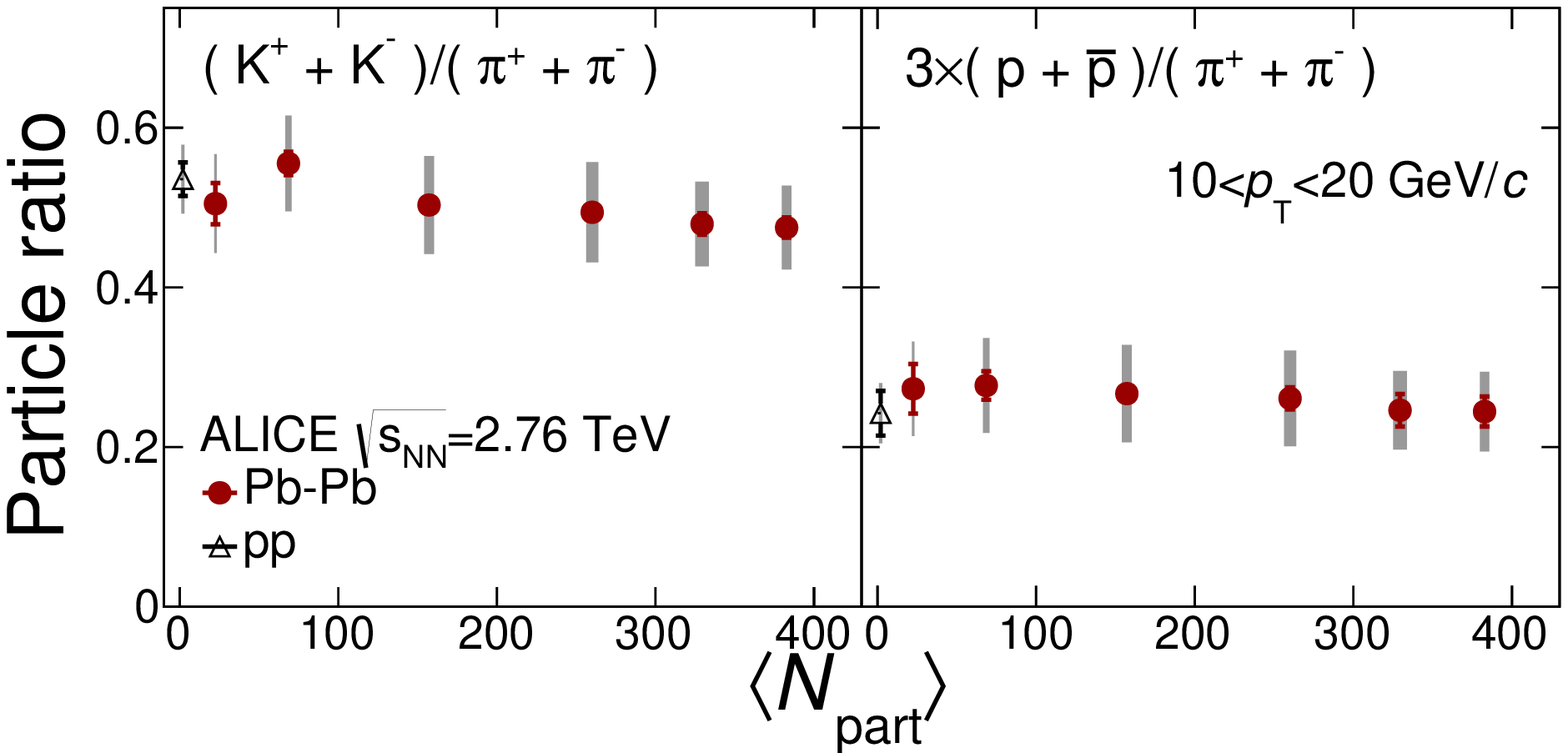}
    \caption{\label{fig:results:ratios_high_pt} (Color online). The integrated
      particle ratios for $\pt > \gevc{10}$ in \pp and \pbpb collisions as a
      function of the number of participants. Left panel: the kaon-to-pion
      ratio. Right panel: the proton-to-pion ratio scaled by a factor of 3 for
      clarity. Statistical and PID systematic uncertainties are plotted as
      vertical error bars and boxes around the points, respectively. Note that
      this kaon-to-pion (proton-to-pion) ``high-\pt'' ratio is ${\approx}4$
      (${\approx}2$) times larger than the bulk ratio~\cite{Abelev:2013vea}.}
  \end{center}
\end{figure}

The proton-to-pion and the kaon-to-pion ratios as a function of \pt are shown
in Fig.~\ref{fig:results:2} and Fig.~\ref{fig:results:3}. The similarity at
high \pt for the \raa implies that the particle ratios there are also the same
in \pp and \pbpb collisions. Since the particle ratios are independent of \pt
in this region, we use the integrated particle ratios for $\pt > \gevc{10}$ to
elucidate the precision with which the suppression of pions, kaons, and
protons is similar, see Fig.~\ref{fig:results:ratios_high_pt}. The advantage
of using particle ratios is that the results for heavy-ion collisions can be shown
separately from the \pp results. Furthermore, in the ratios the systematic
uncertainty associated with the inclusive charged particle \pt spectra
normalization cancels. All the steps in the high-\pt \dedx analysis discussed
in Sec.~\ref{sec:highpT} are done independently for each centrality class
(using disjunct datasets) so one does not expect any direct correlations of
the results. We conclude that all kaon-to-pion (proton-to-pion) ratios as a
function of $N_{\text{part}}$ are consistent within the systematic uncertainty
of ${\approx}10~\%$ (${\approx}20\%$). Measurements with improved precision
using Run 2 and Run 3 LHC data could reveal possible subtle particle-species
differences.

\subsection{The intermediate \pt results}

In the following, the intermediate \pt regions in Fig.~\ref{fig:results:2} and
Fig.~\ref{fig:results:3}, where the proton-to-pion and the kaon-to-pion ratios
are enhanced, are discussed.

The observation of the large proton-to-pion ratio at intermediate \pt at the
RHIC generated numerous speculations that the degrees of freedom in the medium
are constituent quark-like and that they recombine when hadronizing to give
rise to distinct meson and baryon properties. As the $\phi$ meson has a
similar mass to a proton, it is crucial in testing these ideas and results at
the RHIC indeed seemed to confirm this picture~\cite{Abelev:2007rw}, while at
LHC the picture appears to be more
complicated~\cite{Abelev:2014uua,Abelev:2014pua}. Some of the models developed
to describe results at the RHIC have been extended to the LHC energies. One
can, in general, separate recombination models into two classes. In soft
models, recombination only occurs for soft thermal radially-flowing
partons. In~\cite{Abelev:2014uua} ALICE showed calculations for such a
model~\cite{PhysRevC.68.044902} and the prediction is that at the LHC energies
the particle ratios in central collisions are similar to those measured at the
highest RHIC energy. In hard recombination models, jet fragments can recombine
with both partons from the medium and other jets. At LHC energies, the
mini-jet activity is much larger than at RHIC energies, which motivated
predictions for central collisions of particle ratios an order of magnitude
larger ($p/\pi \sim 10$--$20$) than the peak values reported here and
persisting out to much higher \pt~\cite{Hwa:2006zq}. The failure of hard
recombination is in qualitative agreement with the picture where the jet
interacts with the medium as a whole so that the hard fragments of the jet
cannot recombine with partons in the medium or in another jet.
  
EPOS~\cite{PhysRevC.85.064907} is a full MC generator which contains both soft
and hard physics. It incorporates a hydrodynamical phase and additional
hadronization processes at intermediate \pt where the interaction between bulk
matter and quenched jets is considered~\cite{Werner:2012sv}. This interaction
introduces a baryon-meson effect, where fully quenched jets are allowed to
hadronize with flowing medium quarks. When we study the full set of ratios at
all centralities (Fig.~\ref{fig:results:2} and Fig.~\ref{fig:results:3}) EPOS
generally reproduces the centrality dependence well, even for very peripheral
events, where it is known that pure hydrodynamical calculations fail to
describe the data~\cite{Abelev:2013vea}. However, EPOS overpredicts the
magnitude of both the proton-to-pion and the kaon-to-pion peak; it is
therefore critical to understand how important the additional hadronization
processes are, relative to the hydrodynamic flow, when all parameters have
been tuned.

\begin{figure} [htb]
  \begin{center}
    \includegraphics[width=1.38\mylength]{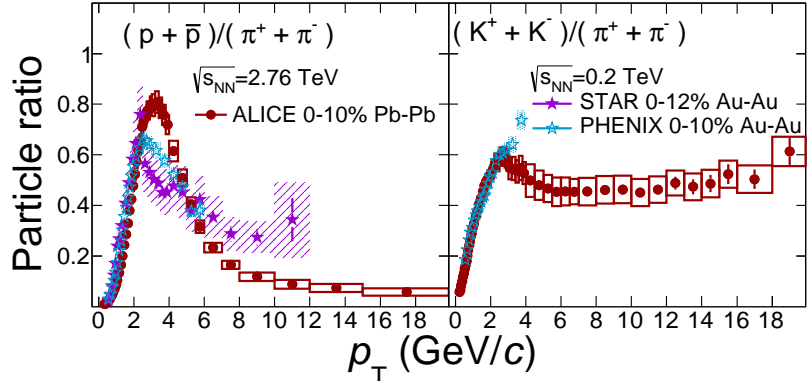}
    \caption{(Color online) ALICE (circles) results from \snnt{2.76} \pbpb
      collisions compared with STAR and PHENIX results for \snng{200} \auau
      collisions. Left panel: the proton-to-pion ratio. Right panel: the kaon-to-pion
      ratio.}
    \label{fig:results:rhic}
  \end{center}
\end{figure}

Figure~\ref{fig:results:rhic} shows a comparison of particle ratios with
results from STAR~\cite{Abelev:2006jr} and PHENIX~\cite{Adare:2013esx} at the
RHIC measured in \auau collision at \snng{200}. In both cases, the results
have been averaged for both charge signs for pions and protons. We use the
STAR feed­down-corrected data for this comparison~\footnote{Values taken from
  \href{https://drupal.star.bnl.gov/STAR/files/starpublications/65/data.html}{https://drupal.star.bnl.gov/STAR/files/starpublications/65/data.html}
  for protons and a similar feed-down correction has been assumed for
  anti-protons.}. The proton-to-pion peak at the LHC is approximately 20\%
larger than at the RHIC, which is consistent with an increased average radial
flow velocity. At high \pt, the systematic uncertainties of the STAR data are
very large and it was noted in a later publication that they might even be
underestimated~\cite{Agakishiev:2011dc}. Interestingly, there is no evidence
for a peak in the kaon-to-pion ratio measured by PHENIX, which is similar to
the ALICE data points for $\pt \leq \gevc{3}$, but continues to rise in the
few data points above this \pt.

Careful modeling of \pt spectra and azimuthal flow is needed to answer the
question of whether there are additional hadronization processes such as soft
recombination at the LHC\footnote{We note that in a recent preprint it is
  shown that soft recombination together with pQCD+quenching can give a good
  description of pion, kaon, and (anti)proton spectra in central heavy-ion
  collisions both at the RHIC and the LHC for $1.5 < \pt <
  \gevc{10}$~\cite{Minissale:2015zwa}.}. Since the multiplicity evolution of
particle ratios in \ppb collisions is similar to what is observed for \pbpb
collisions~\cite{Abelev:2013haa} it would be interesting to include those
results in the modeling, in particular, since there is no indication of jet
quenching~\cite{ALICE:2012mj} which conceptually simplifies the problem.


\section{\label{sec:level8}Conclusion}

We have reported the centrality dependent measurement of charged pions, kaons
and (anti)protons at large transverse momenta in \pbpb collisions at the
LHC. When combined with previously published data at lower \pt, the new
results provide a comprehensive dataset of pion, kaon, and (anti)proton \pt
spectra with unprecedented systematic precision and \pt reach. The spectra are
sensitive to physics mechanisms that differentiate between baryons and mesons,
strange and non-strange, or heavy and light hadrons.

At high \pt ($\pt > \gevc{10}$), particle ratios and nuclear modification
factors allow the study of effects related to jet quenching. The measurements
in this \pt range do not show any difference in the nuclear modification
factor for pions, kaons, and protons. A comparison of the present results with
jet measurements and theoretical calculations establishes that jet quenching
does not introduce large species-dependent modifications for leading
particles. Instead, at high \pt, for all 6 centrality classes and the \pp data
analyzed here, the same kaon-to-pion and proton-to-pion ratios are obtained
within a systematic precision of ${\approx}10$--20\%.

At intermediate \pt, calculations are needed to determine whether models
containing only hydrodynamics and jet quenching can provide a good description
across many observables of the available experimental results or if additional
processes such as recombination are needed. Since the initial geometry of the
collision directly affects both the flow and the energy loss, the centrality
dependence presented in this paper is important for constraining both the
low-\pt hydrodynamics and the high-\pt jet quenching in the calculations.

The results in this paper, taken together with the wealth of other high \pt
and jet results from the LHC, point toward a need for further development of
a microscopic QCD-based picture that explains in detail the relation between
the jet, the medium, and the energy loss.

\newenvironment{acknowledgement}{\relax}{\relax}
\begin{acknowledgement}
\section*{Acknowledgements}

The ALICE Collaboration would like to thank all its engineers and technicians for their invaluable contributions to the construction of the experiment and the CERN accelerator teams for the outstanding performance of the LHC complex.
The ALICE Collaboration gratefully acknowledges the resources and support provided by all Grid centres and the Worldwide LHC Computing Grid (WLCG) collaboration.
The ALICE Collaboration acknowledges the following funding agencies for their support in building and
running the ALICE detector:
State Committee of Science,  World Federation of Scientists (WFS)
and Swiss Fonds Kidagan, Armenia,
Conselho Nacional de Desenvolvimento Cient\'{\i}fico e Tecnol\'{o}gico (CNPq), Financiadora de Estudos e Projetos (FINEP),
Funda\c{c}\~{a}o de Amparo \`{a} Pesquisa do Estado de S\~{a}o Paulo (FAPESP);
National Natural Science Foundation of China (NSFC), the Chinese Ministry of Education (CMOE)
and the Ministry of Science and Technology of China (MSTC);
Ministry of Education and Youth of the Czech Republic;
Danish Natural Science Research Council, the Carlsberg Foundation and the Danish National Research Foundation;
The European Research Council under the European Community's Seventh Framework Programme;
Helsinki Institute of Physics and the Academy of Finland;
French CNRS-IN2P3, the `Region Pays de Loire', `Region Alsace', `Region Auvergne' and CEA, France;
German Bundesministerium fur Bildung, Wissenschaft, Forschung und Technologie (BMBF) and the Helmholtz Association;
General Secretariat for Research and Technology, Ministry of
Development, Greece;
Hungarian Orszagos Tudomanyos Kutatasi Alappgrammok (OTKA) and National Office for Research and Technology (NKTH);
Department of Atomic Energy and Department of Science and Technology of the Government of India;
Istituto Nazionale di Fisica Nucleare (INFN) and Centro Fermi -
Museo Storico della Fisica e Centro Studi e Ricerche "Enrico
Fermi", Italy;
MEXT Grant-in-Aid for Specially Promoted Research, Ja\-pan;
Joint Institute for Nuclear Research, Dubna;
National Research Foundation of Korea (NRF);
Consejo Nacional de Cienca y Tecnologia (CONACYT), Direccion General de Asuntos del Personal Academico(DGAPA), M\'{e}xico, Amerique Latine Formation academique - European Commission~(ALFA-EC) and the EPLANET Program~(European Particle Physics Latin American Network);
Stichting voor Fundamenteel Onderzoek der Materie (FOM) and the Nederlandse Organisatie voor Wetenschappelijk Onderzoek (NWO), Netherlands;
Research Council of Norway (NFR);
National Science Centre, Poland;
Ministry of National Education/Institute for Atomic Physics and National Council of Scientific Research in Higher Education~(CNCSI-UEFISCDI), Romania;
Ministry of Education and Science of Russian Federation, Russian
Academy of Sciences, Russian Federal Agency of Atomic Energy,
Russian Federal Agency for Science and Innovations and The Russian
Foundation for Basic Research;
Ministry of Education of Slovakia;
Department of Science and Technology, South Africa;
Centro de Investigaciones Energeticas, Medioambientales y Tecnologicas (CIEMAT), E-Infrastructure shared between Europe and Latin America (EELA), Ministerio de Econom\'{i}a y Competitividad (MINECO) of Spain, Xunta de Galicia (Conseller\'{\i}a de Educaci\'{o}n),
Centro de Aplicaciones Tecnológicas y Desarrollo Nuclear (CEA\-DEN), Cubaenerg\'{\i}a, Cuba, and IAEA (International Atomic Energy Agency);
Swedish Research Council (VR) and Knut $\&$ Alice Wallenberg
Foundation (KAW);
Ukraine Ministry of Education and Science;
United Kingdom Science and Technology Facilities Council (STFC);
The United States Department of Energy, the United States National
Science Foundation, the State of Texas, and the State of Ohio;
Ministry of Science, Education and Sports of Croatia and  Unity through Knowledge Fund, Croatia.
Council of Scientific and Industrial Research (CSIR), New Delhi, India
\end{acknowledgement}

\bibliographystyle{utphys}   
\bibliography{biblio_CDS_noArxiv}

\newpage
\appendix
\section{The ALICE Collaboration}
\label{app:collab}



\begingroup
\small
\begin{flushleft}
J.~Adam\Irefn{org40}\And
D.~Adamov\'{a}\Irefn{org83}\And
M.M.~Aggarwal\Irefn{org87}\And
G.~Aglieri Rinella\Irefn{org36}\And
M.~Agnello\Irefn{org111}\And
N.~Agrawal\Irefn{org48}\And
Z.~Ahammed\Irefn{org132}\And
S.U.~Ahn\Irefn{org68}\And
I.~Aimo\Irefn{org94}\textsuperscript{,}\Irefn{org111}\And
S.~Aiola\Irefn{org137}
% ...
\renewcommand\labelenumi{\textsuperscript{\theenumi}~}

\section*{Affiliation notes}
\renewcommand\theenumi{\roman{enumi}}
\begin{Authlist}
\item \Adef{0}Deceased
\item \Adef{idp25091132}{Also at: M.V. Lomonosov Moscow State University, D.V. Skobeltsyn Institute of Nuclear, Physics, Moscow, Russia}
\item \Adef{idp26170716}{Also at: University of Kansas, Lawrence, Kansas, United States}
\end{Authlist}

\section*{Collaboration Institutes}
\renewcommand\theenumi{\arabic{enumi}~}
\begin{Authlist}

\item \Idef{org1}A.I. Alikhanyan National Science Laboratory (Yerevan Physics Institute) Foundation, Yerevan, Armenia
\item \Idef{org2}Benem\'{e}rita Universidad Aut\'{o}noma de Puebla, Puebla, Mexico
\item \Idef{org3}Bogolyubov Institute for Theoretical Physics, Kiev, Ukraine
\item \Idef{org4}Bose Institute, Department of Physics and Centre for Astroparticle Physics and Space Science (CAPSS), Kolkata, India
\item \Idef{org5}Budker Institute for Nuclear Physics, Novosibirsk, Russia
\item \Idef{org6}California Polytechnic State University, San Luis Obispo, California, United States
\item \Idef{org7}Central China Normal University, Wuhan, China
\item \Idef{org8}Centre de Calcul de l'IN2P3, Villeurbanne, France
\item \Idef{org9}Centro de Aplicaciones Tecnol\'{o}gicas y Desarrollo Nuclear (CEADEN), Havana, Cuba
\item \Idef{org10}Centro de Investigaciones Energ\'{e}ticas Medioambientales y Tecnol\'{o}gicas (CIEMAT), Madrid, Spain
\item \Idef{org11}Centro de Investigaci\'{o}n y de Estudios Avanzados (CINVESTAV), Mexico City and M\'{e}rida, Mexico
\item \Idef{org12}Centro Fermi - Museo Storico della Fisica e Centro Studi e Ricerche ``Enrico Fermi'', Rome, Italy
\item \Idef{org13}Chicago State University, Chicago, Illinois, USA
\item \Idef{org14}China Institute of Atomic Energy, Beijing, China
\item \Idef{org15}Commissariat \`{a} l'Energie Atomique, IRFU, Saclay, France
\item \Idef{org16}COMSATS Institute of Information Technology (CIIT), Islamabad, Pakistan
\item \Idef{org17}Departamento de F\'{\i}sica de Part\'{\i}culas and IGFAE, Universidad de Santiago de Compostela, Santiago de Compostela, Spain
\item \Idef{org18}Department of Physics and Technology, University of Bergen, Bergen, Norway
\item \Idef{org19}Department of Physics, Aligarh Muslim University, Aligarh, India
\item \Idef{org20}Department of Physics, Ohio State University, Columbus, Ohio, United States
\item \Idef{org21}Department of Physics, Sejong University, Seoul, South Korea
\item \Idef{org22}Department of Physics, University of Oslo, Oslo, Norway
\item \Idef{org23}Dipartimento di Elettrotecnica ed Elettronica del Politecnico, Bari, Italy
\item \Idef{org24}Dipartimento di Fisica dell'Universit\`{a} 'La Sapienza' and Sezione INFN Rome, Italy
\item \Idef{org25}Dipartimento di Fisica dell'Universit\`{a} and Sezione INFN, Cagliari, Italy
\item \Idef{org26}Dipartimento di Fisica dell'Universit\`{a} and Sezione INFN, Trieste, Italy
\item \Idef{org27}Dipartimento di Fisica dell'Universit\`{a} and Sezione INFN, Turin, Italy
\item \Idef{org28}Dipartimento di Fisica e Astronomia dell'Universit\`{a} and Sezione INFN, Bologna, Italy
\item \Idef{org29}Dipartimento di Fisica e Astronomia dell'Universit\`{a} and Sezione INFN, Catania, Italy
\item \Idef{org30}Dipartimento di Fisica e Astronomia dell'Universit\`{a} and Sezione INFN, Padova, Italy
\item \Idef{org31}Dipartimento di Fisica `E.R.~Caianiello' dell'Universit\`{a} and Gruppo Collegato INFN, Salerno, Italy
\item \Idef{org32}Dipartimento di Scienze e Innovazione Tecnologica dell'Universit\`{a} del  Piemonte Orientale and Gruppo Collegato INFN, Alessandria, Italy
\item \Idef{org33}Dipartimento Interateneo di Fisica `M.~Merlin' and Sezione INFN, Bari, Italy
\item \Idef{org34}Division of Experimental High Energy Physics, University of Lund, Lund, Sweden
\item \Idef{org35}Eberhard Karls Universit\"{a}t T\"{u}bingen, T\"{u}bingen, Germany
\item \Idef{org36}European Organization for Nuclear Research (CERN), Geneva, Switzerland
\item \Idef{org37}Excellence Cluster Universe, Technische Universit\"{a}t M\"{u}nchen, Munich, Germany
\item \Idef{org38}Faculty of Engineering, Bergen University College, Bergen, Norway
\item \Idef{org39}Faculty of Mathematics, Physics and Informatics, Comenius University, Bratislava, Slovakia
\item \Idef{org40}Faculty of Nuclear Sciences and Physical Engineering, Czech Technical University in Prague, Prague, Czech Republic
\item \Idef{org41}Faculty of Science, P.J.~\v{S}af\'{a}rik University, Ko\v{s}ice, Slovakia
\item \Idef{org42}Faculty of Technology, Buskerud and Vestfold University College, Vestfold, Norway
\item \Idef{org43}Frankfurt Institute for Advanced Studies, Johann Wolfgang Goethe-Universit\"{a}t Frankfurt, Frankfurt, Germany
\item \Idef{org44}Gangneung-Wonju National University, Gangneung, South Korea
\item \Idef{org45}Gauhati University, Department of Physics, Guwahati, India
\item \Idef{org46}Helsinki Institute of Physics (HIP), Helsinki, Finland
\item \Idef{org47}Hiroshima University, Hiroshima, Japan
\item \Idef{org48}Indian Institute of Technology Bombay (IIT), Mumbai, India
\item \Idef{org49}Indian Institute of Technology Indore, Indore (IITI), India
\item \Idef{org50}Inha University, Incheon, South Korea
\item \Idef{org51}Institut de Physique Nucl\'eaire d'Orsay (IPNO), Universit\'e Paris-Sud, CNRS-IN2P3, Orsay, France
\item \Idef{org52}Institut f\"{u}r Informatik, Johann Wolfgang Goethe-Universit\"{a}t Frankfurt, Frankfurt, Germany
\item \Idef{org53}Institut f\"{u}r Kernphysik, Johann Wolfgang Goethe-Universit\"{a}t Frankfurt, Frankfurt, Germany
\item \Idef{org54}Institut f\"{u}r Kernphysik, Westf\"{a}lische Wilhelms-Universit\"{a}t M\"{u}nster, M\"{u}nster, Germany
\item \Idef{org55}Institut Pluridisciplinaire Hubert Curien (IPHC), Universit\'{e} de Strasbourg, CNRS-IN2P3, Strasbourg, France
\item \Idef{org56}Institute for Nuclear Research, Academy of Sciences, Moscow, Russia
\item \Idef{org57}Institute for Subatomic Physics of Utrecht University, Utrecht, Netherlands
\item \Idef{org58}Institute for Theoretical and Experimental Physics, Moscow, Russia
\item \Idef{org59}Institute of Experimental Physics, Slovak Academy of Sciences, Ko\v{s}ice, Slovakia
\item \Idef{org60}Institute of Physics, Academy of Sciences of the Czech Republic, Prague, Czech Republic
\item \Idef{org61}Institute of Physics, Bhubaneswar, India
\item \Idef{org62}Institute of Space Science (ISS), Bucharest, Romania
\item \Idef{org63}Instituto de Ciencias Nucleares, Universidad Nacional Aut\'{o}noma de M\'{e}xico, Mexico City, Mexico
\item \Idef{org64}Instituto de F\'{\i}sica, Universidad Nacional Aut\'{o}noma de M\'{e}xico, Mexico City, Mexico
\item \Idef{org65}iThemba LABS, National Research Foundation, Somerset West, South Africa
\item \Idef{org66}Joint Institute for Nuclear Research (JINR), Dubna, Russia
\item \Idef{org67}Konkuk University, Seoul, South Korea
\item \Idef{org68}Korea Institute of Science and Technology Information, Daejeon, South Korea
\item \Idef{org69}KTO Karatay University, Konya, Turkey
\item \Idef{org70}Laboratoire de Physique Corpusculaire (LPC), Clermont Universit\'{e}, Universit\'{e} Blaise Pascal, CNRS--IN2P3, Clermont-Ferrand, France
\item \Idef{org71}Laboratoire de Physique Subatomique et de Cosmologie, Universit\'{e} Grenoble-Alpes, CNRS-IN2P3, Grenoble, France
\item \Idef{org72}Laboratori Nazionali di Frascati, INFN, Frascati, Italy
\item \Idef{org73}Laboratori Nazionali di Legnaro, INFN, Legnaro, Italy
\item \Idef{org74}Lawrence Berkeley National Laboratory, Berkeley, California, United States
\item \Idef{org75}Lawrence Livermore National Laboratory, Livermore, California, United States
\item \Idef{org76}Moscow Engineering Physics Institute, Moscow, Russia
\item \Idef{org77}National Centre for Nuclear Studies, Warsaw, Poland
\item \Idef{org78}National Institute for Physics and Nuclear Engineering, Bucharest, Romania
\item \Idef{org79}National Institute of Science Education and Research, Bhubaneswar, India
\item \Idef{org80}Niels Bohr Institute, University of Copenhagen, Copenhagen, Denmark
\item \Idef{org81}Nikhef, Nationaal instituut voor subatomaire fysica, Amsterdam, Netherlands
\item \Idef{org82}Nuclear Physics Group, STFC Daresbury Laboratory, Daresbury, United Kingdom
\item \Idef{org83}Nuclear Physics Institute, Academy of Sciences of the Czech Republic, \v{R}e\v{z} u Prahy, Czech Republic
\item \Idef{org84}Oak Ridge National Laboratory, Oak Ridge, Tennessee, United States
\item \Idef{org85}Petersburg Nuclear Physics Institute, Gatchina, Russia
\item \Idef{org86}Physics Department, Creighton University, Omaha, Nebraska, United States
\item \Idef{org87}Physics Department, Panjab University, Chandigarh, India
\item \Idef{org88}Physics Department, University of Athens, Athens, Greece
\item \Idef{org89}Physics Department, University of Cape Town, Cape Town, South Africa
\item \Idef{org90}Physics Department, University of Jammu, Jammu, India
\item \Idef{org91}Physics Department, University of Rajasthan, Jaipur, India
\item \Idef{org92}Physik Department, Technische Universit\"{a}t M\"{u}nchen, Munich, Germany
\item \Idef{org93}Physikalisches Institut, Ruprecht-Karls-Universit\"{a}t Heidelberg, Heidelberg, Germany
\item \Idef{org94}Politecnico di Torino, Turin, Italy
\item \Idef{org95}Purdue University, West Lafayette, Indiana, United States
\item \Idef{org96}Pusan National University, Pusan, South Korea
\item \Idef{org97}Research Division and ExtreMe Matter Institute EMMI, GSI Helmholtzzentrum f\"ur Schwerionenforschung, Darmstadt, Germany
\item \Idef{org98}Rudjer Bo\v{s}kovi\'{c} Institute, Zagreb, Croatia
\item \Idef{org99}Russian Federal Nuclear Center (VNIIEF), Sarov, Russia
\item \Idef{org100}Russian Research Centre Kurchatov Institute, Moscow, Russia
\item \Idef{org101}Saha Institute of Nuclear Physics, Kolkata, India
\item \Idef{org102}School of Physics and Astronomy, University of Birmingham, Birmingham, United Kingdom
\item \Idef{org103}Secci\'{o}n F\'{\i}sica, Departamento de Ciencias, Pontificia Universidad Cat\'{o}lica del Per\'{u}, Lima, Peru
\item \Idef{org104}Sezione INFN, Bari, Italy
\item \Idef{org105}Sezione INFN, Bologna, Italy
\item \Idef{org106}Sezione INFN, Cagliari, Italy
\item \Idef{org107}Sezione INFN, Catania, Italy
\item \Idef{org108}Sezione INFN, Padova, Italy
\item \Idef{org109}Sezione INFN, Rome, Italy
\item \Idef{org110}Sezione INFN, Trieste, Italy
\item \Idef{org111}Sezione INFN, Turin, Italy
\item \Idef{org112}SSC IHEP of NRC Kurchatov institute, Protvino, Russia
\item \Idef{org113}SUBATECH, Ecole des Mines de Nantes, Universit\'{e} de Nantes, CNRS-IN2P3, Nantes, France
\item \Idef{org114}Suranaree University of Technology, Nakhon Ratchasima, Thailand
\item \Idef{org115}Technical University of Ko\v{s}ice, Ko\v{s}ice, Slovakia
\item \Idef{org116}Technical University of Split FESB, Split, Croatia
\item \Idef{org117}The Henryk Niewodniczanski Institute of Nuclear Physics, Polish Academy of Sciences, Cracow, Poland
\item \Idef{org118}The University of Texas at Austin, Physics Department, Austin, Texas, USA
\item \Idef{org119}Universidad Aut\'{o}noma de Sinaloa, Culiac\'{a}n, Mexico
\item \Idef{org120}Universidade de S\~{a}o Paulo (USP), S\~{a}o Paulo, Brazil
\item \Idef{org121}Universidade Estadual de Campinas (UNICAMP), Campinas, Brazil
\item \Idef{org122}University of Houston, Houston, Texas, United States
\item \Idef{org123}University of Jyv\"{a}skyl\"{a}, Jyv\"{a}skyl\"{a}, Finland
\item \Idef{org124}University of Liverpool, Liverpool, United Kingdom
\item \Idef{org125}University of Tennessee, Knoxville, Tennessee, United States
\item \Idef{org126}University of the Witwatersrand, Johannesburg, South Africa
\item \Idef{org127}University of Tokyo, Tokyo, Japan
\item \Idef{org128}University of Tsukuba, Tsukuba, Japan
\item \Idef{org129}University of Zagreb, Zagreb, Croatia
\item \Idef{org130}Universit\'{e} de Lyon, Universit\'{e} Lyon 1, CNRS/IN2P3, IPN-Lyon, Villeurbanne, France
\item \Idef{org131}V.~Fock Institute for Physics, St. Petersburg State University, St. Petersburg, Russia
\item \Idef{org132}Variable Energy Cyclotron Centre, Kolkata, India
\item \Idef{org133}Vin\v{c}a Institute of Nuclear Sciences, Belgrade, Serbia
\item \Idef{org134}Warsaw University of Technology, Warsaw, Poland
\item \Idef{org135}Wayne State University, Detroit, Michigan, United States
\item \Idef{org136}Wigner Research Centre for Physics, Hungarian Academy of Sciences, Budapest, Hungary
\item \Idef{org137}Yale University, New Haven, Connecticut, United States
\item \Idef{org138}Yonsei University, Seoul, South Korea
\item \Idef{org139}Zentrum f\"{u}r Technologietransfer und Telekommunikation (ZTT), Fachhochschule Worms, Worms, Germany
\end{Authlist}
\endgroup

\end{document}